\documentstyle[11pt,az,psfig]{report}

\def\ledd{L_{\rm Edd}}

\newcommand\prad{P_{\rm rad}}
\newcommand\Tmin{T_{\rm min}}
\newcommand\lbb{l_{\rm bb}}
\newcommand\lnet{\Lambda_{\rm net}}
\newcommand\luv{L_{\rm UV}}
\newcommand\lx{L_{\rm x}}

\newcommand\tauc{\tau_{\rm c}}

\newcommand\fx{F_{\rm x}}
\newcommand\fuv{F_{\rm uv}}
\newcommand\fmag{F_{\rm m}}
\newcommand\fdisk{F_{\rm d}}
\newcommand\ftot{F_{\rm tot}}

\newcommand\pmg{P_{\rm mag}}
\newcommand\va{V_{\rm A}}
\newcommand\pr{P_{\rm rad}}
\newcommand\pg{P_{\rm gas}}
\newcommand\ptot{P_{\rm tot}}

\newcommand\lrel{L/L_{\rm Edd}}
\newcommand\simlt{\lower.5ex\hbox{$\; \buildrel < \over \sim \;$}}
\newcommand\simgt{\lower.5ex\hbox{$\; \buildrel > \over \sim \;$}}

\newcommand\msun{{\,M_\odot}}

\newcommand\sigmax{\sigma_{\rm x}}
\newcommand\sigmauv{\sigma_{\rm uv}}
\newcommand\sigmad{\sigma_{\rm d}}
\def\mean#1{\langle #1 \rangle}
\newcommand\dm{\dot{m}}
\newcommand\dms{\dot{m}_{\rm soft}}
\newcommand\dmr{\dot{m}_{\rm rad}}
\newcommand\erg{\rm erg}
\newcommand\cm{\rm cm}

\newcommand\gm{\rm gm}
\newcommand\tautr{\tau_{\rm tr}}

\newcommand\lc{l_{\rm c}}
\def\cm{{\rm\,cm}}

\def\sec{{\rm\,s}}

\def\gm{{\rm\,g}}

\def\erg{{\rm\,erg}}

\def\>{$>$}
\def\<{$<$}

\def\simlt{\lower.5ex\hbox{$\; \buildrel < \over \sim \;$}}
\def\simgt{\lower.5ex\hbox{$\; \buildrel > \over \sim \;$}}
\def\sqr#1#2{{\vcenter{\hrule height.#2pt
      \hbox{\vrule width.#2pt height#1pt \kern#1pt
         \vrule width.#2pt}
      \hrule height.#2pt}}}

	\title{THE PHYSICS OF ACCRETION DISKS WITH MAGNETIC FLARES}
	\author{Sergei Nayakshin} 
	\director{Fulvio Melia}
\begin{document}
       \bibliographystyle{plain}
       \maketitle
       \begin{specialabstract}
       Rapid progress in multi-wavelength observations of Seyfert Galaxies in
recent years is providing evidence that X-ray emission in these
objects may be produced by magnetic flares occurring above a cold
accretion disk. Here we attempt to develop a physically consistent
model of accretion disks producing radiation via magnetic flares as
well as the optically thick intrinsic disk emission, and apply this
model to observations of Active Galactic Nuclei (AGN) and Galactic
Black Hole Candidates (GBHCs). The following issues are considered:
(1) the pressure equilibrium in the flare region, (2) the reflection
and reprocessing of the X-radiation from flares in the underlying
disk, (3) the spectra of GBHCs in the context of the model, (4) and
the generation of the flares by the disk -- the energy budget of the
corona.

\noindent{Our results show that:}

(1) The temperature of the disk atmosphere near active magnetic flares
in AGN is in the range $1-3\times 10^5$ Kelvin, and that the material
is relatively non-ionized. This temperature is in a good agreement
with the observed rollover energy in the Big Blue Bump (BBB) of
Seyfert 1 Galaxies. We thus suggest that the BBB is simply the X-rays
from magnetic flares reprocessed into the X-ray skin of the accretion
disk.

(2) We suggest an explanation for the recently discovered X-ray
Baldwin effect and the controversy over the existence of BBBs in
quasars more luminous than typical Seyferts.

(3) Due to an ionization instability and much higher X-ray incident
flux, we found that the X-ray skin in GBHCs is nearly completely
ionized. Using an approximate model to describe this effect, we
calculated the reflected/reprocessed spectrum and the resulting corona
spectrum simultaneously. We found that the spectrum of GBHCs in their
hard state may be explained with this model, with basically the same
parameters for magnetic flares as in the AGN case.

(4) The magnetic energy transport is shown to be large enough to
account for the observed amount of X-rays from Seyferts and GBHCs. We
predict that X-ray spectra are hard for accretion rates below the
gas-to-radiation transition, and that they are softer above this
transition.

(5) We collected our results into a diagram that shows how the
observational appearance of accreting black holes changes with the
accretion rate and the mass of the hole, and compared it with
observations of AGN and GBHCs.

Our conclusion is that the agreement between theory and observations
is very encouraging and we suggest that the physics of magnetic flares
is the physics that should be added to the standard accretion disk
theory in order to produce a more realistic description of accretion
flows with large angular momentum.

       \end{specialabstract}
       \begin{statement}
       This thesis has been submitted in partial fulfillment of requirements
for an advanced degree at The University of Arizona and is deposited
in the University Library to be made available to borrowers under
rules of the Library.

Brief quotations from this thesis are allowable without special
permission, provided than accurate acknowledgment of source is made.
Requests for permission for extended quotation from or reproduction of
this manuscript in whole or in part may be granted by the head of the
major department or the Dean of the Graduate College when in his or
her judgment the proposed use of the material is in the interests of
scholarship. In all other instances, however, permission must be
obtained from the author.

       \end{statement}
       \begin{acknowledgement}
       
I thank my graduate adviser, Prof. Melia, for years of support and
guidance. I feel indebted to Dr. Edward E. Fenimore for support,
advising and teaching basics of science ethics while in Los Alamos
National Lab (I can say ``I want to be Ed Fenimore when I grow
up''). The faculty and fellow students of the Moscow Institute of
Engineer Physics are thanked most sincerely, since this is where I was
first introduced to theoretical Physics and where my mind often
returns for a source of motivation and high standards of work.  And, of
course, numerous faculty members, stuff and students of the Department
of Physics, the University of Arizona, are being thanked for providing
me with a friendly and productive work environment.

I am also very grateful to my parents for making me who I am and for
their understanding when I first left my home town to study Physics,
and eventually the country to be a physicist. Finally, my wife, Elena,
has been a constant source of happiness and support for me. I thank
her for putting up with all the late nights, working weekends,
conferences, and for delivering our daughter, Sonia, few hours after
my Ph.D. defense.

       \end{acknowledgement}
        \tableofcontents
%
	\chapter{Introduction}
	\section{Existing Theories of Accretion Disks}\label{sect:1}

Accretion Disks are among the most luminous and ubiquitous sources in
Astrophysics, and they have drawn a good deal of attention from the
observing and theoretical communities since the first complete theory
of such disks was formulated by Shakura \& Sunyaev (1973). The disks
are expected to form whenever an interstellar material or wind from
nearby stars is captured by the gravitational attraction of the
central object (a star or a black hole), but may not accrete via
radial in-fall because of the excess angular momentum. From this brief
description, it is evident that this situation is met in a variety of
astrophysical systems.

In addition, accretion disks in Active Galactic Nuclei (AGN) and
Galactic Black Hole Candidates are believed to harbor a black hole --
a very controversial object, a complete understanding of which should
provide the modern physics with new horizons. In order to put
observational constraints on the black hole physics, we need a
thorough understanding of the physics and spectra of ADs. Yet a
convincing accretion disk theory, capable of explaining spectra from
many types of objects where these disks are expected to form, is still
being searched for. The goal of this work is to expand our theoretical
understanding of one of the several existing theories, and to motivate
future work on that model. Our first task is then to briefly describe
the existing theories of accretion disks, and to point out any
difficulties or unresolved questions.

\subsection{Shakura-Sunyaev (Standard) Theory}\label{sect:sth}

Shakura and Sunyaev (1973), and several other workers (e.g., Novikov
\& Thorne 1973), built an accretion disk theory in which
the viscosity of the disk material was parameterized through a
parameter $\alpha$ -- the so-called viscosity parameter. These authors
employed equations for angular momentum conservation, vertical
pressure balance and energy balance between viscous heating and
vertical radiation transport. The radiation field was assumed to be
local blackbody emission. This theory is still the most widely cited
and successful out of accretion disk theories, since it provides a
fair description of AD observations (e.g., Frank et al. 1992, \S 5.7),
especially when the outer part of the disk is concerned.

However, it is clear that in the innermost accretion disk region,
i.e., within $\sim$ few tens $R_g$ ($R_g$ is the gravitational radius,
i.e., $R_g = 2 GM/c^2$, and $M$ is the black hole mass), the model
fails, since observed spectra deviate substantially from simple
blackbody model of Shakura and Sunyaev (1973). Spectrum of almost any
accretion disk system contains a power-law component up to hard
X-rays/soft $\gamma$-rays. In some objects (see Chapter 4), the hard
X-rays dominate the overall energy output. This fact is impossible to
reconcile with the standard theory. Furthermore, the model is
viscously and thermally unstable for high accretion rates, when the
disk pressure is dominated by the radiation pressure.

An extensive theoretical effort went into search for a better theory,
with inconclusive results so far. A simplest modification to the
theory is to assume that the viscosity law in the disk is different
from the one prescribed by the standard theory. For some viscosity
laws this eliminates the disk instability (e.g., Lightman \& Eardley
1974). This does not help to resolve the issue of the spectrum,
however, and so does not constitute a satisfactory model.

\subsection{Two Temperature Model}\label{sect:sle}

The two-temperature disk model was suggested by Shapiro, Lightman
\& Eardley (1976) to explain Cyg~X-1, a Galactic Black Hole Candidate
(GBHC) that exhibited hard X-ray spectrum up to hundreds of keV,
rather than multi-temperature disk blackbody spectrum. The model is
based on the assumptions that electrons and protons are coupled by
Coulomb collisions only. In this case it turns out to be possible for
protons to be much hotter than the electrons. The proton thermal
pressure dominates over the radiation pressure in this model, and the
model is viscously stable. Electron temperature turns out to be such
that the model may explain the hard X-ray spectrum.  For a recent work
on the model, see Misra \& Melia (1996), and further references there.

However, the model is thermally unstable (for a discussion of the disk
instabilities, see Frank et al. 1992, Chapter 5). Furthermore, there
are serious reasons to doubt the plausibility of the assumption of
Coulomb interactions being the only way through which electrons
receive heat (see \S \ref{sect:adaf} below). Finally, the model cannot
be reconciled with the fact that the cold disk stretches all the way
down to the last stable orbit in AGNs (\S \ref{sect:om}).

\subsection{Advection Dominated Accretion Flows}\label{sect:adaf}

Advection Dominated Accretion Flows (ADAF) have recently received a
considerable attention (e.g., Narayan \& Yi 1994, 1995, Abramowicz et
al. 1995a). The model assumes the same Coulomb-only connection between
electrons and much hotter protons. The latter are nearly virialized,
and the proton pressure is large enough to make the accretion disk
geometrically thick (i.e., the disk scale height $H$ is of the order
of the local radius $R$). The gas is optically thin, and the electron
temperature is assumed to be much smaller than the proton temperature,
and so the disk radiates not as effectively as the optically thick
Shakura-Sunyaev disk. Further, since radial velocity $v_R$ scales as
$v_R\sim \alpha c_s H/R\propto \alpha (H/R)^2$, the advective energy
flux is much greater than it is in the standard theory. Due to these
reasons, the advection of energy into the black hole, rather than
local release of energy through radiation becomes possible. The model
predicts that the radiative efficiency of the accretion process is
very low, in contrast to usual value $\sim 0.1$ for the standard
accretion disk theory.

The model has been applied to a number of accretion disk systems and
observations, and has been claimed to be successful in many of these
cases. However, we will argue that, at least in some cases, the
explanations offered are hardly predictive, and should rather be
considered to prove that the model contains enough parameters to
reproduce the main features of observed spectra, if these parameters
are varied in a way that fits the data.

It is of particular concern to us that the model neglects any electron
heating mechanism but Coulomb collisions (see Bisnovatiy-Kogan \&
Lovelace 1997 for a critique of this assumption; also Begelman \&
Chiueh 1988). The magnetic fields are assumed to be important only for
synchrotron emission, which is hardly justified, since magnetic fields
close to the equipartition value are extremely buoyant (see Chapters 2
and 6). When these fields rise out of the disk into a lower density
corona environment, they may reconnect. If the electrons are to stay
much cooler than the protons, all the reconnection energy must be
channeled to {\it and be retained by} protons, which (to my knowledge)
has never been convincingly demonstrated to be the case. This
reconnection process should lead to additional deposition of energy
into the electrons and thus an emission not taken into account in the
ADAF model. For these reasons, we believe that internal consistency of
this model is yet to be proven.

In addition, in the case of AGNs, where observations offer an
invaluable tool -- the fluorescent iron line -- with which to
determine the structure of the disk in the innermost region, the ADAF
model is clearly ruled out since the cold disk must exist as close as
$3 R_g$ from the black hole (\S
\ref{sect:om}). One may argue that disks in GBHCs do not show such a
line, and thus the material there is hot in the inner disk
region. However, in Chapter 4 we will show that the iron line would
not even be produced if the same physical model, developed for the AGN
case, is applied to GBHCs. Summarizing, we see no reason to believe
that ADAF are either internally self-consistent as a theory, or exist
in Nature, except possibly for very low accretion rates ($\simlt
10^{-4}$ of the Eddington value) in some cases.

\subsection{Accretion Disks with Coronae}\label{sect:adwc}

Liang \& Price (1977) were the first to suggest that the X-rays coming
from Cyg~X-1, and other accreting blackholes, are produced in a hot
tenuous corona above the cold disk. Their model was motivated by
observations of hot corona on the Sun and stars in general.  The
model, in its present day version -- the two-phase patchy corona-disk
model -- is consistent with observations of Seyfert Galaxies.  This
model is the subject of our study here, and is considered in detail in
the next section.

\section{Observational Motivation and Fundamentals of the two-phase 
model}\label{sect:om}

Here we present a short summary of the current state of the two-phase
patchy corona-disk model. By this we mean the purely ``empirical''
two-phase model, i.e., the model suggested by observations of Seyfert
Galaxies with no reference to magnetic flares whatsoever. (We call the
model ``empirical'' because, with the exception of Haardt et
al. (1994), one typically makes many key physical assumptions with no
hint of a physical proof -- see \S \ref{sect:deficiencies}. It is only
when one starts to discuss the physics of the model that the necessity
of magnetic fields becomes evident). The purpose of our discussion
here is to let the reader, possibly not very familiar with current
models of the X-ray observations of Seyferts, to see that the
two-phase model is an excellent explanation of the observations and
that it is actually hard to see how a different physical model can
explain the observational facts. Many arguments mentioned in this
section, as well as further references to the literature, can be found
in excellent reviews by Haardt (1996), Maraschi \& Haardt (1996),
Svensson (1996a,b).

Observations of Radio Quiet Seyfert Galaxies show that most of the
radiation power is contained in the two distinct components: the high
energy part -- a power-law with an exponential rollover at around
several hundred keV, and the broad bump between optical to soft X-ray
energies, frequently referred as the Big Blue Bump (BBB, e.g., Walter
\& Fink 1993). In most cases the power emitted in X-rays is comparable
but not larger than that in the optical-soft X-ray band. This fact
alone means that {\it there has to be two phases} in the inner part of
the accretion disk: the hot phase that emits X-rays and the cold one
that produces UVs, since it is well known that most of the accretion
power is liberated at the smallest radii, where the gravitational
energy per particle is the largest. If the inner accretion disk was
composed of just one hot phase, as several accretion disk theories
predict (e.g., Shapiro, Lightman \& Eardley 1976, Narayan \& Yi 1994,
1995), then the UV component could never be dominant because of its
origin in the outer region of the disk.

There are numerous confirmations to this simple energy argument.
Haardt \& Maraschi (1991, 1993) identified the hot phase with a corona
on the top of the cold phase -- the accretion disk, thought to be
reasonably well described by the standard accretion disk theory (e.g.,
Shakura \& Sunyaev 1973). They showed that if most of the energy is
dissipated in the hot corona rather than in the cold disk, then the
resulting spectrum naturally explains many of the observed features in
these sources. In particular, they argued that since the emission
process is roughly isotropic, about half of the coronal X-ray
radiation is directed towards the cold disk, where it gets absorbed
and re-emitted as UV radiation, which then re-enters the corona and
contributes to the cooling of the electrons. Thus, the coronal gas
cooling rate becomes proportional to its heating rate. It is this
proportionality of heating and cooling that makes the inverse Compton
up-scattering of the UV radiation in the corona to produce an almost
universal X-ray spectral index. This ability to reproduce the observed
narrow range in the X-ray spectral index (e.g., according to Nandra \&
Pounds 1994, $\alpha \simeq 1.95\pm 0.15$ for a sample of Seyfert
Galaxies) is one of the strongest points of the model.

Further, the hardening of the spectrum above about 10 keV (Nandra \&
Pounds 1994) was understood as due to the broad hump centered at
$\sim$ 50 keV (e.g., Zdziarski et al 1995). The shape of the hump is
well described by the Compton reflection of the hard X-rays in the
cold disk (e.g., White, Lightman \& Zdziarski 1988, Pounds et
al. 1990). The inferred solid angle of the cold phase as seen from the
corona is a large fraction of $2\pi$, which points to a geometry of
the X-ray source placed above a plane of cold material. Moreover, the
corona plus cold disk geometry is also supported by the fact that
reprocessing of the X-rays into the UV range in the cold disk can
naturally account for the observation of correlated variability of the
UV and X-rays (e.g., Clavel et al. 1992). Additional and significant
support for this geometry comes from observations of the broad iron
K$\alpha$ lines, since the shape of these lines cannot be easily
understood without invoking a cold accretion disk persisting as close
as $\sim 3$ gravitational radii to the black hole (e.g., Reynolds
\& Begelman 1997 and references there).

However, observationally the X-ray luminosity, $\lx$, can be a few
times smaller than the UV luminosity $\luv$.  This is inconsistent
with the two-phase disk- {\it full} corona model, because the latter
predicts about the same luminosity in both X-rays and UV (due to the
fact that all the UV radiation arises as a consequence of reprocessing
of the hard X-ray flux, which is about equal in the upward and
downward directions). To overcome this apparent difficulty, Haardt,
Maraschi \& Ghisellini (1994) introduced a patchy disk-corona model,
which assumes that the X-ray emitting region consists of separate \lq
active regions\rq (AR) independent of each other. In this case, a
portion of the reprocessed as well as intrinsic radiation from the
cold disk escapes to the observer directly, rather than entering ARs,
thus allowing for a greater ratio of $\luv/\lx$. This model is
commonly called the two-phase {\it patchy} corona-accretion disk
model.

\begin{figure*}
\centerline{\psfig{file=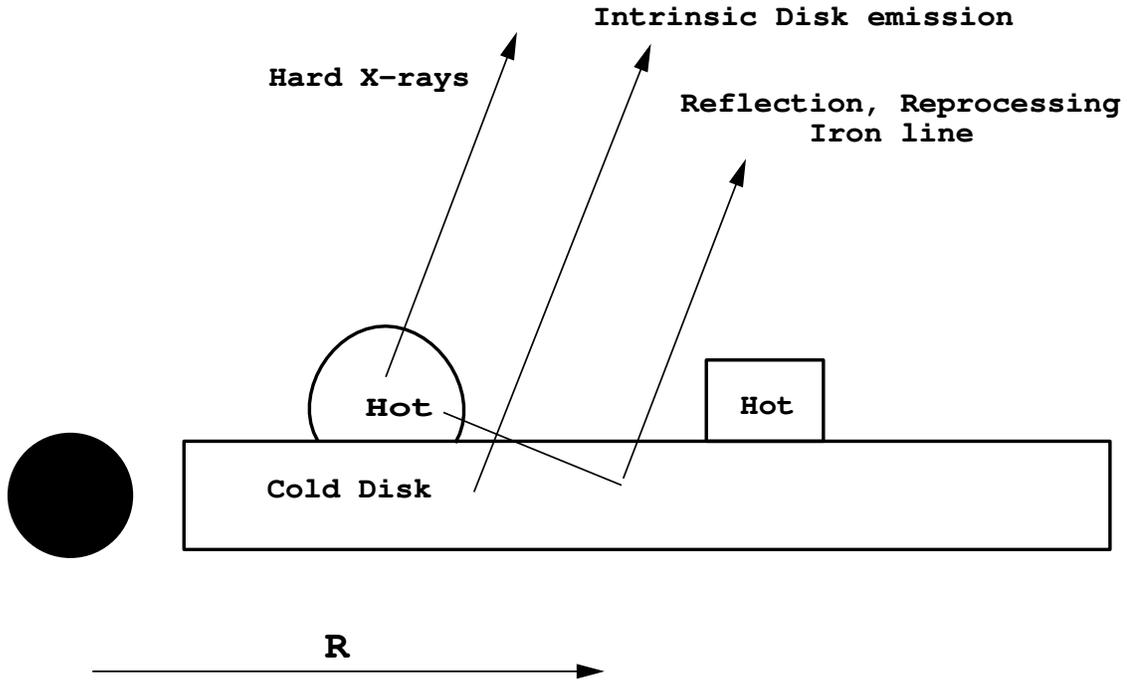,width=1.\textwidth,angle=0}}
\caption{The geometry of the two-phase patchy corona-disk model. The
accretion disk is assumed to be cold up to the last stable orbit, and
X-rays originate in hot active regions above the disk. X-ray flux from
an AR is much greater than the disk intrinsic flux. Reflection of
X-rays off the ``cold'' disk accounts for the Iron K$\alpha$ line and
correlated variability of UV and X-rays.}
\label{fig:2phase}
\end{figure*}

Recently, Stern et al. (1995) and Poutanen \& Svensson (1996) carried
out state of the art calculations of the radiative transport of the
anisotropic polarized radiation, for a range of AR geometries. They
showed that this type of model indeed reproduces the observed X-ray
spectral slopes, the compactness, and the high-energy cutoff {\it if}
the geometry of the source is hemisphere-like rather than a slab.  The
cutoff value is explained as being due to pair equilibria in a hot
mildly relativistic plasma (e.g., Fabian 1994) and requires a high
compactness parameter (for definition see below). The model has very
few parameters, namely, the compactness parameter and the temperature
of the intrinsic/reprocessed radiation from the cold disk.

To summarize this discussion, we show the geometry of the inner
accretion disk learned from the spectral modelling in Figure
(\ref{fig:2phase}). Note that the usual assumption that all the power
is dissipated in the active regions is physically equivalent to saying
that the X-ray flux from the AR substantially exceeds the disk
intrinsic flux, which has to be true only in the immediate vicinity of
the AR. It is then not necessary to transfer most of the disk power to
the corona to reproduce the correct X-ray spectra.

\subsection{Deficiencies of the model}\label{sect:deficiencies}

Despite the considerable (and unmatched by any other theory) success
in the interpretation of observations of Seyfert Galaxies, the
two-phase patchy corona-accretion disk model is often criticized for
its lack of a self-consistent calculation of the physics of the active
regions and the magnetic flux tubes that create these
regions. Basically, it is fair to say that there is no complete and
detailed physical account of how the accretion disk and magnetic
flares can work together. Indeed, even though a very important first
step to provide some base for the model was done by Haardt, Maraschi
\& Ghisellini (1994), who showed that an individual magnetic flare can
sustain a high enough energy release rate, several very crucial
questions were not addressed by the model. In particular, it has never
been questioned whether the two-phase model can provide enough overall
power in X-rays to explain the observations (i.e., enough active
regions at any given time), since the spectral fitting applied to
Seyfert Galaxies addressed the shape of the spectrum, but not its
normalization. In addition, the high energy cutoff of the spectrum is
controlled by the Thomson optical depth of the AR, and the model
assumes that it is given by the pair creation and annihilation
equilibrium. However, this approach avoids consideration of the
pressure balance in the X-ray source, that is, the confinement of the
source. Essentially, one puts particles inside of an artificial rigid
box, which is highly unsatisfying physically (see Chapter
3). Furthermore, due to the fact that covering fraction of the patchy
corona may well be tiny (see \S \ref{sect:cov_frac}), the local X-ray
flux incident on the surface of the cold disk can be larger by several
orders of magnitude than that assumed by all existent X-ray reflection
calculations. This implies that the static X-ray reflection
calculations typically performed are in question when one considers
magnetic flares, and thus the {\it whole} spectral calculation is in
question as well.

\section{Philosophy and Main Goals of This Work}\label{sect:phyl}

As we already discussed, the two-phase patchy corona-disk model enjoys
a considerably success in explaining observations of X-ray bright
Seyfert Galaxies spectra, and yet it is not a fully self-consistent
physical model. The model does not include so far, even though it
urgently needs it, some input of the physics from ``another'' research
field -- the field of magnetic flares. Similarly, the information
gained due to the spectral studies of the two-phase model has not been
appreciated or used by magnetic flare workers. As a matter of fact,
the spectroscopic modelling of X-rays from Seyfert Galaxies and the
theoretical studies of magnetic flares seem to exist independently and
unaware of each other. It is obvious to us that this strange situation
must be changed as soon as possible, if we are to really advance our
understanding of physical processes in accretion disks around black
holes in AGNs and GBHCs. This is the goal of the present work.

We will select problems that are of most interest for current {\it
observations} of accretion disks in AGN and GBHCs. We believe it
should be our primary task to show that the model provides an
appealing framework for many observed phenomena, and that it is time
for a strong theoretical effort to understand magnetic flares in
accretion disks. In line with this plan, we will keep discussion of
the actual magnetic energy release mechanism to a minimum. One reason
for this is that we feel it is quite model dependent, since the
physics of magnetic reconnection is not understood
quantitatively. Second, under certain circumstances, the resulting
spectrum does not depend sensitively on the details of the particle
energising mechanism. This consideration (\S \ref{sect:spectra})
enables us to make some qualitative and quantitative predictions that
are needed in order to compare the theory and observations.

Most of our discussion will be devoted to the connection between
magnetic flares and the accretion disk, since this connection is most
essential when issues of the global spectral behavior of accretion
disks are concerned. Further, we should be honest to note that neither
we nor anybody else for that matter can build the theory of magnetic
flares in accretion disks starting from first principles at this
time. It is advisable and promising to start with observations, and
attempt to understand whether we can see what characteristics the
flares must posses in order to explain these observations. Then, once
we have those constraints, we will try to develop the theory taking
those constraints into account, which will lead to testable
theoretical predictions. Since we plan to use constraints from
observations of such diverse objects as AGN (blackhole mass of $\sim
10^8$ or more Solar masses $\msun$) and GBHCs (blackhole mass of $\sim
10\msun$), our hope is that this will allow us to find any strong and
weak points of the theory.

	\chapter{Magnetic Flares in Accretion Disks: Preliminaries}
        \section{Basics of Magnetic Flare Physics}
\label{sect:chapter2}

One of the important facts learned from observations of turbulent,
differentially rotating fluids is that they generate magnetic fields
(Parker 1979, Priest 1982, Tajima \& Shibata 1997). Another surprising
observation is that these fields are not distributed uniformly in the
fluid, but tend to concentrate into strong magnetic flux tubes, with
magnetic pressure of the order of the ambient gas pressure. For
example, Solar observations show that as much as $90$\% of the overall
Solar surface magnetic energy is in the form of magnetic flux tubes
(see references in Parker 1979, \S 10.1). This concentration of the
field to the flux tubes is truly amazing, since the {\it volume}
average of the magnetic pressure outside the flux tubes is smaller
than the gas pressure (which is about the maximum that magnetic
pressure in the flux tubes can attain) by a factor probably as large
as $\sim 10^6$!

The next well understood (qualitatively, if not quite quantitatively
yet) feature of the magnetic flux tubes in astrophysical plasmas is
that these tubes are buoyant with respect to the fluid that contains
less magnetic field (Parker 1955). The magnetic buoyancy is somewhat
similar to convection. Convection is caused by the fact that a parcel
of gas hotter than its surroundings is less dense due to the pressure
equilibrium between the parcel of the gas and the ambient gas. This
parcel of gas is then lighter and is buoyant with respect to the rest
of the fluid. Similarly, a magnetic field provides pressure, but not
the mass density, thus making the gas possessing the field to be
buoyant. In principle one could construct equilibria such that
magnetic buoyancy is balanced by other forces, e.g., magnetic tension,
but these situations are often found unstable, so that magnetic
buoyancy is always important, as soon as strong magnetic fields exist
(Parker 1979, chapter 13).

\begin{figure*}
\centerline{\psfig{file=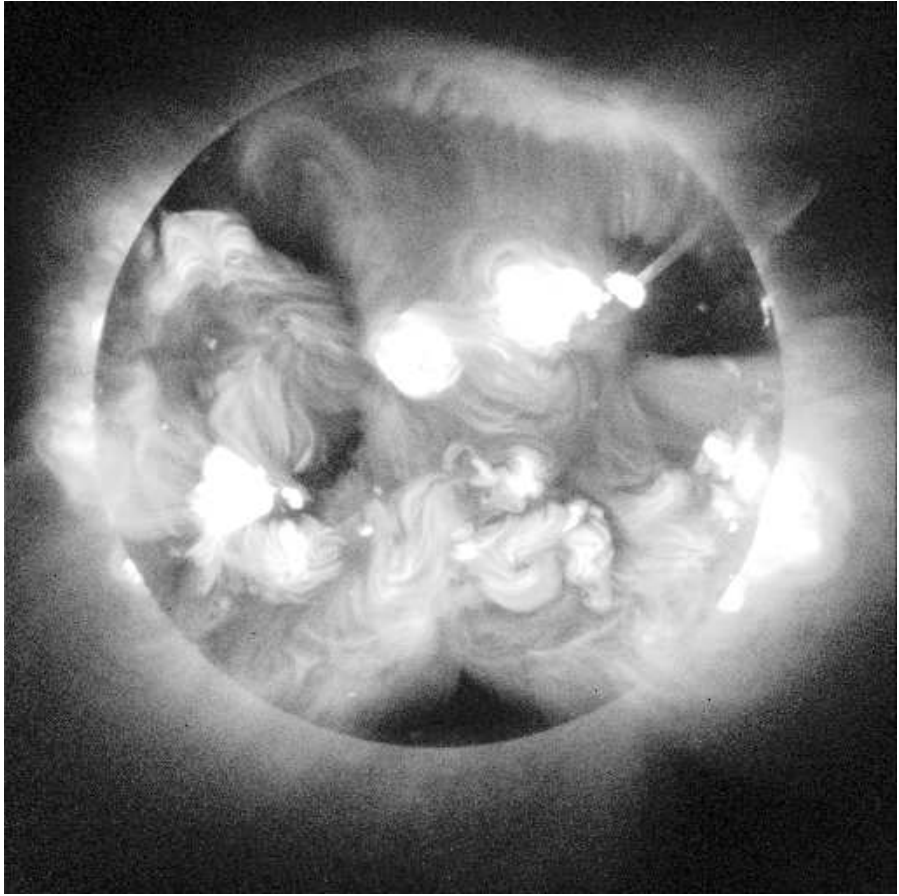,width=.8\textwidth,angle=0}}
\caption{This is an X-ray image of the sun taken at 07:33
UT on 12 November 1991 and is described in the article "The Yohkoh
Mission for High- Energy Solar Physics", by L. Acton, et. al., Science
vol. 258 , 23 Oct. 1992 pp. 618-625.  Picture brightness scales as the
logarithm of intensity. A thin aluminum filter restricted the
instrument bandpass to the 3 to 40 Angstrom wavelength interval ($\sim
$ 0.3 -- 4 keV). The hot ionized gases in the solar atmosphere which
emit in this interval trace the solar magnetic fields extending up
into the corona.}
\label{fig:solar_flare}
\end{figure*}

It is thought that the phenomenon of buoyantly rising magnetic flux
tubes explains appearance of strong and concentrated magnetic fields
above the Solar surface, and eventually, magnetic flares that are
observed (e.g., Tsuneta 1996, Tajima \& Shibata 1997, \S 3.3). Figure
(\ref{fig:solar_flare}) shows an X-ray image of the Sun obtained with
the {\it Yohkoh} Solar mission. Several active magnetic flares are
clearly seen. Note the localized and turbulent nature of the X-ray
emitting corona, which consists of many magnetic loops. Magnetic
flares may be defined as a rapid transfer of magnetic energy to the
gas trapped inside the flux tubes, leading to radiation with photon
energies up to hard X-rays. To understand why the energy release
happens above the Solar or accretion disk surface, and not where the
fields were originally produced, one should notice that Solar or
accretion disk plasmas are ideal in the MHD sense to a large degree,
and thus the magnetic flux (energy) is conserved, that is, it cannot
be transferred to particles. Above the disk, however, the gas density
is very low, and there is a possibility for the reconnection process
(breakdown of ideal MHD). Both theory and observations of reconnection
process (see, e.g., Parker 1979, Priest 1982, Tajima \& Shibata 1997
\S 3.3), the reconnection rate is (see Parker 1979, Priest 1982,
Tajima \& Shibata 1997) proportional to the the Alfv\'en velocity $\va
\equiv B/\sqrt{4\pi \rho}$, where $B$ is magnetic field intensity and
$\rho$ is the gas density. Therefore, inside the flux tubes above the
disk where the gas density is low the reconnection can happen at a
much higher rate than in the mid-plane.

\section{Magnetic Fields and Flares in Accretion Disks}
\label{sect:mfad}

Since it is well established that magnetic flares do occur on the Sun,
it is useful to draw a parallel between Solar and the accretion disk
physical conditions. We shall discuss some of the differences in
radiation mechanisms in Solar and accretion disk flares in \S
\ref{sect:spectra}, but for now it is interesting to just question why
we would expect substantial magnetic field effects in accretion disks
at all. It is believed that plasma {\it differential} motions generate
magnetic fields (e.g., Tajima \& Shibata 1997, \S 3.1). A useful
number is then the ratio of the gas differential velocity $v_d$ to the
sound speed $c_s$. To estimate that, one can take the differential
velocity to be the change in the gas rotational velocity between
points separated by the length scale typical of the region where the
fields are produced. In the case of the Sun, we should take 0.1 of the
Solar radius, since this is probably the depth of the convection
zone. With that we obtained $v_d/c_s\sim 0.01$. In the accretion disk
case, the appropriate typical dimension is the disk scale height (see
Galeev et al. 1979). The standard accretion disk theory then yields
$v_d/c_s\sim 1$. Accordingly, we should expect that accretion disks
are much more magnetically active astrophysical objects than stars.

Galeev, Rosner \& Vaiana (1979) were the first to show that magnetic
flares are likely to occur on the surface of an accretion disk, since
the internal dissipative processes are ineffective in limiting the
growth of magnetic field fluctuations. As a consequence of magnetic
buoyancy, magnetic flux should be expelled from the disk into a
corona, consisting of many magnetic loops, where the energy is stored.
Galeev et al. (1979) also speculated that just as in the Solar case,
the magnetically confined, loop-like structures produce the bulk of
the X-ray luminosity. The X-rays were assumed to be created by Compton
upscattering of the intrinsic disk emission or by bremsstrahlung.

Since then, several workers have elaborated on this subject (e.g.,
Kuperus \& Ionson 1985; Burm 1986; Burm \& Kuperus 1988; Stepinski
1991; de Vries \& Kuijpers 1992; Volwerk, van Oss \& Kuijpers 1993;
van Oss, van Oord \& Kuperus 1993, Field \& Rogers 1993).
Unfortunately, all of these models were very much more complicated
than simpler plasma models, e.g., Poutanen \& Svensson (1996), that
take into account the detailed interaction of particles and radiation
but leave out the question of how the plasma is confined and energy is
supplied. Consequently, there were no agreement even among the
accretion disk magnetic flare theorists as to what spectrum will
result from magnetic flares. Thus, although the magnetic flares above
the cold accretion disk were a recognized possibility for the X-ray
emission from accretion disk, the model lacked predictive power and
was not popular among the observing community.

An important step forward was done by Haardt, Maraschi \& Ghisellini
(1994), who for the first time attempted to connect the physics of
magnetic flares with the observational need for localized active
regions above the disk. These authors showed that physical conditions
for the gas trapped inside a magnetic flare may well be similar to
those required by the two-phase patchy corona-disk model. This really
brought the magnetic flare model on a new, testable level, since with
this work it became clear that due to high compactness parameter (see
\S \ref{sect:tpmodel}) of the plasma in the active regions, the
dominant emission mechanism is Comptonization (see Fabian 1994), which
always leads to a power-law plus exponential roll-over with X-ray
spectral indexes (for this type of geometry) close to those actually
observed in Seyfert Galaxies.  However, the model was still too vague
and had many unresolved questions (e.g., \S \ref{sect:deficiencies}).

\section{Cold Accretion Disk Structure}\label{sect:adstr}

The magnetic flares are ``slaves'' to the underlying accretion disk,
they are created and controlled by it. Our first goal then is to
introduce some sort of accretion disk model that would be compatible
with the well understood accretion process physics (e.g., Frank et
al. 1992) and the presence of magnetic flares. In general it is an
impossible task, but we may approximate the situation by noting that
as long as vertically and time averaged disk quantities are concerned,
the magnetic flares are just another energy transport mechanism (in
addition to the usual radiation flux). We will find that the volume
average magnetic pressure is likely to be rather small compared to the
gas pressure in the disk, so we need not worry about magnetic pressure
effects. The additional energy transport, on the other hand, should be
included in the vertical energy balance equation as an additional
cooling. We found the approach of Svensson \& Zdziarski (1994; SZ94
hereafter) to be the most practical here. These authors considered a
uniform corona above the standard accretion disk, and allowed a
fraction $f\le 1$ of the total local gravitational energy release to
be channeled to the magnetic energy transport, and the rest,
i.e. $1-f$ to be transported via the usual radiation diffusion energy
flux. Since the heating of the disk interior by the incident X-rays is
negligible in both static and flaring corona, the disk structure
should be adequately described by this formalism. The main results of
studies conducted by SZ94 is that such disk plus corona system is
depicted by the standard accretion disk theory ``corrected'' by the
factor $1-f$; the accretion disk is cooler (because energy is vented
away by the flares) and that the disk may be more stable to viscous
and thermal instabilities than the standard disk is for the same
$\dm$. $\dm$ here is the dimensionless accretion rate, defined as $\dm
= \eta
\dot{M}c^2/L_{\rm Edd}$, where $\eta = 0.06$ is the efficiency of
gravitational energy conversion into radiation, $\dot{M}$ is the
actual accretion rate in the physical units, and $\ledd$ is the
Eddington luminosity. Note that in this definition $\dm =1$
corresponds to the total luminosity equal to $\ledd$, and that our
definition of $\dm$ is that of SZ94 times $\eta$.

For our discussions throughout this paper, we will often need typical
numbers for the gas mid-plane temperature $T_d$, the disk effective
temperature $T_{\rm eff}$, ratio of the pressure scale height $H$ to
the radius $R$, and the disk intrinsic flux $F_d$. We write these
quantities below using corresponding equations of SZ94. In writing
down the quantities mentioned above, we will choose $R = 6 R_g$ as a
typical radius where the flares occur. The gas-dominated solution
yields:
\begin{equation}
{H\over R} = 2.25\times 10^{-3} \left(\alpha M_8\right)^{-1/10}
\dm^{1/5} \left[\zeta(1-f)\right]^{1/10}
\label{hoverr}
\end{equation}
\begin{equation}
\fdisk = 2. \times 10^{16} \dm (1-f)M_8^{-1} {\rm erg\ cm}^{-2}
\;{\rm sec}^{-1}
\label{fdisk}
\end{equation}
\begin{equation}
T_d = 2.2\times 10^6 \left(\alpha M_8\right)^{-1/5} \dm^{2/5}
\left[\zeta(1-f)\right]^{1/5}\; {\rm K}
\label{tmid}
\end{equation}
\begin{equation}
T_{\rm eff} = 1.4 \times 10^5 \left[\dm(1-f)\right]^{1/4} 
M_8^{-1/4}\; {\rm K}
\label{teffc}
\end{equation}
and the ``critical'' accretion rate, at which the transition from the
gas-dominated to the radiation-dominated solution takes place, is
$\dm_{\rm cr}$, where
\begin{equation}
\dm_{\rm cr} = 2.2 \times 10^{-3}  \left(\alpha M_8\right)^{-1/8}
\left[\zeta(1-f)\right]^{-9/8}{\rm ,}
\label{mcrit}
\end{equation}
where $\fdisk$ is the disk intrinsic radiation flux, i.e., the one
transported by the usual radiation transport. The parameter $\zeta$
here describes the uncertainty in the radiation flux from the
accretion disk in the vertical direction. This uncertainty is caused
by the usual approximate averaging of the disk equations in the
$z$-direction instead of finding the exact vertical disk
structure. Different authors choose $\zeta$ to lie between $2/3$ and $2$
(see SZ94). We will assume that $\zeta=1$, but will keep in mind that
certain quantities, most notably $\dm_{\rm cr}$, depend rather
sensitively on this poorly determined parameter.

The-radiation-dominated solution gives
\begin{equation}
{H\over R} = 0.3 \dm\,\left(1-f\right)
\label{hoverrr}
\end{equation}

\begin{equation}
T_d = 2.1 \times 10^5 \left(\alpha M_8\right)^{-1/4} 
\left[\zeta(1-f)\right]^{-1/4}\; {\rm K}
\label{tmid}
\end{equation}

\begin{equation}
{\prad\over \pg} = 2.0\times 10^5 \left(\alpha M_8\right)^{1/4}
\, \dm^2 \left[\zeta(1-f)\right]^{9/4} {\rm .}
\label{prpgsz}
\end{equation}

\section{Physical Constraints on the Two-Phase Model}
\label{sect:tpmodel}

As we detailed in \S \ref{sect:deficiencies}, the two-phase patchy
corona-disk model does not provide a full description of the physics
of the active regions. In particular, the plasma in the ARs should be
confined during the active phase, otherwise the energy will be lost to
the expansion of the plasma rather than producing the X-rays. Not
confined, the source would expand at the sound speed, which may be a
fraction of $c$ for these conditions. It is not clear that the
spectrum from such an expanding and short lived source can resemble
anything studied thus far in the literature. The familiar
gravitational confinement, operating in the main part of the accretion
disk, does not work here because there is no mechanism for counter
balancing a side-way expansion of the plasma. Therefore, since there
seems to be no other reasonable possibility for confinement of the AR
plasma, it may be argued that a magnetic field is required to provide
the bounding pressure. Without a magnetic field, the AR would expand
side-ways at least, and form a slab like corona, which was shown to be
incompatible with observations of Seyferts by Haardt et al. (1994) and
Poutanen \& Svensson (1996).

One of the most restrictive and important parameters of the ARs in the
two-phase model is the compactness parameter
\begin{equation}
l\equiv\fx\sigma_T \Delta R/m_e c^3\,{\rm ,}
\label{defcom}
\end{equation}
where $\fx$ is the radiation energy flux at the top of the AR and
$\Delta R$ is its typical size. The compactness parameter is an
indication of the total energy content of the magnetic flare, given
its size. Note that this definition is for the local compactness,
i.e., the one that characterizes the local properties of the plasma,
unlike the global compactness $l_g\equiv L\sigma_T/R' m_e c^3$, where
$L$ is the total luminosity of the object and $R'$ is the typical size
of the region that emits this luminosity.  It is the latter that
should be compared to the observed compactness rather than the former.

A local compactness much larger than unity is required by current
two-phase thermal models of ARs (e.g., Svensson \& Poutanen 1996) in
order to provide a large enough Thomson optical depth due to
electron-positron pairs. However, we note there is no a priori reason
why pair production must be important, other than the fact that under
some conditions it could explain the observed electron temperature
(e.g., Fabian 1994, but see chapter 3). Maciolek-Niedzwiecki,
Zdziarski \& Coppi (1995) have shown that the annihilation of {\it
thermal} electron-positron pairs always produces a broad spectrum of
the resulting photons, and that it is very hard, if possible at all,
to single out this component from the total spectrum on the background
of the dominant Comptonized spectrum. Thus, we really cannot tell
based on the spectra alone whether pairs are abundant on not in the
X-ray producing regions of AGN or GBHCs. In fact, due to spectral
constraints on the X-ray reprocessing, we find (Chapters 4 \& 5) that
the pairs are not likely to be important.

A more stringent constraint on the value of $l$ is the free-free
emission from the active region to be negligible (not to destroy the
two-phase energy balance). The compactness parameter due to
bremsstrahlung emission may be estimated using equation (5.15b) of
Rybicki \& Lightman (1979):
\begin{equation}
l_{\rm brem}\sim 3. \times 10^{-3} \,\tau_T^2\,\Theta^{1/2} {\rm ,}
\label{fab}
\end{equation}
where $\Theta$ is electron temperature in the units of $m_e c^2/k_B$.
For the typical values $\Theta\sim 0.3$, and $\tau_T\sim 1$, this
requires that $l \gg 10^{-2}$. Finally, since the two-phase model was
built under the assumption that the disk intrinsic flux $\fdisk$ is
much smaller than the X-ray flux from the active regions, the
conditions for the applicability of the model are:
\begin{equation}
\fx\gg\fdisk {\rm ,}
\label{cond1}
\end{equation}
\begin{equation}
l\gg 0.01 {\rm .}
\label{cond2}
\end{equation}

\section{Magnetic Flares}\label{sect:mfintro}
\subsection{Geometry}\label{sect:geometry}

\begin{figure*}
\centerline{\psfig{file=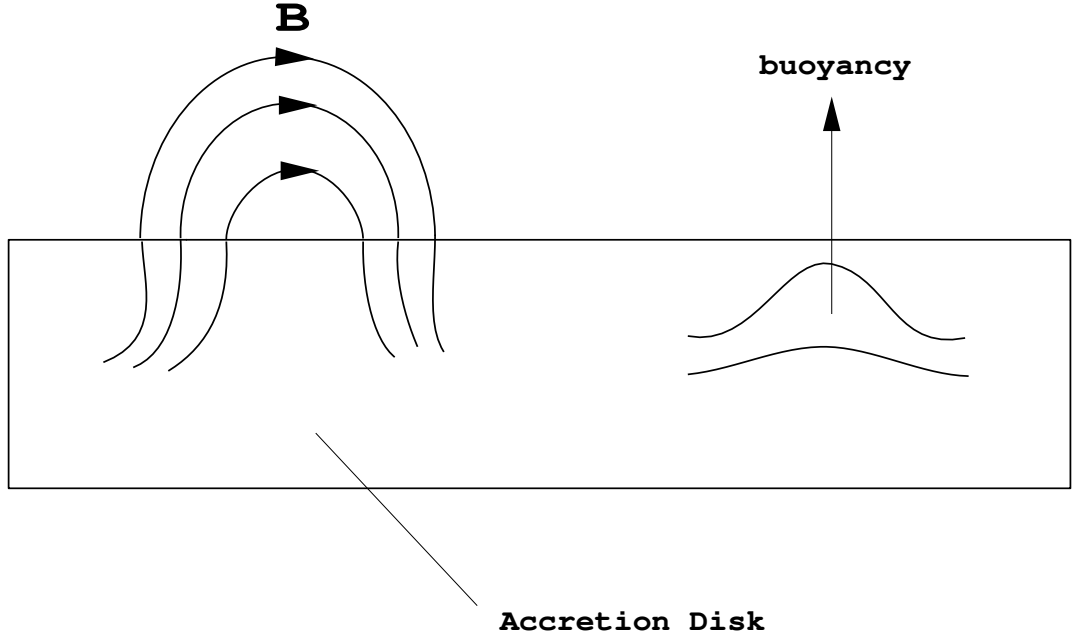,width=.95\textwidth,angle=0}}
\caption{A simple sketch of a magnetic flux tube that broke through
the accretion disk surface due to buoyancy. Notice that magnetic flare
size cannot be much larger than the accretion disk scale height. The
flux tube is ``fat'', since there is little bounding pressure above
the disk, and the field takes a quasi-potential configuration.}
\label{fig:flare}
\end{figure*}

We sketch a typical magnetic flux tube after it has broken up through
the surface of the disk in Figure (\ref{fig:flare}). The part of the
flux tube that is above the accretion disk is the one that produces an
active region. On the right of the Figure (\ref{fig:flare}) we also
show a part of the submerged magnetic flare, which has just started to
develop a buoyancy-unstable region. X-ray Solar observations show that
magnetic flux tubes can be rather quiet for a relatively long time and
then suddenly become active, when they release energy comparable to
the total magnetic energy of the tube. It is this release of magnetic
energy into radiation that is called a ''magnetic flare''. Note that
the geometry of the flares is very similar to what the ARs should look
like (compare Figures \ref{fig:2phase} and \ref{fig:flare}).

\subsection{Compactness of Magnetic Flares}\label{sect:compactness}

We can estimate the maximum compactness of the magnetic flares by the
following considerations (following Haardt et al. 1994). The magnetic
field is limited by the equipartition value in the mid-plane of the
disk. The size of the AR, $\Delta R$, is of the order of one turbulent
cell, which is at most equal the disk scale height $H$ (e.g., Galeev
et al. 1979). Let us assume that the field annihilation occurs on a
time scale $t_{\rm l}$ equal to the light crossing time $\Delta R/c$
times some number $b\sim 10$. This may be justified by noting that the
Alfv\'en velocity can be close to $c$ for these conditions. We will
also assume that the flare occurs at 6 gravitational radii, where the
accretion disk energy generation per unit area has a maximum. Using
the definition of the local compactness parameter and the results of
Svensson \& Zdziarski (1994), we obtain:
\begin{equation}
l \simlt 400 \, {{\dm} \over \alpha b}\; {\varepsilon_{\rm m}
\Delta R \over \varepsilon_{\rm d} H} {\rm ,}
\label{compm}
\end{equation}
where $\alpha$ is the standard $\alpha$-parameter of Shakura-Sunyaev
viscosity prescription, $\varepsilon_{\rm m}$ is the magnetic energy
density in the AR, which must be smaller than the disk mid-plane
energy density $\varepsilon_{\rm d}$. Note that the estimate of the
compactness parameter (\ref{compm}) does not depend on $f$.

For future convenience, we will re-write the above equation as
\begin{equation}
l \simeq 40\,\dm\alpha^{-1}\zeta{\rm ,}
\label{comps}
\end{equation}
where we have collected parameters that we cannot accurately calculate
at this time in a single quantity $\zeta$, which we expect to be of
order of unity. However, due to a very approximate nature of our
method to estimate the compactness parameter, we will use this
expression as a guide which can tell us how $l$ scales with accretion
rate, $\alpha$ and geometry, rather than an exact equation.

\subsection{The X-ray flux}\label{sect:xflux}

For the sake of completeness, and for future reference, we should also
compare the X-ray flux $\fx$ generated by a magnetic flare with
$\fdisk$, the flux from the underlying cold disk, as given by equation
(\ref{fdisk}). A major assumption of the two-phase patchy corona model
is that the X-ray flux greatly exceeds the intrinsic disk flux. If
this assumption does not hold, then the spectrum will be steeper than
the observed Seyfert spectra. The magnetic flare X-ray flux can be
determined from the definition of the compactness parameter (equation
{\ref{defcom}), and equation (\ref{comps}). This yields
\begin{equation}
\fx = 1.\times 10^{17} l \dm^{-1/5}\,M_8^{-9/10}\, 
\alpha^{1/10} (1-f)^{-1/10} {H\over \Delta R}
\,{\rm erg \ cm}^{-2}\, {\rm sec}^{-1} {\rm .}
\label{flux}
\end{equation}
One can see that the flux from an active region is very much larger
than the disk intrinsic flux, which parenthetically means that the
total area covered by magnetic flares should be much smaller than the
total inner disk area. This is similar to the Solar X-ray emission,
where X-rays come from localized magnetic flares rather than uniformly
from the whole disk surface (see Figure \ref{fig:solar_flare})

\subsection{Number of Flares and Variability}\label{sect:number}

There are a few other simple but quite valuable estimates that we can
make based on the simple energy budget argument. One of these is the
average number of active magnetic flares, $N$. To determine it, we
will require that $N$ times the luminosity of a single flare equals
the overall X-ray luminosity observed from a source. If we assume that
almost all X-rays impinging on the cold disk are reprocessed into the
UV range, which is a good approximation for AGNs (see Chapters 4 \&
5), the X-ray luminosity of the corona-disk system is approximately
equal to (Haardt \& Maraschi 1991, 1993)
\begin{equation}
L_{\rm x} = (f/2) \, L
\label{lx}
\end{equation}
where $L$ is the bolometric luminosity of the source, and $f$ is the
fraction of power transfered to the corona. To find the luminosity
$L_1$ of a typical flare, we can use the definition of the compactness
parameter (equation
\ref{defcom}), and express $L_1$ as $L_1\sim \Delta R^2 \fx$. Working
through some simple algebra, one obtains
\begin{equation}
N\sim {f L\over 2 L_1} \sim 10^3 \left({\dm\over 10^{-3}}\right)
\left({R \over 10^{3} \Delta R }\right) l^{-1} (f/2)
\label{numb}
\end{equation}
Here we have scaled $\dm$ and $\Delta R/R$ on values typical for these
quantities in Seyfert Galaxies. The often observed short time scale
variability of the X-ray continuum from Seyferts by a factor of 2 can
be explained by random fluctuations in the number of flares if $N\sim
10$ (e.g., Haardt et al. 1994), which would then suggest that $l\sim
100$. However, one should not forget that flares are controlled by the
accretion disk, so, it is possible that the disk modulates the
appearance of the flares and thus flares may be not statistically
independent events. Thus, the estimate of $N$ based on X-ray
variability needs to be put to a serious test before we could
constrain $l$ based on it.

Furthermore, the continuum variations happen on time scales from a few
hundred to $\sim 10^5$ sec (e.g., Done \& Fabian 1989), which may be a
very short time scale for a massive AGN. For example, the light
crossing time of one gravitational radius is $R_g/c = 10^3 M_8$ sec,
where $m_8\equiv M/10^8 \msun$. Since any model of X-ray emission from
AGNs should be able to reproduce variations on the observed time
scales, large scale (i.e., $\sim R_g$) emission regions are ruled
out. The magnetic flare model is in a better shape here, since the
emission regions are very small in size ($H/R\sim 10^{-3}$). To
estimate the typical flare life time $t_l$, we can write $b H/c \simlt
t_l \simlt 2\pi R/v_{\phi}$, where $v_{\phi}$ is the Keplerian
velocity. The lowest limit here is equal to $b\geq 10$ light crossing
times of the flare region, and the upper limit is equal to one
dynamical time scale, i.e., the orbital time scale. This reasoning
yields
\begin{equation}
60 \;{\rm sec}\; M_8 \,\simlt\, t_l \, \simlt 2\times 10^4 \; {\rm
sec}\; M_8 {\rm .}
\label{tlife}
\end{equation}
This estimate shows that the typical life time of a flare is in the
range of observed variability time scales (see also Galeev et al.
1979 and de Vries \& Kuijpers 1992).

\subsection{Covering Fraction}\label{sect:cov_frac}

The other useful number which we may obtain from simple energy budget
considerations is the covering fraction of the magnetic flares, i.e.,
the fractional area $f_c$ of the inner accretion disk surface that is
covered by active magnetic flares at any time. This fraction may be
found by noting that the product of the area covered by magnetic
flares $A_m$ and the typical X-ray flux should be equal to $L_{\rm x}
= (f/2) L$, whereas the product of the disk intrinsic flux $F_d$ and
the disk area $A_d$ should be equal to the disk luminosity $(1-f)
L$. Thus,
\begin{equation}
f_c\equiv {A_m\over A_d}\simeq {f\over 2(1-f)} \;{F_d\over F_x} =
10^{-1} l^{-1} \, \dm^{6/5} \, f \,(1-f)^{1/10}\;\left(\alpha
M_8\right)^{-1/10}\,{\Delta R\over H} {\rm .}
\label{fcov}
\end{equation}
Thus, the covering fraction may be quite small for small accretion
rates.

\subsection{Spectra from Magnetic Flares}\label{sect:spectra}

As a first guess, one would think that it is extremely challenging to
accurately compute the spectrum from such a complicated phenomenon as
a magnetic flare.  The spectrum will depend on the geometry and the
unknown distributions of the gas density and temperature in the
flaring region. These distributions are dependent on the model assumed
for the reconnection mechanism and other factors, and are {\it
absolutely impossible} to uniquely determine at the present time. In
our opinion, this apparent uncertainty in the spectrum is the foremost
important reason why the magnetic flares have not been firmly
established as a source of X-rays from accretion disks after decades
of theoretical studies.

However, the X-ray spectrum from accretion disk magnetic flares should
be similar to that of a static active region of the two-phase model of
the same size and compactness, as long as the lifetime of the flare
exceeds several light-crossing time scales. The repeated inverse
Compton upscattering mechanism produces {\it always a featureless}
X-ray spectrum -- a power-law with a quasi exponential roll-over --
the form of the intrinsic active region spectrum used to fit the
observations of Seyfert Galaxies by Zdziarski et al. (1995).  To
emphasis how well Comptonized spectra hide the nature of physical
processes creating them (and the geometry of the source), we point out
that the featurlessness of the spectrum arising from Comptonization
was even cited as a principal problem in inferring the shape of the
underlying electron distribution via comparison of the spectra
produced by thermal and non-thermal Comptonization (Ghisellini, Haardt
\& Fabian 1993). This is even more so if the observed spectrum
consists of many separate contributions from flares with different
parameters. Thus, the spectrum of the magnetic flares is adequately
described by the two-phase patchy corona model with a correctly chosen
geometry and compactness parameter.

The major difference between the Solar magnetic flares and those
occurring on the surface of accretion disks is the compactness
parameter. For the Solar case, $l\ll 1$, and can be as small as $l\sim
10^{-6}$, as one can check using typical time scales for the energy
release and the overall energy content of magnetic flares (e.g.,
Priest 1982). For the accretion disk magnetic flares, the compactness
parameter can be considerably larger, $l\gg 0.01$ (see section \S
\ref{sect:compactness}). According to \S \ref{sect:tpmodel}, this implies that
bremsstrahlung is far more important than Comptonization for the Solar
flare spectra.  The other important difference is the Alfv\'en
velocity, which is substantially higher for magnetic flares in disks.

	\chapter{Pressure Equilibrium and Containment}
	\section{Observational Motivation}\label{sect:obstau1}

Haardt et al. (1994) suggested that the active regions (ARs) may be
magnetic flares occurring above the accretion disk's atmosphere and
showed that their compactness $l$ may be quite high ($\sim 30$), so
that pairs can be created. In principle, it is possible to obtain
still larger values for the compactness parameter ($l\sim$ few
hundred), thus creating enough pairs to account for the observed
$\tau_T\simeq 1.0^{+0.4}_{-0.2}$ (Zdziarski et al. 1996), where
$\tau_T$ is the Thomson optical depth of the AR. This explanation for
the observed value of $\tau_T\sim 1$ based on the pair equilibrium
condition, however, relies on the assumption that the particles are
confined to a rigid box, so that no pressure constraints need to be
imposed. This is unphysical for a magnetic flare where the particles
are free to move along the magnetic field lines. Therefore, as far as
the two-phase model {\it without} a proper pressure equilibrium
condition is concerned, the Thomson optical depth is a parameter,
rather than a calculable quantity.

Here we will consider the pressure equilibrium during an intense
magnetic flare occurring above the surface of a cold accretion
disk. Assuming that the heating source for the plasma trapped in the
flaring region is the energy transported by magnetohydrodynamic waves
or energetic particles with group velocity close to the speed of
light, we show that under certain conditions the pressure equilibrium
constrains the Thomson optical depth $\tau_T$ of the plasma to be in
the range $1-2$. We suggest that this pressure equilibrium may be
responsible for the observed value $\tau_T\sim 1$ in Seyfert Galaxies.
We also consider whether current data can distinguish between the
spectrum produced by a single X-ray emitting region with $\tau_T\sim
1$ and that formed by many different flares spanning a range of
$\tau_T$. We find that the available observations do not yet have the
required energy resolution to permit such a differentiation.  Thus, it
is possible that the entire X-ray/$\gamma$-ray spectrum of Seyfert
Galaxies is produced by many independent magnetic flares with an
optical depth $0.5<\tau_T<2$.

\section{The Connection Between The Energy Supply Mechanism and
Pressure Equilibrium}\label{sect:prt}

The `universal' X-ray spectral index of Seyfert Galaxies (e.g., Nandra
\& Pounds 1994) suggests that the emission mechanism is thermal
Comptonization with a $y$-parameter close to one (Haardt \& Maraschi
1991; Haardt \& Maraschi 1993; Fabian 1994). The $y$-parameter is
defined here as the average photon fractional energy gain times the
average number of scatterings that the photon suffers before it
escapes to infinity (e.g., Rybicki \& Lightman 1979, Chapter 7). For
the emission to be dominated by Comptonization, the compactness
parameter needs to be large (e.g., Fabian 1994, and \S
\ref{sect:tpmodel}). Observations of Seyfert Galaxies point to a
global compactness parameter $\sim 1-100$ (Svensson 1996 and
references cited therein).

For electrons and protons at a single temperature $T_e$ and an
electron number density $n_e$ with the assumption of neutrality, the
gas pressure is $2 n_e kT$.  The radiation energy density may be
recast in terms of the luminosity $L$ of the source, and therefore its
compactness parameter $l$, under the assumption that the typical
photon escape time is given by the light crossing time multiplied by
$1+\tau_T$ (see Rybicki \& Lightman 1979). The total pressure is
\begin{equation}
P = {m_e c^2\over \sigma_T R}\,\left [ 2\tau_T \,\Theta_e +
l (1+\tau_T)/3\,\right ]{\rm ,} 
\label{ple}
\end{equation}
where $\Theta_e\equiv k T_e/m_e c^2$ is the dimensionless electron
temperature. The Compton $y$-parameter for thermal electrons is $y=
4\Theta_e \tau_T (1+\tau_T)(1+4\Theta_e)$  and is of order 1 (e.g., Haardt \& Maraschi 1991). Thus, since
the dimensionless gas pressure in Equation (\ref{ple}) is always
smaller than $y$, the radiation pressure dominates over the gas
pressure in a one-temperature plasma when $l\gg 1$.

One consequence of this is that the amount of energy escaping from the
source even during one light crossing time is larger than the total
particle thermal energy. Thus, there must be an agent that energizes
the particles to enable them to radiate at this high rate, and the
presence of this agent must be dynamically consistent with the state
of the system.  We foresee two possibilities for the nature of this
`agent': (i) the gravitational field, and (ii) an external flow of
energy into the system. These two cases are quite distinct physically.

Insofar as the first possibility is concerned, the gravitational
potential energy of the plasma (primarily that of the protons) is
dissipated as the gas sinks deeper into the well of the black hole.
The gravitational field does not provide a pressure, but it does
compress the gas.  However, this leads to an internal (radiation plus
gas) pressure that varies from source to source as the physical
conditions change.  There does not appear to be a scale that sets
$\tau_T$ to have a value of $1$.  For example, in standard accretion
disk theory, the inner radiation pressure-dominated regime has an
optical depth that depends on several parameters, such as the
accretion rate and the $\alpha$-parameter (Shakura \& Sunyaev 1973).
The $\alpha$-parameter reflects the rate at which the protons `use up'
their potential energy, and so a change in this rate leads to a change
in the equilibrium optical depth.  It is even less obvious why
$\tau_T$ should be $\sim 1$ in the gas pressure dominated regimes
since there the pressure has no reference to $\tau_T$ at all.  It
seems that when the pressure equilibrium is dictated by the
gravitational field (e.g., due to a compression of the X-ray emitting
region), the Thomson optical depth should span a range of values
depending on the source geometry, the specific parameter values and
the particle interactions assumed to operate in the source.

This is not so when the energy is supplied to the X-ray emitting
region by an inflow of energy, e.g., via a magnetic field.  The
principal difference between the two cases is that the dynamic portion
of the magnetic field supplies a ``ram'' pressure that is related in a
known way to its energy density. If the magnetic energy flux into the
X-ray emitting region is known, this also constrains the inwardly
directed momentum flux (the compressional force) into the system.
Thus, the compressional force exerted on the active region by the
magnetic field is expected to correlate with the source luminosity.
What makes this useful in terms of setting the optical depth of the
system is that a similar correlation exists between the luminosity and
the outwardly directed radiation pressure in the emitting region.  But
in this case, the pressure also depends on $\tau_T$.  Assuming a
spherical geometry for simplicity, the radiation pressure is $P_r
\simeq \tau_T F_r/c$, where $L\approx 4\pi R^2 F_r$ in terms of the
source radius $R$ and radiation flux $F_r$.  Thus, since all the
balance equations are to first order linear in $F_r$, it is
anticipated that the pressure and energy equilibria of the system
point to a unique value of $\tau_T$.  We explore this possibility in
the next section.

\section{Pressure Equilibrium For Externally Fed Sources}
\label{sect:efs}

Let us first suppose that the X-ray source is a sphere with Thomson
optical depth $\tau_T$, and that the energy is supplied radially by
magnetohydrodynamic waves.  The waves carry an energy density
$\varepsilon$ and propagate with velocity $v_A$.  For definitiveness,
we assume that these are Alfv\'en waves, in which case the momentum
flux that enters the X-ray source is $(1/2)\,\varepsilon$.  The
magnetic energy of the Alfv\'en waves is in equipartition with the
oscillating part of the particle energy density, and so we can
estimate the gas pressure as being of the same order as the ram
pressure of the oscillating part of the magnetic field,
i.e. $(1/2)\,\varepsilon$.  Finally, we assume that all of the wave
energy and momentum are absorbed by the source.

The energy equilibrium for the AR is then given by 
\begin{equation}
F_r = \varepsilon v_A\;,
\end{equation}
whereas in pressure equilibrium 
\begin{equation}
P_r\simeq \tau_T F_r/c \simeq \varepsilon\;.
\end{equation}
Dividing the latter equation by the former, one obtains for the
equilibrium Thomson optical depth:
\begin{equation}
\tau_T \simeq {c\over v_A}
\end{equation}
This value does not depend on luminosity, but it does of course depend
on the geometry and $v_A$. To explain observations, we need $v_A\sim
c$ (see \S \ref{sect:influx} below).

Suppose now that the geometry is not perfectly spherically symmetric,
and that instead the Alfv\'en waves can enter the X-ray source through
an area $A_a$, but the radiation leaves through an area $A_r \simgt
A_a$, which is plausibly just the total area of the AR. This situation
may occur if part of the X-ray source is confined by other than the
Alfv\'en wave ram pressure, e.g., by the underlying (non-dynamic)
large-scale magnetic field (see below). In this case, since the energy
balance is now $F_r A_r = \varepsilon v_A A_a$, the equilibrium
$\tau_T$ is changed to
\begin{equation}
\tau_T \simeq \left({c\over v_A}\right)
\left({A_r\over A_a}\right)\;.
\end{equation}

To understand the scale represented by the bracketed quantities in
this equation, let us consider the physical conditions that are likely
to be attained during a short-lived and very energetic magnetic flare
above the standard $\alpha$-disk.  The magnetic field energy density
is a fraction of the underlying disk energy density and the typical
size $\Delta R_a$ of the flare is expected to be of the order of the
disk scale height (Galeev et al. 1979; Haardt et al. 1994).  Now, the
confinement of the plasma inside the flare, and the observed condition
$l\gg 1$, require that $B^2/8\pi \gg P_r
\gg P_g$. Since the magnetic stress is much larger than any other
stress, the magnetic flux tube adjusts to be in a stress-free vacuum
configuration. The tube is thick (meaning that its cross sectional
radius is of the order of its length), since the pressure in the
disk's atmosphere is insufficient to balance the tube magnetic field
pressure.  This is due to the fact that the magnetic field is
presumably anchored in the disk's mid-plane, where the pressure is much
greater than the atmospheric pressure.  The magnetic waves propagate
upwards along the magnetic flux tube, while radiation pressure from
the AR is pushing the gas along the magnetic lines, i.e., downwards to
the disk. This downward direction of the radiation pressure arises
naturally in a two-phase model (unlike the situation within the
accretion disk) since here most of the energy is released above the
disk's atmosphere (see also Nayakshin \& Melia 1997b)

With this in mind, we may now describe heuristically how the magnetic
flare develops and how pressure equilibrium is established. As is well
known (Parker 1979; Galeev et al. 1979), magnetic flux tubes are
buoyant in a stratified atmosphere, and so they rise to the surface of
the accretion disk. As the tube is rising, the particles slide along
the magnetic field lines downward to the disk in response to
gravity. The magnetic flux tube becomes more and more particle-free,
$v_A$ is increasing, and so the conditions become more and more
favorable for the dissipation of magnetic field energy. We assume that
magnetohydrodynamic waves are generated and propagate up to the top of
the flux tube, where they are absorbed and produce highly energetic
particles.  The particles in turn produce X-radiation by up-scattering
the UV radiation from the disk.  Since the radiation pressure $P_r$ is
very much smaller than the stress in the underlying magnetic flux
tube, we may neglect the sideways expansion of the flux tube. We need
to consider the pressure equilibrium along the magnetic field lines,
however, since the plasma can in principle move freely in that
direction.  The balance of radiation pressure with the magnetic ram
pressure then sustains the AR optical depth as discussed above.  Since
the flux tube is geometrically thick, the corresponding ratio
$A_r/A_a$ is probably of order $\sim $ one to a few, and with $v_A\sim
c$, we therefore expect
\begin{equation}
\tau_T\sim 1 - 2\;.
\label{etv}
\end{equation}
The lowest values of the equilibrium $\tau_T$ can be reached due to
the fact that $A_r$ in this equation is not necessarily the total area
of the source, because some of the X-ray flux can be reflected by the
underlying disk and re-enter the AR. Some of this re-entering flux can
be parallel to the incoming Alfv\'en energy flux, and thus the
effective $A_r$ is smaller than the full geometrical area of the
source. Furthermore, we have assumed a one-temperature gas, and have
neglected the gas pressure in our calculation. It is possible that the
protons are much hotter than the electrons, and that they account for
a sizable fraction of the total pressure in the AR, which then leads
to a reduction in the value of the equilibrium $\tau_T$ as compared to
Equation (\ref{etv})

\subsection{Influx of relativistic particles as the energy supply mechanism}
\label{sect:influx}

In the analysis developed here, there is nothing specific to Alfv\'en
waves.  We could have instead invoked the influx of energy by other
waves or even energetic particles accelerated by a magnetic
reconnection process.  All that matters is that they have a
well-defined relationship between their momentum and energy densities,
and that their propagation speed is close to $c$. For example, if the
reconnection process takes place in the apex of a magnetic flux tube,
where the gas density is very low, the gas may be accelerated to
relativistic velocities. These relativistic particles will most likely
travel downwards (e.g., Field \& Rogers 1993) to the flare foot-points
(along magnetic field lines). If pairs number is not too great, the
mass density will be dominated by the protons, and so will be the
energy and momentum densities of the reconnection flow. Now, the
protons do not interact efficiently with radiation from the
disk. Thus, the {\it bulk} relativistic flow of particles does not
radiate its energy until it impacts the higher density regions in the
foot-points. There, the gas bulk motion will be stopped and the energy
will be deposited in the gas random thermal motions. Electrons and
protons will most likely thermalized and the electrons will reach
equipartition with the protons (the electrons and protons are likely
to interact through electromagnetic fields rather than by Coulomb
collisions, since the flux tube magnetic field is very large compared
to the gas thermal energy). The electron thermal energy may now be
radiated away by the most efficient emission mechanism, which is
Comptonization of the disk radiation if $l\gg 10^{-2}$. In this
scenario, the active region is squeezed between the incident energetic
particles and the underlying denser layer of the disk that effectively
acts as a wall, since the gas pressure rapidly increases in the
downward direction in the disk.

Furthermore, there are independent arguments that favor the second
physical setup. Let us estimate the Alfv\'en velocity $v_A$ for a
magnetic flare with compactness parameter $l$, by requiring that the
total energy radiated away during the flare life time $b\Delta R/c$ is
smaller than the total magnetic energy content in the volume $\sim
(4\pi/3) \Delta R^3$ (cf \S \ref{sect:compactness}):
\begin{equation}
{v_A\over c}\simeq 0.1 \times \,\left [ {b\over 10} \; {l}\;
\tau_T^{-1}\right]^{1/2}\; .
\label{va}
\end{equation}
It is seen that $v_A$ may be quite close to $c$ only if $l\sim 100$
(if $v_A$ as given by equation \ref{va} exceeds $c$, the relativistic
corrections, which we did not include here, will permit it to approach
$c$ only). At the same time, by considering X-ray reflection
calculations, we found relatively strong spectral constraints that
limit the compactness parameter to values not greater than $\sim 1$
(see \S \ref{sect:sconstraints}). 

We do not see how to remedy this problem for the case of energy
transport by MHD waves. On the other hand, for the case of the active
region situated at the flux tube foot-points, there might be a natural
reason why $v_A/c$ is larger than $\sim 0.1$. Indeed, the point here
is that the gas density of the particles in the apex of the tube may
be much lower than that in the active region, so that $\tau_T\ll 1$
{\it in the apex}, where the reconnection takes place. Equation
(\ref{va}) then allows larger values of $v_A/c$, and we see no
fundamental reason why $\tau_T$ cannot be as small as $\sim 10^{-2}$,
thus permitting $v_A\sim c$. We believe this second scenario may be
more realistic, since it is in agreement with Solar magnetic flare
theory and observations, where it is believed that magnetic flares are
powered by a reconnection process rather than by some sort of MHD wave
heating from below (e.g., Innes et al. 1997, Klimchuk 1997, Blackman
1997).

\section{The Range in $\tau_T$ Permitted by Current Observations} 
\label{sect:rtau}

Zdziarski et al. (1996) produced a fit of the average {\it Ginga}/OSSE 
spectrum of Seyfert 1 galaxies assuming that the active regions form 
hemispheres above the disk.  They found that the radial optical 
depth of the hemispheres is $\tau_T\sim 1$. Here, we will
examine whether the Seyfert spectrum can be due to a combination
of spectral components from flares with different $\tau_T$, but the
same $y$-parameter (set arbitrarily at $1.3$). The latter assumption is
introduced to ensure that the X-ray spectral index does not vary considerably 
from flare to flare. A constant $y$-parameter is a natural consequence of 
the fixed geometry of the flare, in the sense that the cooling
of the plasma is fixed by how much of the X-ray flux re-enters the emitting
region after it is reflected from the disk (see Haardt \& Maraschi 1991).

As an illustration of the method, we first compute the spectrum from
flares with a range of Thomson optical depths assuming that they all
have the same luminosity.  We then convolve these spectra with a
Gaussian probability distribution that a flare occurs with $\tau_T$.
The composite spectrum $F(E)$ (in energy/sec/keV) is
\begin{equation}
F(E) = \int_{0}^{\infty}\, d\tau_T\; \exp\left [-
{(\tau_T - \tau_0)^2\over \tau_{\sigma}^2}\right ]
F(E, \tau_T)\,{\rm ,}
\end{equation}
where $F(E, \tau_T)$ is the spectrum from a single flare with
$\tau_T$.  We take $\tau_0 = 1.14$ and adopt several values of
$\tau_{\sigma}={\rm const}$ to represent the possible spread in
$\tau_T$ between different flares.  The individual spectra are
computed assuming a slab geometry using an Eddington
frequency-dependent approximation for the radiative transfer, using
both the isotropic and first moments of the exact Klein-Nishina cross
section (Nagirner \& Poutanen 1994). Although this geometry is clearly
different from that of a realistic flare, our point here is to test
the possibility of co-adding spectra with different $\tau_T$, in order
to see what range in $\tau_T$ may be permitted by current
observations.  We expect that a more accurate calculation with the
correct geometry will yield qualitatively similar limits on $\tau_T$,
though the exact values should be inferred using a $\chi^2$ fit to the
data.

\begin{figure*}[h]
\centerline{\psfig{file=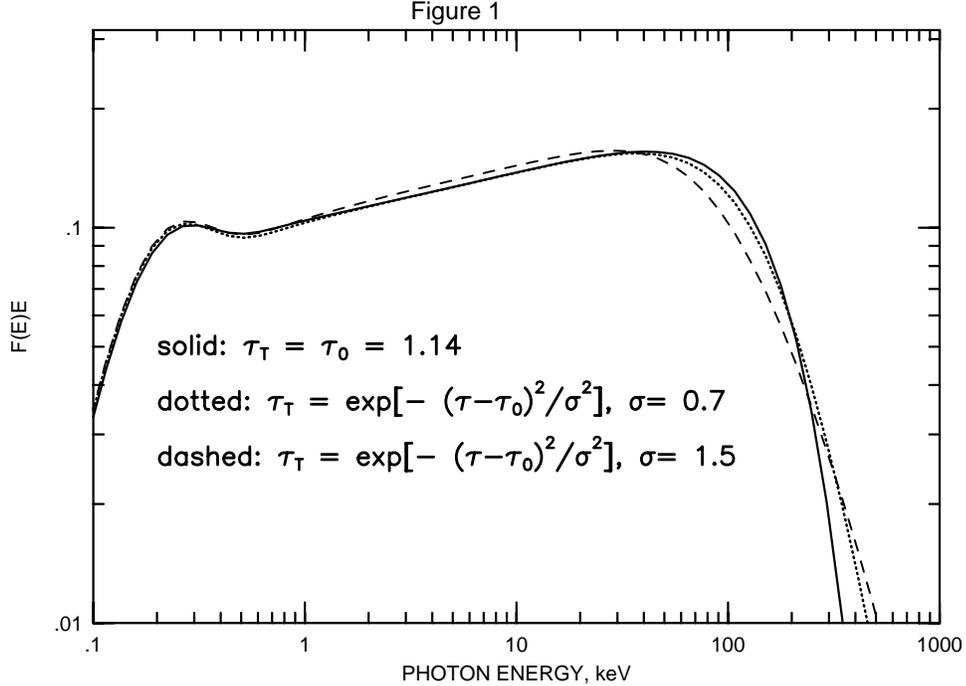,width=.85
\textwidth,angle=90}}
\vskip 0.7cm
\caption{The spectrum resulting from co-adding the
components due to different magnetic flares with a Gaussian
distribution in the Thomson optical depth (see text), centered on
$\tau_0 = 1.15$ with a width $\tau_{\sigma}=0.7$ (long-dashed curve)
and $1.5$ (short-dashed curve). The spectrum from a single flare with
optical depth $\tau_T =\tau_0$ is shown as a solid curve.  The curves
are normalized to the same integrated flux.  These spectra demonstrate
that current observations may not be able to differentiate between a
single-flare spectrum and one comprised of many different flares if
their optical depth is in the range $\sim 0.5 - 2$.}
\label{fig:1}
\end{figure*}

Figure (\ref{fig:1}) shows the results of our calculation for $\tau_0
= 1.14$ and two values of $\tau_{\sigma}$: $0.7$ and $1.5$. The
spectrum from a single flare (solid curve) with $\tau_T = 1.14$ is
also shown for comparison. It can be seen that the plot for
$\tau_{\sigma} = 0.7$ is hardly distinguishable from that for $\tau_T
= 1.14$ (i.e., $\tau_{\sigma}=0$).  Moreover, these curves differ the
most above $100$ keV, where the OSSE data typically have error bars
larger than this difference (see, e.g., Fig. 1 in Zdziarski et al.
1996). On the basis of this simple test, we would expect that Seyfert
spectra may be comprised of contributions from many ARs encompassing a
range ($0.5-2$) of $\tau_T$.  We note, however, that a broader range
in $\tau_T$ is unlikely because of the considerable flattening to the
spectrum for $\tau_{\sigma}>1.5$.

The conclusion that 
$\tau_T$ is allowed to vary within the range of $0.5-2$ is very important
for the magnetic flare model, since it is otherwise difficult to see how different
flares could produce exactly the same $\tau_T$. It may also happen that
a flare evolves through many phases and that its Thomson optical depth
therefore varies with time.  However, these calculations demonstrate that
as long as that variation is restricted to the range $\sim 0.5-2$, the 
resulting spectrum is consistent with the observations.

\section{Conclusions}\label{sect:contau}

We have considered the consequences of imposing a pressure equilibrium
on the active regions of Seyfert Galaxies, in addition to the more
often studied energy equilibrium, under the assumption that the
emission arises within energetic magnetic flares above the surface of
a cold disk. We showed that if the energy is supplied to the
X-radiating plasma by magnetohydrodynamic waves with a group velocity
$\sim c$, then $\tau_T$ probably falls within the range $1-2$. The
current X-ray/$\gamma$-ray observations are consistent with this range
of Thomson optical depths. We conclude that magnetic flares on the
surface of the cold disk remain a viable explanation for the spectra
observed in Seyfert Galaxies. Alternative explanations, based on a
gravitational confinement of the ARs, cannot account for the observed
`universality' in the value of $\tau_T$.

	\chapter{Pressure-Ionization Instability in X-ray Reflection}

\section{Abstract}\label{sect:abstr}

  The spectrum of Seyfert 1 Galaxies is very similar to that of
  several Galactic Black Hole Candidates (GBHCs) in their hard state,
  suggesting that both classes of objects have similar physical
  processes. While it appears that the two phase accretion disk corona
  (ADC) model is capable of explaining the observations of Seyfert
  galaxies, recent work has shown that this model is problematic for
  GBHCs. To address the differences in spectra of Seyferts and GBHCs,
  we consider the structure of the ionized X-ray skin near an active
  magnetic flare. We show that the X-ray skin is subject to a thermal
  instability, similar in nature to the well known ionization
  instability of quasar emission line regions.
 
  We find that for Seyfert Galaxies, the X-ray skin is allowed to
  reside on either the cold ($T\sim 10^5$ K) or the hot ($T\simgt
  10^7$ K) stable branches of the solution, and that observations show
  that the former is the one that is chosen in reality. However, due
  to the much higher ionizing X-ray flux in GBHCs, the only stable
  solution for the upper layer of the accretion disk is that in which
  it is highly ionized and is at the Compton temperature ($\sim $ few
  keV).  Using numerical simulations for a slab geometry ADC, we show
  that the presence of a transition layer, here modeled as being
  completely ionized, with an optical depth $\tautr \simgt 1$
  dramatically alters the reflected spectrum from that predicted by
  ADC models having a discontinuity between a cold disk and a hot
  corona. Due to the higher albedo of the disk, the thermal blackbody
  component is reduced, giving rise to a lower Compton cooling rate
  within the corona.  Therefore, higher coronal temperatures and a
  corresponding harder X-ray spectrum, as compared to the standard ADC
  slab geometry models, are possible. A transition layer also leads to
  a reduction in other observable reprocessing features, i.e., the
  iron line and the X-ray reflection hump.  We conclude that it is
  possible that the differences between the X-ray spectrum of GBHCs
  such as Cyg~X-1 and that of a typical Seyfert Galaxy can be
  explained within a unifying model in which X-rays come from magnetic
  flares above a cold accretion disk.

\section{Introduction}\label{sect:intro4}

The X-ray spectra of Seyfert Galaxies and Galactic Black Hole
Candidates (GBHCs) indicate that the reflection and reprocessing of
incident X-rays into lower frequency radiation is an ubiquitous and
important process. For Seyfert Galaxies, the X-ray spectral index
hovers near a ``canonical value'' ($\sim 0.95$; Pounds et al. 1990,
Nandra \& Pounds 1994; Zdziarski et al. 1996), after the reflection
component has been subtracted out of the observed spectrum. It is
generally believed that the universality of this X-ray spectral index
may be attributed to the fact that the reprocessing of X-rays within
the disk-corona of the two-phase model leads to an electron cooling
rate that is roughly proportional to the heating rate inside the
active regions (AR) where the X-ray continuum originates (Haardt \&
Maraschi 1991, 1993; Haardt, Maraschi \& Ghisellini 1994; Svensson
1996). It has been suggested that the ARs are probably magnetically
dominated structures, i.e., magnetic flares (Haardt et al. 1994; see
also Galeev, Rosner \& Vaiana 1979).

Although the X-ray spectra of GBHCs are similar to that of Seyfert
galaxies, they are considerably harder (most have a power-law index of
$\Gamma \sim 0.7$), and the reprocessing features are less prominent
(Zdziarski et al. 1996). Dove et al. (1997) recently showed that a
Rossi X-ray observation of Cygnus X-1 shows no significant evidence of
reflection features (if the continuum is modeled as a power-law with
an exponential cutoff). It is the relatively hard power law (and
therefore the required large coronal temperature) and the weak
reprocessing/reflection features that led Dove et al.  (1997, 1998),
Gierlinski et al. (1997) and Poutanen, Krolik \& Ryde (1997) to
conclude that the two-phase accretion disk corona (ADC) model, in both
patchy and slab corona geometry cases, does not apply to Cygnus X-1.

One of the main problems with this model is that no self-consistent
coronal temperature is high enough (for a given coronal optical depth)
to produce a spectrum as hard as that of Cyg~X-1 (Dove, Wilms, \&
Begelman 1997). This result is sensitive to the assumption that the
accretion disk is relatively cold, such that $\sim 90$\% of the
reprocessed coronal radiation is re-emitted by the disk as thermal
radiation (with a temperature $\sim 150$ eV). It is this thermal
radiation that dominates the Compton cooling rate within the corona.
However, if the upper layers of the accretion disk were highly
ionized, creating a ``transition layer,'' a smaller fraction of the
incident coronal radiation would be reprocessed into thermal radiation
(i.e., the albedo of the disk would be increased), and therefore the
Compton cooling rate in the corona would be reduced. Furthermore, as
shown by Ross, Fabian \& Brandt (1996; RFB96 hereafter), Auger
destruction of the fluorescent iron line photons may explain the
weakness of observed iron line features in Cyg X-1.

In this Chapter, we extend the earlier work of Nayakshin \& Melia
(1997a), who investigated the X-ray reflection process in AGNs
assuming that the ARs are magnetic flares above the disk. We show that
for parameters appropriate for both Seyfert galaxies and GBHCs, there
should be a thermal instability at the surface of the cold disk. For
AGNs, this thermal instability drives the gas in the X-ray skin down
to temperatures $T \sim$ few$ \times 10^5$ K. For GBHCs, however, the
instability leads to the gas climbing up to $T\sim$ a few $\times
10^7$ K, the Compton temperature with respect to the coronal radiation
field.  This high temperature then explains why the X-ray skin of
GBHCs should be much more strongly ionized as compared to AGN.

In \S \ref{sect:threephase}, we explore the ramifications of this
highly ionized transition layer on the energetics of the corona, and
investigate how it alters the spectrum of the escaping radiation. We
also discuss whether slab geometry ADC models, when transition layers
are included, can account for the observed spectra of GBHCs. Our
conclusions are such that, although a transition layer does allow for
higher coronal temperatures, global two-phase, slab-geometry ADC
models still cannot have coronal temperatures high enough to explain
the data.  However, a model having a patchy corona rather than a
global corona appears very promising. Thus, it is possible that due to
the thermal instability of the surface of the accretion disk, which
leads to different endpoints for GBHCs and Seyfert galaxies, the X-ray
spectra from these two types of objects can be explained by a single
unifying ADC model.

\section{Why a Transition Layer?}\label{sect:why}

\begin{figure*}
\centerline{\psfig{file=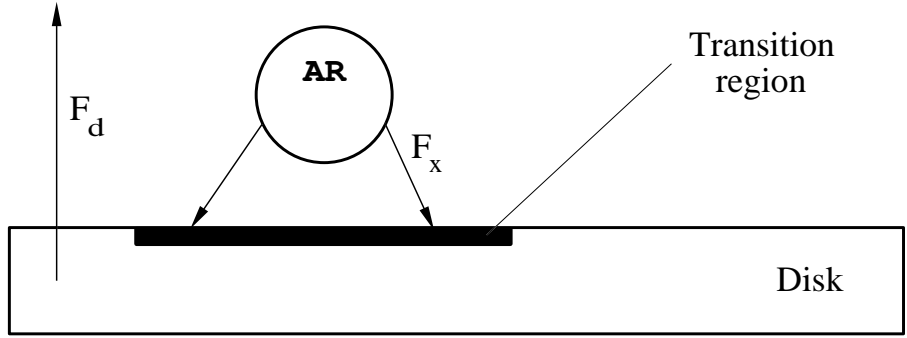,width=.8\textwidth,angle=0}}
\caption{The geometry of the active region (AR) and
the transition layer. Magnetic fields, containing AR and supplying it
with energy are not shown. Transition region is defined as the upper
layer of the disk with Thomson depth of $\sim$ few, where the incident
X-ray flux is substantially larger than the intrinsic disk flux}
\label{fig:active_region}
\end{figure*}

We aim to determine the ionization structure of the disk atmosphere
for the case when the X-ray flux originates in a magnetic flare.  The
relevant geometry is show in Figure (\ref{fig:active_region}). Since
the flux of ionizing radiation from the active region is proportional
to $1/d^2 \times \cos i \propto d^{-3}$, where $i$ is the angle
between the normal of the disk and the direction of the radiation and
$d$ is the distance between the active region and the position on the
disk, the ionization state of the disk surface will vary across the
disk, and consequently only the regions near the active regions (with
a radial size $\sim$ a few times the size of the active region,
situated directly below the active region) may be highly ionized. To
distinguish these important X-ray illuminated regions from the
``average'' X-ray skin of the accretion disk (i.e., far enough from
active magnetic flares), we will refer to these regions as transition
layers or regions. Most reprocessed coronal radiation will take place
in these regions, and, in addition, most radiation emitted by the disk
that propagates through the active regions will have been emitted in
their vicinity. Therefore, in this Chapter, we will only consider the
structure of the cold disk in the transition layer and only solve the
radiation transfer problem for these regions as well.

Although a proper calculation of the ionization state of the
transition layer is preferred, the complexity of this problem is not
matched by any of the X-ray photoionization codes currently available
in the literature. The difficulty is that the density of the
transition layer is coupled to the radiation field, and therefore the
ionization structure, thermal structure, and the radiation field must
be solved self-consistently. Such a problem is outside the scope of
this paper. Instead, we simply provide an order of magnitude estimate
of the properties of the ionization layer to motivate the importance
of the problem for a more elaborate future study.

\section{Pressure Equilibrium}\label{sect:peq}

In this section, we show that the radiation pressure due to coronal
radiation from an active region is very large, and then estimate the
resulting pressure of the transition region.  For transient flares, as
opposed to a static corona, the X-ray flux from magnetic flares can
only persist for a disk hydrostatic time scale (roughly one Keplerian
rotation). Using the model of Svensson \& Zdziarski (1994; SZ94
hereafter), we find that the photon diffusion time across the disk is
much longer than the hydrostatic time scale for both radiation and
gas-dominated disks. Therefore, no thermal equilibrium can be
established between the underlying cold disk and the incident
X-radiation during the flare. Nevertheless, since the optical depth of
the X-ray skin is small compared to total optical depth through the
disk, the radiation diffusion time scale and the atomic processes time
scales are much shorter than the disk hydrostatic time scale
(RFB96). Accordingly, the skin itself will be in quasi-equilibrium
with the incident X-radiation.

The two-phase model with magnetic flares was put forward by Haardt \&
Maraschi (1991, 1993) and Haardt et al. (1994) to explain spectra of
Seyfert Galaxies. The key assumptions of the model are (i) during the
flare, the X-ray flux from the active region greatly exceeds the disk
intrinsic flux, and (ii) the compactness parameter $l$ of the active
region is large, so that the dominant radiation mechanism is
Comptonization of the disk thermal radiation.  The free-free
compactness parameter of the particles in the AR is $l_{\rm ff}\simeq
3\times 10^{-3}\tau_T^2 \Theta^{1/2}$, where the Thomson optical depth
of the active region $\tau_T\simlt 1$, and the dimensionless electron
temperature $\Theta\sim 0.3$ are reasonable numbers to explain X-ray
observations of either GBHCs or Seyfert Galaxies. Therefore, assuming
the luminosity due to bremsstrahlung radiation is negligible is
equivalent to assuming $l \gg 0.01$.

The compactness parameter of the active regions is defined as
\begin{equation}
l\equiv {\fx\sigma_T \Delta R\over m_e c^3},
\label{compact}
\end{equation}
where the size of the active region $\Delta R$ is thought to be of the
order of the accretion disk height scale $H$ (e.g., Galeev et al.
1979), estimated here from the gas pressure dominated solution of SZ94,
\begin{equation}
\frac{H}{R} = 7.5\times 10^{-3} (\alpha M_1)^{-1/10} r^{1/20}
 [\dot{m}J(r)]^{1/5}[\zeta(1-f)]^{1/10},
\end{equation}
where $\alpha$ is the viscosity parameter, $M_1\equiv M/10\msun$ is the
mass of the black hole, $f$ is the fraction of accretion
power dissipated into the corona (averaged over the whole disk), $r$ is
the radius relative to the Schwarzschild radius, $J(r) = 1-(3/r)^{1/2}$,
and $\zeta$ is a constant of order unity (see \S \ref{sect:adstr}).
Therefore, the X-ray flux is approximated by
\begin{equation}
\fx \,= \, 3.6 \times 10^{23} \, l\,\alpha^{1/10} M_1^{-9/10}
\left({\dm\over 0.05}\right)^{-1/5} (1-f)^{-1/10} {\rm erg \ cm}^{-2}\,{\rm sec}^{-1}.
\label{xflux}
\end{equation}
The coronal energy dissipation fraction $f$ should be thought of as
the {\it surface-average} fraction of energy transferred from the cold
disk into the patchy corona above it.  For the case of Cyg X-1, most
of the bolometric luminosity seems to be in the hard X-ray band (e.g.,
Gierlinski et al. 1997). The corona then needs to process most of the
disk power, i.e., $f\sim 1$ (Haardt and Maraschi 1991; Stern et
al. 1995).  In deriving equation (\ref{xflux}), we took the location
of the flare to be at $R= 6R_g$ from the black hole, where the energy
generation rate of the disk is a maximum.  Throughout this Chapter,
this position will be assumed implicitly.

We now compare the X-ray flux incident on the accretion disk with the
intrinsic disk flux $\fdisk$. The intrinsic flux is given by equation
(\ref{fdisk}). We assume the dimensionless accretion rate $\dm =
\eta\dot{M}c^2/\ledd \sim 0.05$, a value thought to be appropriate for
Cyg X-1. Here, $\dot{M}$ is the accretion rate, $\eta =0.056$ is the
efficiency for the standard Shakura-Sunyaev disk, and $\ledd$ is the
Eddington luminosity. Note that this definition of $\dm$ is different
by factor $\eta$ from that used by SZ94 (i.e., $\dm
\simeq 17 \times \dm_{\rm SZ94}$). We obtain for the disk intrinsic
flux
\begin{equation}
\fdisk = 1.0 \times 10^{22} M_1^{-1} \left({\dm\over 0.05}\right) (1-f)
\;{\rm erg \ cm}^{-2}\,{\rm sec}^{-1}{\rm .}
\label{diskflux}
\end{equation}

The illuminating X-ray flux is much larger than the intrinsic disk
emission at regions that are near an active magnetic flare if $1-f\ll
1$ and the compactness parameter $l\gg 0.01$. For future reference, we
also define the disk compactness parameter as $\lbb\equiv
\fdisk\sigma_T H/ ( m_e c^3)$:
\begin{equation}
\lbb = 0.03 \left({\dm\over 0.05}\right)^{6/5} \left(1-f\right)^{11/10}
\left(\alpha M_1\right)^{-1/10}{\rm .}
\label{lbb}
\end{equation}
(Note that instead of finding $\fx$ through its connection with $l$,
we could have required the magnetic flares to have a small covering
fraction $f_c\ll 1$ and $1-f\ll 1$, and then we would have been able to
deduce $\fx\sim \fdisk f/(1-f) f_c^{-1}\gg \fdisk$ and $l\gg \lbb$.
In other words, to describe magnetic flares, one specifies either 
$l$ or $f_c$.)

We first consider the pressure of the disk surface layer before the
occurrence of a flare (or, equivalently, far enough from the flare).
If we assume that the upper layer of the disk with Thomson optical
depth $\tau_x\sim 3$ is in hydrostatic equilibrium with the vertical
gravitational force, the pressure $P_0$ in that region is approximated
as
\begin{eqnarray}
  P_0 &\simeq& {G M m_p\over R^2} \tau_x\, {H\over R} = 6.2\times 10^{10}\;
  M_1^{-11/10} \alpha^{-1/10}\nonumber \\ & &\times \tau_x\,\left({\dm\over
      0.05}\right)^{1/5} (1-f)^{1/10} \;{\rm erg \ cm}^{-3},
\label{p0}
\end{eqnarray}
where $R = 6R_{\rm g}$ (SZ94). When the flare turns on, the ratio of the
incident radiation ram pressure to the unperturbed accretion disk
atmosphere pressure is
\begin{equation}
{\fx\over c P_0}\, = \, 2. \times 10^2 \; l \,\tau_x^{-1}\,\left(\alpha
M_1\right)^{1/5} \,\left({\dm\over 0.05}\right)^{-2/5}
(1-f)^{-1/5}{\rm ,}
\label{prat}
\end{equation}
i.e., much higher than unity. Note that this conclusion is also applicable
to flares in AGN. Thus, due to the equation (\ref{prat}) and the fact that
$\fx\gg \fdisk$, the dynamical properties of the disk atmosphere will be
strongly affected by the irradiating flux, as long as there is an active
magnetic flare nearby, and this flux should be taken into account when
solving the disk ionization structure.

It is possible that a wind is induced by the X-ray heating.  However,
the maximum gas temperature obtained due to the X-ray heating is the
Compton temperature ($\simlt 10^8$ K) and is still far below the gas
virial temperature ($k T_{\rm vir} \simeq GM/R$) for $R\simlt 10^5
R_g$. Therefore, as shown by Begelman, McKee \& Shields (1983), a
large scale outflow cannot occur for $R\simlt 10^4 R_g$. On the other
hand, a local expansion of the gas is still possible, since the
Compton temperature is higher than the the disk temperature. The
maximum energy flux due to the wind, local evaporation or any
mechanical expansion of the gas is $F_{\rm ev} \sim P c_s$, where $P$
is the gas pressure in the transition region, and $c_s$ is its sound
speed. Since $P\simlt
\fx/c$, and $c_s \simlt 3\times 10^{-3}\, c$, we obtain
${F_{\rm ev}/\fx} \simlt 3 \times 10^{-3}$. Therefore, mechanical
processes cannot cool the gas efficiently, and we neglect the
influence of the possible wind on the energetics of the two-phase
model. Note that evaporation of the material from the transition
region could obscure the flare, but the large radiation flux from the
flare is likely to push the gas laterally, away from the flare. This
effect needs to be quantified in the future, but for now we assume
that the flare is not obscured by the evaporation of the material.
Below, we estimate the structure of the transition layer using the
pressure and energy equilibrium conditions.

\section{The Thermal Instability}\label{sect:instability}

Thermal instability was discovered by Field (1965) for a general
physical system. He introduced the ``cooling function'' $\lnet$,
defined as the difference between cooling and heating rates per unit
volume, divided by the gas density $n$ squared. Energy equilibria
correspond to $\lnet = 0$. He argued that a physical system is usually
in pressure equilibrium with its surroundings. Thus, any perturbations
of the temperature $T$ and the density $n$ of the system should occur
at a constant pressure. The system is unstable when
\begin{equation}
\left({\partial \Lambda_{\rm net}\over \partial T}\right)_{P} < 0,
\label{field}
\end{equation}
since then an increase in the temperature leads to heating increasing
faster than cooling, and thus the temperature continues to increase.
Similarly, a perturbation to a lower $T$ will cause the cooling to
exceed heating, and $T$ will continue to decrease.

In the context of the transition layer stability, however, the
constant pressure condition may be of lesser importance than the
energy equilibrium in the following sense. Since the energy is being
supplied by the coronal radiation, the thermal time scale $t_{\rm th}$
is the light crossing time of the transition region. The hydrostatic
time, i.e., the time scale for balancing out pressure perturbations
$t_{\rm h}$ in the layer, is given by the sound crossing time, which
is at least a factor of $10^3$ longer than the thermal time scale
(since $T < 10^8$ K). Thus, in any perturbation, the energy
equilibrium condition $\lnet = 0$ is reached very quickly, whereas the
pressure balance may be perturbed. The instability will exist if
perturbing the gas density $n$ to a higher value will be followed by a
{\em decrease} in pressure in that region, since then the gas will
contract further because of the pressure imbalance with its
surroundings. In other words, the condition for the transition layer
instability is
\begin{equation}
\left({d P\over d n}\right)_{\lnet = 0} < 0 {\rm ,}
\label{pcond}
\end{equation}
where the derivative is taken with condition $\lnet = 0$ satisfied due
to the short thermal time scale. Working through some simple algebra,
one finds that
\begin{equation}
\left ({d P\over d n}\right)_{\lnet = 0} = {P\over n}\,
\left({\partial \Lambda_{\rm net}\over \partial T}\right)_{P}
\left({\partial \Lambda_{\rm net}\over \partial T}\right)_{n}^{-1}
{\rm .}
\label{conn}
\end{equation}
It is a rare occasion that the last multiplier on the right hand side
is negative, and under most circumstances the conditions given by
equations (\ref{pcond}) and (\ref{field}) are equivalent.

When studying ionization balance, it is convenient to define two
parameters. The first one is the ``density ionization parameter''
$\xi$, equal to (Krolik, McKee \& Tarter 1981)
\begin{equation}
\xi = {4\pi \fx\over n} \, .
\label{xid}
\end{equation}
The second one is the  ``pressure ionization parameter'', defined as
\begin{equation}
\Xi = {\fx\over 2 c n k T} \equiv \frac{\prad}{P},
\label{xip}
\end{equation}
where $P$ is the gas pressure. This definition of $\Xi$ is the one
used in the ionization code XSTAR (see below), and is different by a
factor $2.3$ from the original definition of Krolik et al. (1981), who
used the hydrogen density instead of the electron density. Ionization
equilibria depend most sensitively on the density ionization parameter
$\xi$, rather than $\fx$ or $n$ separately. Solving the ionization and
energy equilibrium gives the functions $T(\xi)$ and $\Xi(\xi)$. The
latter one can be written as $\Xi(T)$ using the former. In terms of
these variables, one can write $(dP/dn)_{\Lambda=0} = (d\Xi/dT)\,(\xi/
\Xi)\; (\partial T/
\partial \xi)$. In our calculations, we always found that $(\partial T/
\partial \xi)>0$, so that the instability criterion (\ref{pcond}) is
equivalent to
\begin{equation}
\left( {d\Xi\over d T}\right)_{\lnet=0}\, < 0 .
\label{fcond}
\end{equation}

We now apply the X-ray ionization code XSTAR (Kallman \& McCray 1982,
Kallman \& Krolik 1986) the problem of the transition layer. A truly
self-consistent treatment would involve solving the radiation transfer
in the optically thick transition layer, and in addition, finding the
distribution of the gas density in the transition layer that would
satisfy pressure balance. Since the radiation force acting on the gas
depends on the opacity of the gas, this is a difficult non-linear
problem. Thus, we defer such a detailed study to future work, and
simply solve (using XSTAR) for the local energy and ionization balance
for {\em an optically thin layer} of gas in the transition region. We
assume that the ionizing spectrum consists of the incident X-ray
power-law with the energy spectral index typical of GBHCs in the hard
state, i.e., $\Gamma = 1.5-1.75$, exponentially cutoff at 100 keV, and
the blackbody spectrum from the cold disk below the transition
layer. If the energy and ionization balance is found to be unstable
for this setup, the transition layer will also be unstable.

When applying the code, one should be aware that it is not possible
for the transition region to have a temperature lower than the
effective temperature of the X-radiation, i.e., $\Tmin =
(\fx/\sigma)^{1/4}$.  In the spirit of a one zone approximation for
the transition layer, we should use an average X-ray flux $\mean{\fx}$
as seen by the transition region, which we parameterize as $\mean{\fx}
= 0.1
\fx/q_1$, where $q_1 = q/10$, and $q$ is a dimensionless number of
order 10 (see figure \ref{fig:active_region}; $\fx$ is the X-ray flux
{\em at} the active region). Using equation (\ref{xflux}),
\begin{equation}
\Tmin\simeq 5.0 \times 10^6\; l^{1/4} \, q_1^{-1/4}\, \left({\dm\over
0.05}\right)^{-1/20}\, \alpha^{1/40} \, M_1^{-9/40} \,
\left[1-f\right]^{-1/40} .
\label{tmincyg}
\end{equation}
The reason why simulations may give temperatures lower than $\Tmin$
for a low ionization parameter $\xi$ is that in this parameter range
XSTAR neglects certain de-excitation processes, which leads to an
overestimate of the cooling rate (Zycki et al. 1994; see their section
2.3).

\begin{figure*}
\centerline{\psfig{file=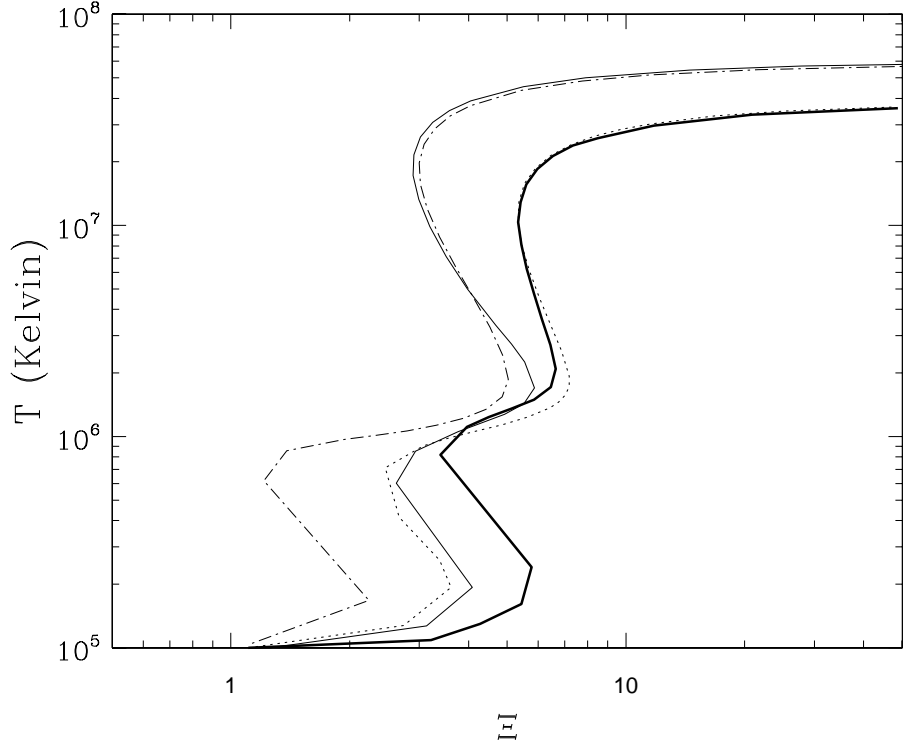,width=.8\textwidth,angle=0}}
\caption{Gas temperature versus the pressure ionization parameter $\Xi$ --
  the ionization equilibrium curves for parameters appropriate for
  GBHCs.  The incident spectrum is approximated by a power-law of
  photon index $\Gamma$, exponentially cutoff at $100$ keV, and the
  reflected blackbody with equal flux and temperature $\Tmin$
  (equation \ref{tmincyg}). Values of the parameters are: $\Gamma =$
  1.5, 1.75, 1.75, 1.7 and $k \Tmin = $ 200, 100, 200, 400 eV,
  corresponding to the fine solid, thick solid, dotted and dash-dotted
  curves, respectively. The ionization equilibrium is unstable when
  the curve has a negative slope. In addition, there exist no solution
  below $\Tmin$.}
\label{fig:scurve}
\end{figure*}

Figure \ref{fig:scurve} shows the results of our calculations for
several different X-ray ionizing spectra. A stable solution for the
transition layer structure will have a positive slope of the curve,
and also satisfy the pressure equilibrium condition. As discussed in
\S
\ref{sect:peq}, $P \leq \fx/c$ (i.e., $\Xi\geq 1$). In addition, if
the gas is completely ionized, the absorption opacity is negligible
compared to the Thomson opacity. Because all the incident X-ray flux
is eventually reflected, the net flux is zero, and so the net
radiation force is zero.  In that case $P$ adjusts to the value
appropriate for the accretion disk atmosphere in the absence of the
ionizing flux (see also Sincell \& Krolik 1996), which is given by
equation (\ref{p0}). Therefore, the pressure ionization parameter
should be in the range
\begin{equation}
1< \Xi < 2\times 10^2 \,l \left(\alpha
    M_1\right)^{1/5} \,\left({\dm\over 0.05}\right)^{2/5} (1-f)^{-1/5}.
\label{xip}
\end{equation}
With respect to the ionization equilibria shown in Figure
(\ref{fig:scurve}), the gas is almost completely ionized on the upper
stable branch of the solution (i.e., the one with $T\simgt 10^7$ K),
and thus the pressure equilibrium for such temperatures requires $\Xi
\sim \fx/c P_0\gg 1$.

In addition to the Compton equilibrium state, for some curves, there
is a smaller stable region for temperatures in the range between $100$
and $200$ eV. The presence of this region is explained by a decrease
in {\em heating}, rather than an increase in cooling (cf. equation
\ref{field} and recall $\lnet=$ cooling -- heating). The X-ray heating
decreases in the temperature range $100-200$ eV with increasing $T$
because of consecutive destruction (ionization) of ions with
ionization energy close to this temperature region. Note that it is
highly unlikely that the transition region will stabilize at the
temperature $100$ -- $200$ eV, because the effective temperature
$\Tmin$ is at or above this temperature range.

\begin{figure*}
\centerline{\psfig{file=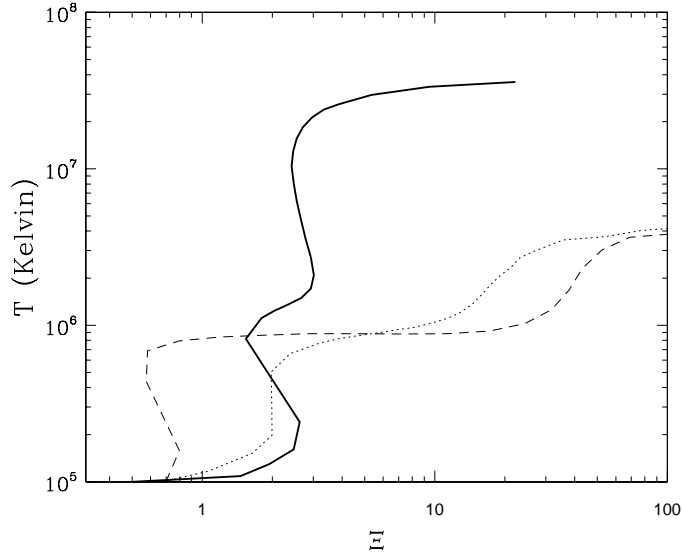,width=.6\textwidth,angle=0}}
\caption{Gas temperature versus the pressure ionization parameter $\Xi$.
The thick solid curve is same as that in Figure \ref{fig:scurve} and
is appropriate for the hard state of a GBHC, whereas the two other
curves are relevant to the soft state in GBHCs, or at large depth in
the transition layer (see text). Values of the parameters are: $\Gamma
=$ 2.1, 2.1 and $k
\Tmin = $ 200, 400 eV for the dotted and dashed curves, respectively.}
\label{fig:ssoft}
\end{figure*}

Thus, although a fully self-consistent treatment of the pressure and
ionization balance of the transition layer is needed to obtain exact
results, it is very likely that the transition layer is highly ionized
in GBHCs {\em in the hard state} for $\tau_{\rm x} \sim$ 1. The upper
limit of $\tau_{\rm x}$ can only be found by a more exact treatment.
In addition, the transition layer may be heated by the same process
that heats the corona above it, albeit with a smaller heating rate.
Furthermore, Maciolek-Niedzwiecki, Krolik \& Zdziarski (1997) have
recently shown that the thermal conduction of energy from the corona
to the disk below may become important for low coronal compactness
parameters and substantially contribute to the heating rate of the
transition layer. Thus, the transition layer may be even hotter than
that found by photoionization calculations.

Eventually, the X-rays are down-scattered and the radiation spectrum becomes
softer as one descends from the top of the transition layer to its bottom.
We can qualitatively test the gas ionization stability properties by
allowing the ionizing spectrum to be softer than the observed spectrum of
GBHCs in the hard state. In Figure \ref{fig:ssoft} we show two examples
of such calculations. The slope of the ionization equilibrium curve becomes
positive everywhere above $k T\sim$ 100 eV, so that these equilibria are
stable, and thus the gas temperature may saturate at $T\sim \Tmin$, far
below few keV, the appropriate temperature for the uppermost layer of the
transition region.  Thus, we know (see also \S \ref{sec:discussion}) that the
transition layer should terminate at some value of $\tau_x \sim$ few.

We also note that the thermal instability is not apparent in studies
where the gas density is fixed to a constant value, regardless of its
value. As shown by Field (1965), the thermal instability for the case
with $n=$ const is always weaker than it is for the case of the system
in pressure equilibrium. Following Field (1965), we argue that the
assumption of a constant gas density is not justified for real
physical systems, and that one always should use the pressure
equilibrium arguments to determine the actual gas density and the
stability properties of the system.

\section{More Accurate Pressure Equilibrium}\label{sect:morepeq}

Summarizing the results learned from Figure (\ref{fig:scurve}), the
transition layer is most likely at the Compton equilibrium state,
although there is a narrow temperature range $T \sim 100-200$ eV (the
``island'' stable state) where the equilibrium state is possible if
$\Tmin < 200$ eV. Let us now examine the pressure equilibrium
arguments in more detail to discuss the stability properties of this
state. Consider the pressure applied by incident X-rays on an
optically thin layer of gas, $\tau_x\equiv n_e\sigma_T z
\ll 1$:
\begin{equation}
P_x = n_e z \left(\sigma_T + \sigma_a\right) {\fx\over c} =
\left(\tau_x + \tau_a\right) {\fx\over c} \ .
\label{eqp1}
\end{equation}
Here, $z$ is the vertical coordinate pointing up, $n_e$ is the local
electron number density, $\sigma_a$ and $\tau_a$ are absorption cross
section and optical depth, correspondingly ($\tau_a <1$ is
implied). In the case of large optical depth, however, every photon
incident on the layer interacts with the gas, thus passing its
momentum to the layer. The radiation pressure then saturates at the
radiation ram pressure $\fx/c$. To take this effect into account, we
can approximate the radiation pressure for all optical depths by
\begin{equation}
P_x = {\tau_x + \tau_a\over 1 + \tau_x + \tau_a}\,\, {\fx\over c} 
\ .
\label{eqp2}
\end{equation}

Let us now come back to the transition layer problem. Within the
optically thin layer, the pressure equilibrium condition is
\begin{equation}
{d\pg\over d z} = -g_0 n_p m_p - n_e \left\{\mean{\sigmax} -
\mean{\sigmauv}
\right\} {\fx\over c} + n_e \mean{\sigmad}{\fdisk\over c}\ .
\label{peq} 
\end{equation}
Here, $g_0$ is the local gravity, which is approximately constant
throughout the transition layer, equal to
\begin{equation}
g_0 = {G M\over R^2}\;{H\over R}\ .
\label{g0}
\end{equation}
We also define the ``Roseland mean'' cross sections for the three
components of the radiation field in the transition region, i.e., the
incident X-ray flux, the reflected UV (or soft X-ray in GBHC case)
flux $\fuv$, and the intrinsic disk emission $\fdisk$. In particular,
each of these cross sections is defined as
\begin{equation}
\mean{\sigma_i}\equiv {1\over F_i}\, \int d E {d F_i\over d E}
\, \sigma_i(E) \ ,
\label{rmc}
\end{equation}
where $E$ is the photon energy, and $i$ stands for either x, uv, or
d. In the case at hand, $\fx=\fuv$ (although a part of the ''UV'' flux
can actually come out in the hard X-ray range, e.g., $10-20$ \%
may come out as the X-ray reflection component -- see Magdziarz
\& Zdziarski 1995, for example), and $\fx\gg \fdisk$, so that
we will neglect $\fdisk$ for now. We will consider situations when
$\fdisk$ is non-negligible in \S \ref{sect:xbaldwin}.

In the spirit of a one-zone approximation, we may integrate equation
(\ref{peq}) in the $z$-direction to obtain 
\begin{equation}
\pg \simeq P_0 + A {\fx \over c}\ ,
\label{peqt}
\end{equation}
where $P_0$ is the unperturbed pressure of the accretion disk
atmosphere, given by equation (\ref{p0}) and we introduce the
dimensionless constant $A$, which describes the effects of the
radiation pressure on the transition layer. Following the discussion
just before equation (\ref{eqp2}), $A$ is
\begin{equation}
A = {\tau_x + \tau_x'\over 1 + \tau_x + \tau_x'}\, -\,
{\tau_x + \tau_{uv}\over 1 + \tau_x + \tau_{uv}} \ ,
\label{adef}
\end{equation}
where $\tau_x' \equiv \mean{\sigmax} n_e z$, $\tau_{uv} \equiv
\mean{\sigmauv} n_e z$. Note that this expression now takes into
account the most important features of the pressure equilibrium for
the transition layer.  In particular, $A$ is always smaller than
unity; and, most significantly, it also takes into account the fact
that when $\mean{\sigmauv} > \mean{\sigmax}$, the radiation pressure
actually points up rather than down. Thus, the pressure equilibrium of
the transition layer can be much more restrictive than that given by
the simple estimate $\sim \fx/c$, and we should compute the gas
opacities to treat the pressure equilibrium correctly. We can also
write down the pressure ionization parameter found using the estimate
(\ref{peqt}) for the gas pressure:
\begin{equation}
\Xi = \left[{P_0 c\over \fx} + \, A\right]^{-1}\ .
\label{xim}
\end{equation}

\begin{figure*}
\centerline{\psfig{file=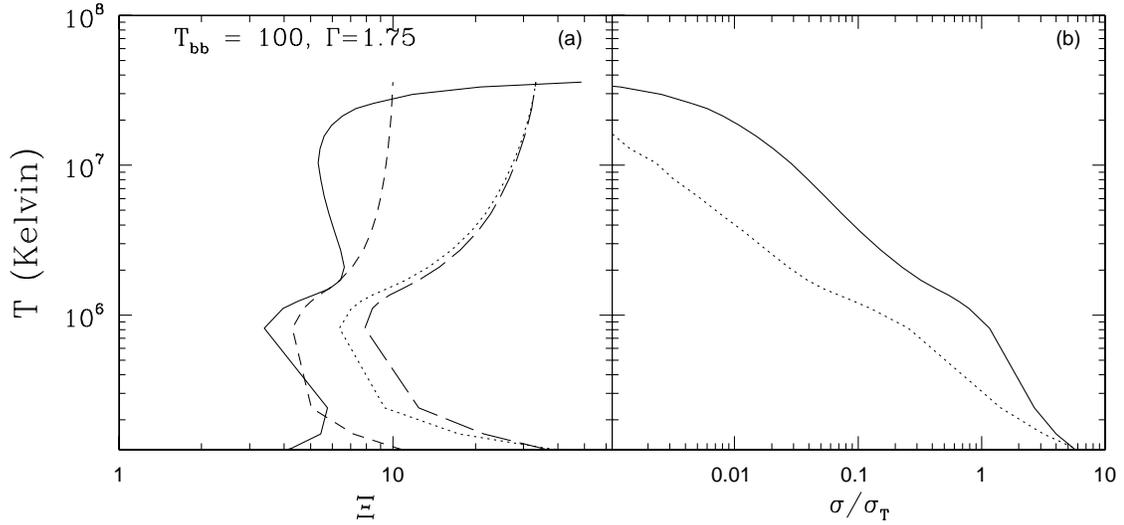,width=.99\textwidth,angle=-90}}
\caption{(a): Solid curve shows the ionization equilibrium curve. Using
that equilibrium, the pressure equilibrium argument (equation
\ref{xim}) is used to estimate the corresponding ionization parameter. 
For the three pressure equilibrium curves the parameters are: $\tau_x
= 0.5$, 1, 2 and $P_ 0 c/\fx = $ $0.1$, 0.03 and 0.03 for the dashed,
dotted and long-dashed curves, correspondingly. A stable configuration
of the transition layer is achieved at the location where the
ionization and pressure equilibrium cross. The gas absorption
opacities are shown in panel (b). The solid curve is the X-ray opacity
$\mean{\sigmax}$, whereas the dotted curve depict the UV-opacity
$\mean{\sigmauv}$.}
\label{fig:gbhc1}
\end{figure*}

Figure (\ref{fig:gbhc1} (a)) shows the ionization equilibrium together
with approximate equilibrium given by equations
(\ref{adef},\ref{xim}), for one of the curves shown in
(\ref{fig:scurve}). We tested several different values of $\tau_x$ and
$P_0c/\fx$. A stable solution is obtained when the solid curves
intercepts one of the other curves, at which point ionization, energy
and pressure equilibrium conditions are satisfied. As previously,
regions with negative slope of the solid curve are thermally unstable,
and temperatures lower than $\Tmin = 100$ eV are forbidden.  Notice
that only for relatively low Thomson depths of the transition layer
the pressure equilibrium curve intercepts the island stable state.  We
doubt that a transition layer with such a low value of $\tau_x\simeq
0.5$ could be formed in reality, and thus we do not expect that the
island state is truly stable, at least for magnetic flares in
GBHCs. The X-ray spectra of GBHCs in the hard state contain most of
the energy in the very hard X-ray band. Since the X-ray absorption is
negligible above $\sim 10$ keV, these photons will not scatter until a
Thomson depth of $\simlt 1$ is reached. After the first scattering,
when these hard photons loose a considerable (almost all) fraction of
their energy, they become vulnerable to the X-ray absorption below 10
keV, and can be finally absorbed. Therefore, $\tau_x$ less than 1 -- 2
seems to be unrealistic.

The two curves that do not intercept the island state, with $\tau_x =$
1 and 2, meet with the ionization equilibrium curve only at the
Compton equilibrium state. Panel (b) of Figure (\ref{fig:gbhc1}) shows
the gas temperature and the X-ray and UV mean absorption cross
sections in units of the Thomson cross section for the same tests. One
can observe that X-ray absorption is larger than the UV absorption,
and thus the net radiation pressure force points down to the cold
disk, but both of these absorption opacities become very small for
$T\simgt 200$ eV. To a first approximation, since material is highly
ionized for these conditions, the opacity is given by the Thomson
opacity only. Since the reflected UV flux is equal to the ionizing
X-ray flux, this implies that these two fluxes almost cancel each
other in terms of the radiation pressure on the gas. So, the
coefficient $A$ in equation (\ref{xim}) is very small and the effect
of gravity is actually larger than the X-radiation pressure for these
conditions. Thus, the pressure equilibrium curves saturate at $\Xi =
\fx/cP_0$ for high temperatures.

Rounding this discussion up, we believe that the only stable
configuration available for the transition layer of GBHCs in the hard
state is the one at the local Compton temperature. Future work should
concentrate on finding not only the exact value of $\tau_x$, but the
exact distribution of gas temperature, density and ionization state in
the atmosphere of the accretion disk as well. For now, however, we
will treat $\tau_x$ as a free parameter and numerically investigate
the ramifications of the transition layer on the spectrum of escaping
radiation and the physical properties of the corona in the next
section.

\section{``Three-Phase'' Model}\label{sect:threephase}
\subsection{Physical Setup}

As is clear from the foregoing discussion, there is an urgent need to
explore the structure of the ionized transition region and how it
affects the X-ray spectrum from magnetic flares. Since one depends on
the other, this is a non-linear problem, and a very difficult one. At
this point, however, we feel it will be useful to make a parameter
search even in the context of a simplified model, where the accretion
disk below the flare is broken into two regions: (i) the completely
ionized transition region, and (ii) the cold accretion disk, which
emits blackbody radiation at a specified temperature. By computing the
X-ray spectrum from a magnetic flare above the transition layer with a
range of $\tau_{\rm trans}$, we will try to determine whether there is
a value of $\tau_{\rm trans}$, physically plausible enough, which
would lead to the X-ray spectrum at least qualitatively close to the
observed hard spectrum of Cyg X-1. If this turns out to be the case,
then a future more detailed and accurate comparison of spectra from
magnetic flares and Cyg X-1 spectrum will be forthcoming.

Gierlinski et al. (1997) have attempted to fit the broad-band spectrum
of Cyg X-1 with active regions above a cold accretion disk. From this
work, and the analysis below, it can be seen that the most difficult
issue for the two-phase model is the too small observed lack of
significant reprocessed soft X-radiation. For example, Zheng et al.
(1997) showed that Cyg X-1's luminosity in the hard state below 1.3
keV is about $5\times 10^{36}$ erg/s, whereas the luminosity above 1.3
keV is $\sim 3-4\times 10^{37}$ erg/s. This is impossible in the
context of the simple two-phase corona-disk model, since about half of
the X-radiation impinges on the cold disk and gets reprocessed into
blackbody radiation. Accordingly, the minimum luminosity in soft
X-rays below 1.3 keV should be about that of the hard component.
Thus, the focus of our attention here will be the reprocessed
radiation and not the active region intrinsic spectrum. This allows us
to first use a simple radiation transfer code for the AR and yet
retain most of the physics. Even though the geometry of the AR is
probably closer to a sphere or a hemisphere than a slab, we shall
adopt the latter for numerical convenience, neglecting the boundary
effects.  Experience has shown that spectra produced by Comptonization
in different geometries are usually qualitatively similar (i.e., a
power-law plus an exponential roll-over), and it is actually the
fraction of soft photons entering the corona that accounts for most of
the differences in the various models, because it is this fraction
that affects the AR energy balance.

Following the standard practice in ionization/reflection calculations,
we model the reflecting medium as being one dimensional, with its only
dimension being the optical depth into the disk (measured from the
top). The X-radiation enters the transition region through its top. In
this region, the only important process taken into account is Compton
interactions. After being down-scattered (but not absorbed, since iron
ions are assumed to be completely ionized!), X-radiation is
``incident'' on the cold accretion disk from the bottom of the
transition layer.  The incident spectrum is reflected and reprocessed
in the standard manner, and then re-enters the transition layer from
below. Specifically, the reflected spectrum is given by the reflection
component (Magdziarz \& Zdziarski 1995) and the blackbody component
due to the disk thermal emission. The blackbody emission is normalized
such that the incident flux from the transition region is equal to the
sum of the fluxes from the reflection component and the blackbody.
The optically thick cold disk is held at a temperature $T_{\rm bb} =
2.4 \times 10^6$ Kelvin, and the blackbody spectrum is renormalized to
produce the correct energy conserving flux. 

The gas in the active region is heated uniformly throughout the region
with a given heating rate (which is normalized to give the assumed
compactness for the region). The gas is cooled by Compton interactions
with radiation re-entering the active region from below (Compton
interactions are the dominant cooling mechanism for these
conditions). To crudely take the geometry into account, we permit only
a part of the reprocessed radiation to re-enter the corona, and fix
this fraction at $0.5$ (cf. Poutanen \& Svensson 1996). The Thomson
optical depth of the corona is fixed at $\tau_{\rm c} = 0.7$ to avoid
complications with pair production, which will be included in later
work.  Further, we shall see below that the compactness of the active
regions is limited to a relatively small number, i.e., $l\simlt$ few,
so that pair production may simply be irrelevant for real flares.

We employ the Eddington (two-stream) approximation for the radiative
transfer in both the AR and the transition layer, using the zero
(isotropic) and the first order moments of the exact Klein-Nishina
scattering kernel (Nagirner \& Poutanen 1994).  The transition region
and the AR are divided into some number of zones, such that the
Thomson optical depth of each zone does not exceed 0.1. The energy
balance for the transition region within the scope of our approximate
treatment is given by requiring the gas to be at the Compton
temperature, and is solved separately for each zone. The latter is
important for a moderately optically thick transition region, since
then we find the region to be stratified in temperature due to a
change in the local radiation field as one moves from the bottom to
the top of the layer.  Since our code is time-dependent, we simply
start with some ``reasonable'' initial conditions and then allow the
Active Region, the transition region and the radiation field to come
into equilibrium.

According to the physical setup of this problem, the observed spectrum
consists of the direct component, emerging through the top of the AR,
and a fraction of the reflected radiation that emerges from the
transition layer and does not pass through the corona on its way to
us.  This fraction is chosen to be 0.5 as well.  Physically, it
accounts for the fact that, as viewed by an observer, a part of the
transition region itself is blocked by the active region. The overall
setup of the active region - disk connection is very similar to the
one used by Poutanen \& Svensson (1996), except for the addition of a
transition layer above the cold disk.

\subsection{Results}\label{sect:results}

\begin{figure*}
\centerline{\psfig{file=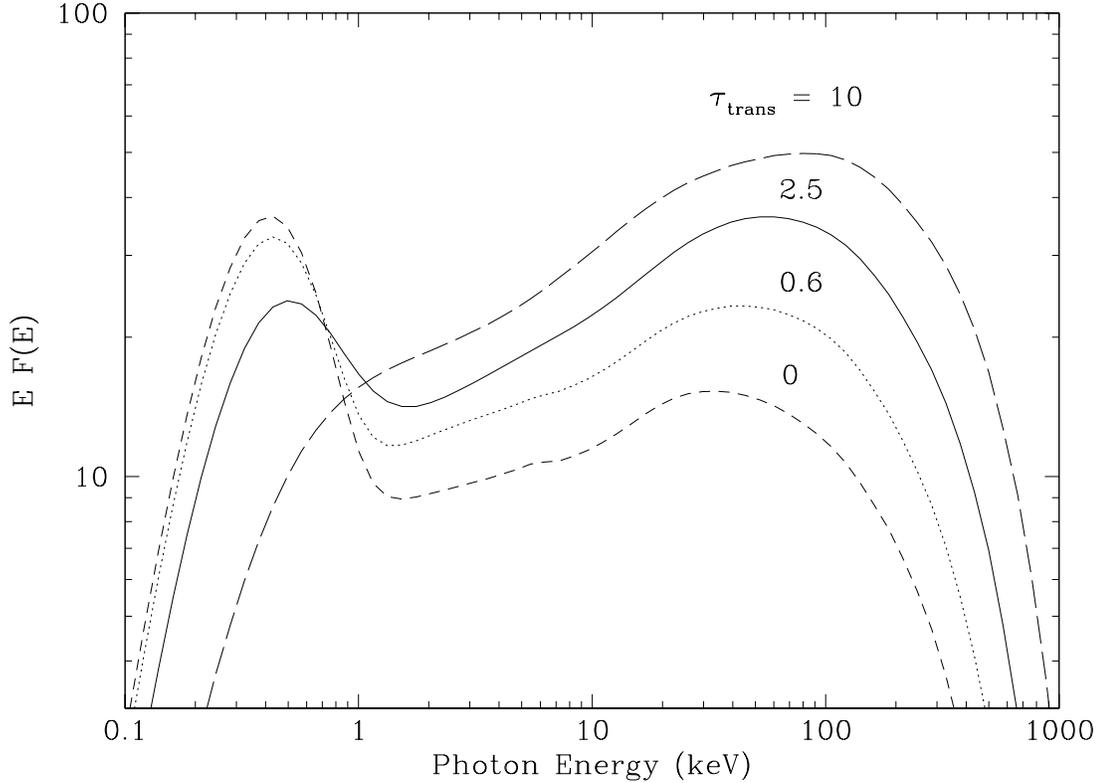,width=.97
\textwidth,angle=-90}}
\caption{Resulting spectrum from the patchy corona disk model as a
function of the Thomson optical depth $\tau_{\rm trans}$ of the
transition layer. The layer is assumed to be completely ionized.
Notice that higher values of $\tau_{\rm trans}$ lead to a harder
spectrum with the disk blackbody component progressively smaller.}
\label{fig:sequence_of_spectra}
\end{figure*}

Figure (\ref{fig:sequence_of_spectra}) shows the ``observed'' spectrum
for several values of $\tau_{\rm trans}$: $0$, $0.6$, $2.5$, and $10$,
with $\Omega=0.5$.  It can be seen that the spectrum hardens as
$\tau_{\rm trans}$ increases.  To help explain why this happens, we
plot in Figure (\ref{fig:albedo}) the integrated albedo $a$ for
photons with energy $E> 1$ keV as a function of $\tau_{\rm
trans}$. The albedo is simply the ratio of the returning flux in the
given energy range to the incident one. The returning flux is the one
that emerges from the top of the transition layer. As $\tau_{\rm
trans}$ increases, a large fraction of the photons from the AR are
reflected before they have a chance to penetrate into the cold disk
where the blackbody component is created. Therefore, a smaller flux of
energy is deposited below the transition layer, which leads to a
decreased cooling from the Comptonization of soft, reprocessed
radiation.  For a moderate optical depth $\tau_{\rm trans}$, this
result is quite insensitive to the temperature in the transition layer
as long as Fe is highly ionized.  We checked this by simply setting
the transition temperature at the arbitrarily chosen values of $1.5$
and $6$ keV, instead of the self-consistent temperature distribution
calculated above, which varied (with optical depth into the transition
layer) from about $2$ to $4$ keV for the respective values of
$\tau_{\rm trans}$.  We found that the relative variations in the
spectrum and the albedo resulting from this were less than about
$3\;\%$.  For higher optical depths ($\tau_{\rm trans}\simgt 4$),
pre-Comptonization of the soft disk radiation becomes important and
additionally decreases the Compton cooling of the corona by this
component, so that the temperature of the transition layer becomes
essential.

\begin{figure*}
\centerline{\psfig{file=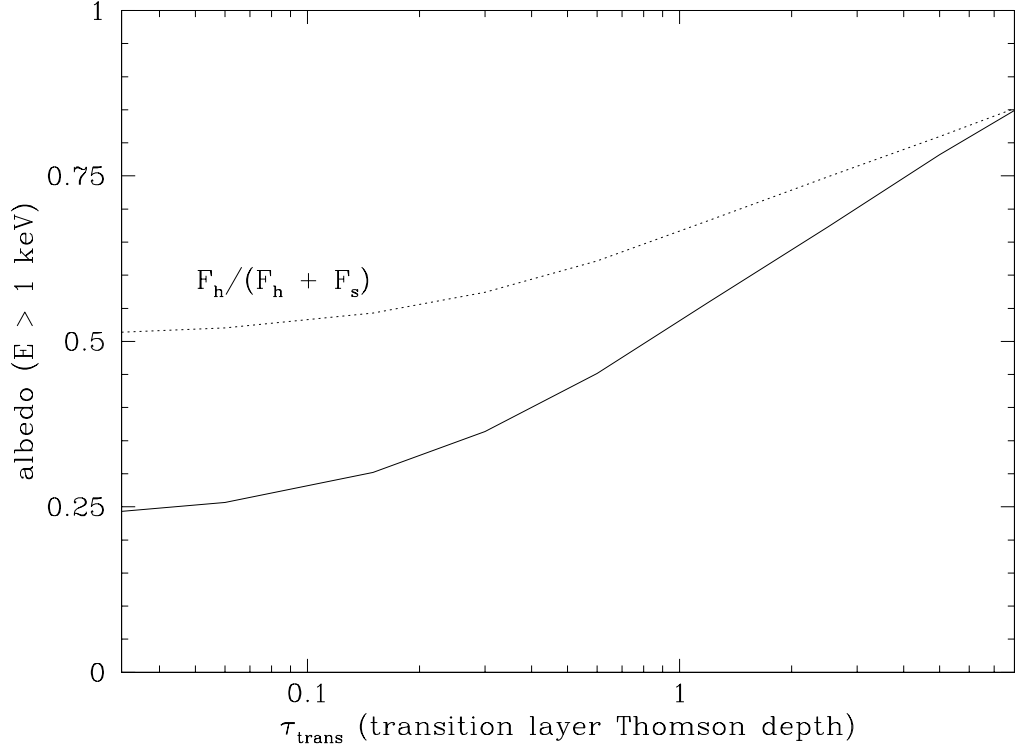,width=.9
\textwidth,angle=-90}}
\caption{Integrated albedo (reflected fraction) as a function of the
transition layer optical depth, $\tau_{\rm trans}$, for photons with
energy $> 1$ keV.  Also plotted (dotted curve) is the ratio of the
observed hard luminosity (above $2$ keV) to the observed total
luminosity.}
\label{fig:albedo}
\end{figure*}

\begin{figure*}
\centerline{\psfig{file=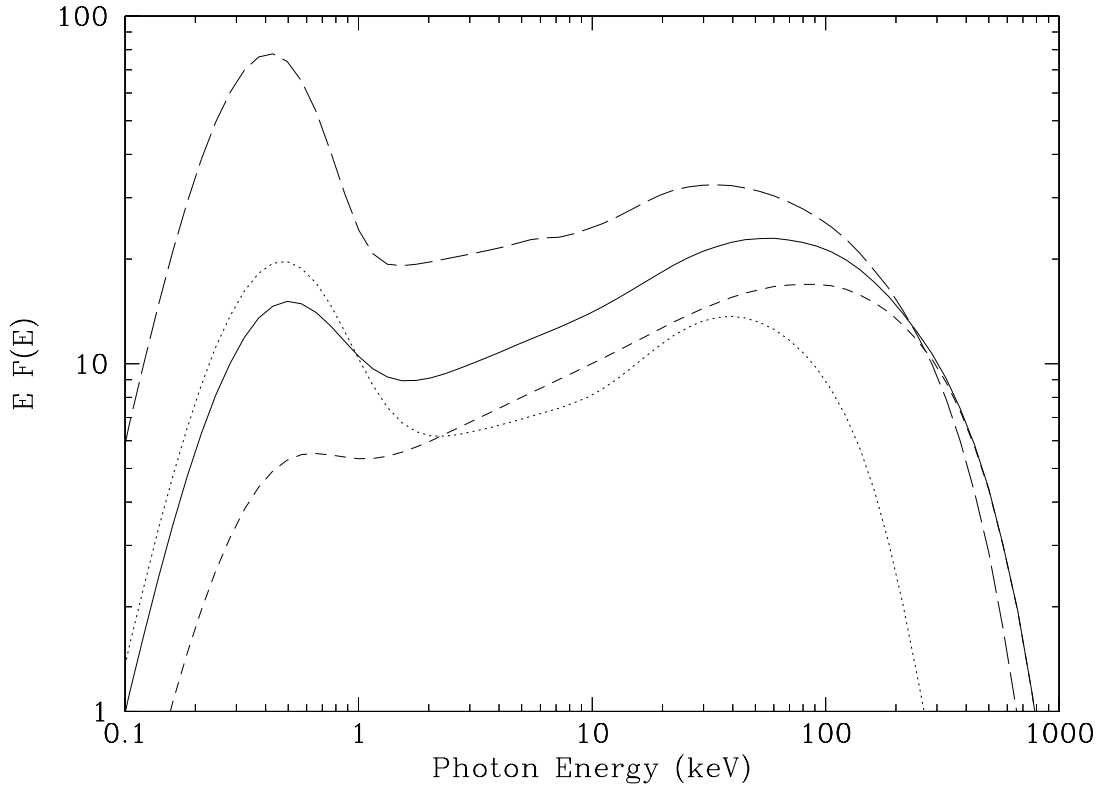,width=.97
\textwidth,angle=-90}}
\caption{Decomposition of the observed spectrum (solid curve) into its
essential components: the intrinsic AR spectrum (short-dash) plus the
reflected component (emerging from the top of the transition layer;
dotted curve) multiplied by $\Omega = 0.5$. The observed spectrum for
the case of $\tau_{\rm trans} = 0$ is also shown by the long-dashed
curve for comparison (shifted for clarity). See text for a further
discussion.}
\label{fig:decomposition}
\end{figure*}

Figure (\ref{fig:decomposition}) shows the observed spectrum (solid
curve), comprised of the intrinsic AR spectrum (short-dash) and the
reflected component (emerging from the top of the transition layer;
dotted curve) multiplied by $\Omega = 0.5$. Also shown for comparison
is the observed spectrum for the case of $\tau_{\rm trans} = 0$
(long-dash).  All the intensities propagate in the upward
direction. Notice that due to the presence of the transition layer,
the reflected component is much harder than the reprocessed component,
which would be the ``normal'' reflection/reprocessing component
without this layer. Notice also that the bump around $\sim 40$ keV
normally attributed to the reflected component is broad (compared with
the long-dashed curve), and so the reflected component is here less
noticeable.

Furthermore, the combined power below $2$ keV accounts for only
$25\;\%$ of the total, whereas the corresponding fraction is about
$50\;\%$ in the standard (static) two-phase model. This large power
coming out in low energy photons was the main reason why the standard
two-phase corona-disk model failed to account for the observations
of Cyg~X-1 (e.g., Gierlinski et al. 1997)

\section{Tests with a Non-Linear Monte Carlo routine}

The Eddington approximation for radiative transfer in the corona and
the transition layer is rather inaccurate for optically thin cases.
It means that the results presented in the previous section cannot be
trusted quantitatively, i.e., one may not use them to produce a fit to
some observed spectra. We thus would like to check our approximate
code with one of the best existing codes on the Comptonization and
energy balance of the corona.  Namely, we use the slab-geometry ADC
model of Dove, Wilms, \& Begelman (1997), which uses a non-linear
Monte Carlo (NLMC) routine to solve the radiation transfer problem of
the system. The free parameters of the model are the seed optical
depth $\tau_e$ (the optical depth of the corona excluding the
contribution from electron-positron pairs), the blackbody temperature
of the accretion disk and its compactness parameter, $\lbb$, and the
heating rate (i.e., the compactness parameter), $\lc$, of the ADC.
The temperature structure of the corona is determined numerically by
balancing Compton cooling with heating, where the heating rate is
assumed to be uniformly distributed.  The $e^-e^+$-pair opacity is
given by balancing photon-photon pair production with annihilation.
Reprocessing of coronal radiation in the cold accretion disc is also
treated numerically. For a more thorough discussion of the NLMC
routine, see Dove, Wilms, \& Begelman (1997). The transition layer is
treated identically to the corona, accept here the heating rate is set
to zero. Therefore, the transition layer, numerically modeled using 8
shells, each with equal optical depth ${\rm d}\tau = \tau_{tr}/8$,
will obtain the Comptonization temperature due to the radiation field
from both the corona and the accretion disk.
  
\begin{figure*}
\centerline{\psfig{file=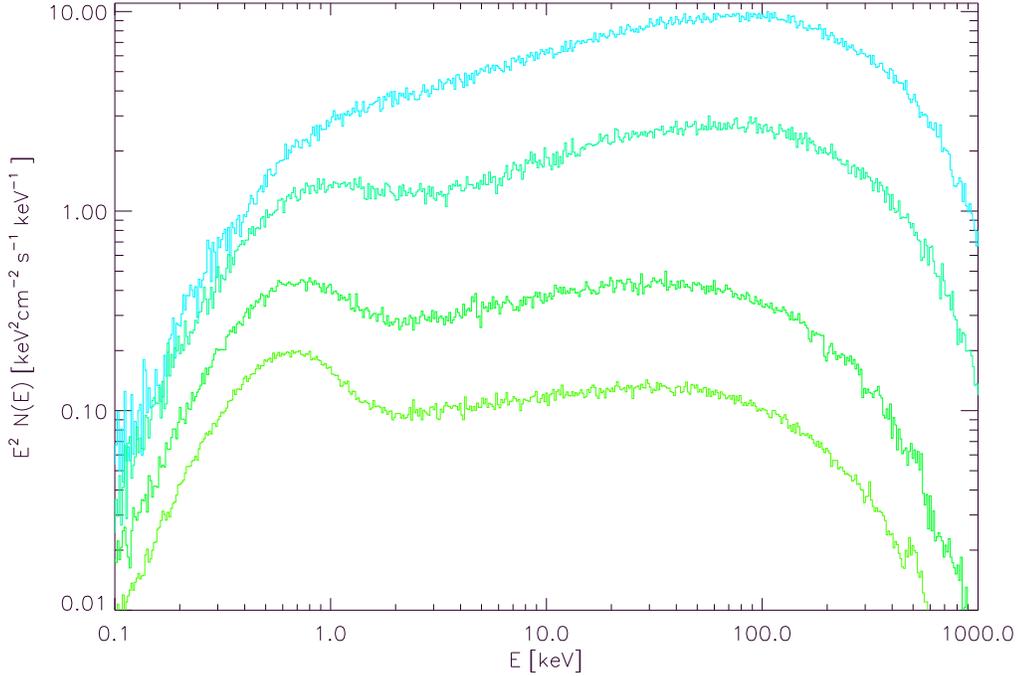,width=.9\textwidth}}
\caption{The predicted spectrum for various values of the transition
layer optical depth computed with the non-linear Monte-Carlo
code. Planar geometry is assumed. From top to bottom, $\tautr = 10, 5,
2.5,$ and $1.0$.}
\label{fig:sequence-of-spectra-dove}
\end{figure*}

The model contains three regions: (1) A cold accretion disk, assumed
here to have a temperature $kT_{\rm min} = 150$ eV, (2) the transition
layer, situated directly above the cold disk, and (3) the corona,
situated directly above the transition layer. Plane parallel geometry
is assumed.  In Figure \ref{fig:sequence-of-spectra-dove}, we show the
resulting broad band spectra for the model parameters tested. This
figure is to be compared with figure (\ref{fig:sequence_of_spectra})
obtained with the Eddington approximation code. We found that the
latter was systematically off in the energy balance for the corona,
i.e., it was always hotter than found by the NLMC code. However, the
qualitative behavior of the system is the same in both codes, which is
what we expected. We plan to use the NLMC code in future work to
attempt to fit some actual spectra of GBHCs and AGN.  For all models
in Figure \ref{fig:sequence-of-spectra-dove}, $\lc = 2$, $\lbb =
0.01$, $kT_{\rm min} = 150$ eV, and $\tau_c = 0.3$. These parameters
correspond to the model producing the maximum corona temperature. In
contrast to models in which $\tau_{\rm tr} = 0$, the corona
temperature for a given value of $\tau_c$ is not simply a function of
$\lc/\lbb$. To see this, consider the case where $\tau_{tr} \gg
1$. Here, the albedo of the disk is essentially unity, and therefore
all of the soft photons emitted will be from the intrinsic flux of the
disk (no reprocessing). Therefore, setting $\lbb
\ll 1$ yields the maximum coronal temperatures possible.  Note that
the maximum temperature levels out as $\tau_{tr} \rightarrow
\infty$. Although, in this limit, there is no reprocessing of hard
X-rays in the cold disk, there is still ``reprocessing'' in the
transition layer.  As $\tau_{tr}$ increases, more coronal radiation is
down-scattered to the Compton temperature of the transition layer,
which is $kT_{tr} \sim 1 - 4$ keV. Even at these temperatures, Compton
cooling of this ``reprocessed'' radiation in the corona is very
efficient.

It is interesting to note that, only for $\tau_{\rm tr} \simgt 10$,
the coronal temperature is high enough such that the corresponding
spectrum of escaping radiation is hard enough to describe the
observations of Cyg~X-1. (The canonical value of the photon power-law
of Cyg~X-1 is $\Gamma = 1.7$; for $\tau_c = 0.3$, this power law
corresponds to $kT_c \sim 150$ keV). It is probably unphysical,
however, to assume the transition layer is completely ionized for such
large optical depths. In fact, the numerical model for $\tau_{\rm tr}
= 10$ predicts a temperature of $kT_{\rm tr} \sim 500$ eV near the
bottom of the layer.  Therefore, even with the advent of transition
layers, it still appears unlikely that a global slab geometry ADC
model can have self-consistent temperatures high enough to reproduce
the observed hard spectra of Cyg X-1 and other similar GBHCs.

\section{Discussion}\label{sec:discussion}

By considering the irradiated X-ray skin close to an active magnetic
flare above a cold accretion disk, we have shown that solutions for
the skin equilibrium structure found with the usual assumption of a
constant gas density are unstable in a broad range of parameter state.
When the pressure equilibrium is taken into account, one finds that
two stable states (one cold and one hot) are possible. For the case of
GBHCs, we showed that the low temperature equilibrium state is
forbidden due to a high value of the ionizing flux. Thus, the X-ray
irradiated skin of GBHCs must be in the hot equilibrium configuration,
where the gas is at the local Compton temperature ($k T\sim$ few keV)
with the radiation field from both the corona and the cold disk.  In
fact, even for global ADC models of GBHCs, such a transition layer is
found to be likely.

The transition layer may be thought of as a partially transparent
mirror. Crudely, some of the photons are scattered back without a
change in energy, and the rest proceed to the cold disk and suffer the
usual transformation into soft disk photons. This effect changes the
integrated albedo of the transition layer, so that a smaller fraction
of the incident X-rays is used to create the soft radiation that is
the dominant source of cooling in the two-phase model. In addition,
the Compton down-scattering in the transition layer does not
contribute to the cooling because the energy gained by an electron in
the layer is later used to up-scatter the softer photons coming from
the cold disk (whereas without the transition layer this energy would
be used to produce the soft radiation).

This reduction in X-ray reprocessing yields a lower Compton cooling
rate within the corona, and higher coronal temperatures than previous
ADC models are allowed. Using the NLMC routine, we have found that for
global ADC models with $\tauc \sim 0.3$, the coronal temperature can
be as high as $\sim 150$ keV {\em if} the optical depth of the
transition layer is $\tautr \simgt 10$. These coronal properties are
what is needed to explain the X-ray observations of GBHCs such as
Cyg~X-1. In addition, for $\tau_{\rm trans}\gg 1$, the predicted
reprocessing features as well as the thermal excess should be
substantially smaller than that of previous ADC models in which the
transition layer was not considered.  This reduction of the
reprocessing features is crucial for the model being consistent with
the observations of GBHCs (e.g., Gierlinski et al.  1997, Dove et
al. 1998).

It is interesting to note that if magnetic flares have the same
geometry and compactness in GBHCs as they do in AGN, the existence of
the transition layer means that GBHC spectra should be harder than
those in typical Seyfert 1s (where the layer is non-ionized, see
Chapter 5).  Further, for $\tau_{\rm trans}\gg 1$, the spectrum may be
somewhat different from that of single cloud Comptonization plus a
{\it standard} cold reflection component, because the reflection
component here is broadened by scatterings in the transition layer. It
is possible that this effect explains Gierlinski et al.'s (1997)
finding that the Cyg~X-1 spectrum cannot be fit with one component.

The observed soft X-ray excess should contain comparatively less power
than the hard component, in contrast to Seyfert 1s. The difference is
caused by the difference in the X-ray reflection albedo $a$, since the
fraction of the energy reflected as the soft disk radiation is
$(1-a)$. The albedo is only $10-20$ \% in Seyferts (e.g., Magdziarz \&
Zdziarski 1995), whereas we found the albedo for GBHCs to be as large
as $a\sim 0.75$, which still may be not the highest possible, since a
further testing with a range of geometries is needed.  The relatively
small disk emission is consistent with observations of Cyg~X-1, but
was listed as one of the primary problems with the two-phase model by
Gierlinski et al. (1997), who used the standard X-ray reflection
formula (i.e., with $a \sim 0.1-0.2$).

Note also that the Thomson optical depth of the flares should be
similar in GBHCs and Seyfert Galaxies, and therefore so should the
electron temperature within their ARs (see Chapter 3; this aspect of
the model does not depend on $M$).

Gierlinski et al. (1997) found that an anisotropy break (that is not
seen in Cyg~X-1 data) is always present in the patchy two-phase
model. As was discussed in Poutanen \& Svensson (1996), the anisotropy
break occurs where the second order scattering (of the disk radiation
in the corona) peaks. However, as we found from our numerical results,
the anisotropy break disappears as the optical depth of the transition
layer increases. The reason for this is the following: the reprocessed
continuum is no longer the blackbody emission (which was assumed by
Gierlinski et al. 1997, and Poutanen \& Svensson 1996) and is quite
broad. Compton scattering broadens any initial photon distribution,
and therefore the second order scattering of the reprocessed continuum
becomes a very diffuse function, with a shallow peak.  Among other
effects that should reduce the anisotropy break is the variance of the
temperature of the disk emission with distance from the flare (which
we neglected here in one-zone approximation).  The cold disk emission
will then be a sum of blackbodies with different temperatures, and
will be even broader than what we obtained in our
calculations. Finally, since the overall spectrum is a sum from flares
with a distribution of temperatures and the optical depths (see
\S \ref{sect:rtau}), second order scattering will mean different 
amplification factors for the soft photons entering different flares,
which will further dilute the break.  Thus, we believe that earlier
contrasting results found by Gierlinski et al. (1997) are due to an
over-simplification of the magnetic flare model physics.

The reflected component in the observed spectrum must be less
pronounced or not observable, depending on the transition layer
optical depth, which is again consistent with observations (e.g.,
Zdziarski et al. 1996).

Most of the hard X-rays do not penetrate through the transition layer,
and the spectrum gets softer as it approaches the cold disk. The Iron
K$\alpha$ line, small to start with due to the small amount of
reprocessing of coronal radiation, is completely smeared out by the
time the radiation escapes the system.  No line photons are created in
the transition layer itself, because we found that the Compton
equilibrium state typically resides at the ionization parameter $\xi
\simgt 10^4$, whereas no fluorescent iron line emission is produced
for $\xi \simgt 5\times 10^3$ (Matt, Fabian \& Ross 1993, 1996). The
observed weak K$\alpha$ line may then be arising from the cold outer
disk. The same is true for the Fe edge. Note that observationally, it
is very hard to detect a broad Fe edge in the case of Cygnus X-1
(Ebisawa 1997, private communication).

Thus, as far as we can see, observations of the hard state of the
GBHCs do not rule out magnetic flares as the source of X-rays, and
instead support this theory. Earlier findings on the contrary were
affected by the use of assumptions that magnetic flares or the X-ray
reflection process in GBHCs cannot deliver. Furthermore, it appears
that the observed X-ray spectrum of Cyg~X-1 can be explained by the
transition optical depth of $\sim 3$, which is physically plausible,
and that, apart from the self-consistent difference in the structure
of the transition layer, the same parameters for magnetic flares might
be used in both AGN and GBHCs to explain their spectra.

The spectral calculations reported here could also be appropriate for
the static patchy corona model if the upper layer of the disk were
hotter than usually assumed. Indeed, it is not hard to imagine that
the upper layer is being heated in a way similar to heating of the
Solar corona. If the temperature of the layer is few keV and its
Thomson optical depth is close to $\sim 3$, then the spectra produced
in this situation may be close to the observed hard spectra of the
GBHCs. This possibility does not appear to have been explored by
previous workers. At the same time, even if such a static model could
remove the problems for the GBHCs spectra, one would need to explain
why the upper layer of the disk in Seyfert Galaxies is not being
heated in a similar manner. Thus, the real strength of the magnetic
flare model of the active regions is in the fact that this is the same
physics that explains the spectra of both Seyfert Galaxies and GBHCs.

	\chapter{X-ray reflection in AGNs and The BBB}
	\section{BBB in Seyfert Galaxies and the Transition Layer}
\label{sect:bbbs}

The UV to soft X-ray spectrum of many Active Galactic Nuclei (AGNs)
may be decomposed into a non-thermal power-law component and the
so-called Big Blue Bump (BBB), which cuts off below about 0.6 keV
(e.g., Sanders et al. 1989).  A major obstacle in constraining the
characteristics of the BBB has been that it lies in the difficult to
observe EUV and very soft X-ray region.  In recent years, however,
there has been considerable progress in this direction (e.g., Walter
\& Fink 1993; Walter et al. 1994; Zhou et al. 1997). The observed
spectral shape of the bump component in Seyfert 1's hardly varies,
even though the luminosity $L$ ranges over 6 orders of magnitude from
source to source.  Walter et al.  (1994) concluded that the cutoff
energy $E_c$ of the BBB (when fitted as a power-law with exponential
rollover) is very similar in different sources whose luminosities vary
by a factor of 10$^4$. Note that we here will refer to the results of
the second method of fitting the spectral shape of the bump suggested
by these authors, i.e., using prescription (B) (see their \S 4.2). The
first prescription (A) assumes that the ultraviolet to the far
ultraviolet spectral slope remains constant, which is contrary to what
one would expect based on our model of the accretion disk emission. In
this model, the far-UV component is due to reprocessing in the
transition region of the X-rays from magnetic flares (as elaborated
below in this Chapter), and has no relation to the intrinsic disk
emission, which should show up at the disk effective temperature of
$\sim $ few eV. Thus, it is important to allow the reprocessed
spectrum and the disk intrinsic emission to vary with respect to one
another in the fits, and prescription (B) satisfies this criterion
better than prescription (A). Although the data of Walter et
al. (1994) were not precise enough to distinguish between different
emission mechanisms, Walter et al. (1994) pointed out that if the
variations in the ratio of the soft X-ray excess to UV flux from one
object to another are interpreted as a change in the temperature of
the BBB, then this change is smaller than a factor of 2.  Confirming
conclusions follow from the work of Zhou et al. (1997).

Early theoretical work on the BBB spectrum focused on the role of
optically thick emission from the hypothesized accretion disk
surrounding the central engine (e.g., Shields 1978; Malkan \& Sargent
1982; Czerny \& Elvis 1987; Laor \& Netzer 1989).  However, this
mechanism is now facing several obstacles (e.g., Barvainis 1993;
Mushotzky et al. 1993).  An alternative model, in which the BBB is
interpreted as thermal, optically thin free-free radiation, has been
proposed by Antonucci \& Barvainis (1988), Barvainis \& Antonnuci
(1990), Ferland et al. (1990), and Barvainis (1993). There are strong
arguments against this emission mechanism as well (Malkan 1992).

It seems to us that the observations Walter \& Fink (1993) and Walter
et al. (1994) are difficult to interpret in the terms of {\it
accretion disk} thermal emission. AGNs are thought to accrete both
from their nearby environments via the Bondi-Hoyle process and from
the tidal disruption of stars, though over time, the former is
dominant (e.g., Melia 1994).  At least initially, the accretion rate
is therefore $\dot M\propto M^2$, where $M$ is the black hole mass,
but this constitutes a runaway process in the sense that
$L/L_{Edd}\propto t$, where $t$ is the time, and $L_{Edd}$ is the
Eddington luminosity.  When $L\rightarrow L_{Edd}$, the outward
radiation pressure presumably suppresses the inflow, with the effect
that $L$ saturates at the value $\sim L_{Edd}\propto M$.  A second
argument in favor of the supposition that the ratio $L/L_{Edd}$ is
relatively independent of $M$ is the fact that we observe very similar
X-ray spectra for objects of very different luminosities (e.g.,
Zdziarski et al.  1996), for otherwise the disk structure would differ
from source to source, giving rise to different spectra. As a
statistical average, we thus expect that $L\propto M$.

In view of this, let us next examine how the various different
emission mechanisms fare in their prediction of the BBB cutoff energy
$E_c(L)$. For any radiation process, the flux $F$ scales as $L$ over
the emitting area, which itself scales as $M^2$. Thus, in general we
expect that $F\propto L^{-1}$. The blackbody flux is $F_{\rm
bb}=\sigma T^4$, where $T$ is the effective temperature, and so
$T\propto L^{-1/4}$.  Thus, when $L$ varies by 4 orders of magnitude,
it is expected that $T$ ought to itself vary by a factor of 10. This
is not consistent with the observations discussed above. Further, the
disk effective temperature is
\begin{equation}
T_{\rm eff}\simeq 3 \times 10^4 \, M_8^{-1/4}\, \dm_{-2}^{1/4}
\left[(1-f)\right]^{1/4}\, {\rm K} \ ,
\label{deft}
\end{equation}
where $\dm_{-2}\equiv \dm/0.01$. This temperature is considerably
smaller than the roll-over energies ($E_c \sim 50-80$ eV) in the BBB
emission found by Walter et al. (1994).

A more sophisticated treatment of the disk structure in its inner
region shows that the scattering opacity may dominate over the
absorptive one, and thus the emission spectrum may differ from that of
a blackbody. For example, the disk may radiate as a `modified
blackbody' (Rybicki \& Lightman 1979), for which the flux is then
given by
\begin{equation}
F_{\rm mb}\sim 2.3 \times 10^7 T^{9/4}\rho_d^{1/2}
\erg\;\cm^{-2}\;\sec^{-1}{\rm ,}
\end{equation}
where $\rho_d$ (in $\gm\;\cm^{-3}$) is the disk mass density and $T$
is in Kelvins. For accretion disks, $\rho_d\sim M^{-1}$ (e.g., \S
\ref{sect:adstr}), and so $T\sim M^{-2/9}$, which again is not
consistent with the data, if $L$ is in general proportional to $M$. We
believe that more complicated emission mechanisms, i.e., accretion
disk atmosphere calculations, are unpromising as well because the
emission will always be characterized by some sort of temperature,
either equal to $T_{\rm eff}$ or some function of it, and we see no
physical reason why this temperature would not vary with $\dm$ and
$M$. These last two quantities most likely vary by orders of magnitude
for different sources in any large sample of QSO or Seyferts. Our
conclusion is thus that the BBB emission does not come from the
accretion disk intrinsic emission.

\section{The Thermal Ionization Instability for AGN}
\label{sect:inst_agn}

We now discuss the thermal instability (Chapter 4) of the surface
layer for AGN.  The most important distinction from GBHC case is the
much higher mass of the AGN, and thus an ionizing X-ray flux is
smaller by $\sim 7$ orders of magnitude (see equation
\ref{xflux}). The minimum X-ray skin temperature is again approximated
by setting the blackbody flux equal to the incident flux, assumed to
be $q\sim 10$ times less than the escaping coronal flux.  The gas
pressure dominated solutions gives
\begin{equation}
T_{\rm min}\simeq 1.5\times 10^5\, l^{1/4} \,\alpha^{1/40} \,M_8^{-9/40}
\left[{\dm\over 0.005}\right]^{-1/20} \,\left(1-f\right)^{-1/40} \,
\left({q\over 10}\right)^{-1/4}\,
{\rm ,}
\label{tminsg}
\end{equation}
whereas the radiation-dominated one yields
\begin{equation}
T_{\rm min}\simeq 7.\times 10^4\, l^{1/4} \,M_8^{-1/4}
\left[{\dm\over 0.05}\right]^{-1/4} \,\left(1-f\right)^{-1/4}\,
\left({q\over 10}\right)^{-1/4}\, \ .
\label{tminsr}
\end{equation}

These estimates show our main point right away: the lower X-ray flux
density in AGN may allow the transition layer to saturate at either
the cold equilibrium state or the ``island'' state, whereas that was
not possible for GBHCs. To investigate this idea, we ran XSTAR as
described in Chapter 4, but for parameters appropriate for the AGN
transition layer. In particular, we accepted that the disk blackbody
emission temperature is $6$ eV (cf. equations \ref{tminsg}
\& \ref{tminsr}), and that the gas density is $10^{17}$ cm$^{-3}$.
The X-rays illuminating the transition region are assumed to mimic the
typical Seyfert hard spectra, i.e., a power-law with photon index
$\Gamma = 1.9$ and the exponential roll-over at 100 keV. The ratio of
the X-radiation ram pressure to the X-ray skin unperturbed pressure is
chosen to be correspondingly higher $\fx/cP_0\sim 10^2 - 10^3$ (see
equation \ref{prat}; we also believe that $\dm \simlt 0.05$ for
typical Seyferts, as explained in Chapter 6). As explained earlier in
\S \ref{sect:instability}, XSTAR produces inaccurate results below 
$T\sim \Tmin$, so that these regions of the ionization equilibrium
curve should be disregarded. To test sensitivity of our results to the
parameters of the X-ray spectrum, we also ran a case for $k T_{\rm bb}
= 12$ eV and the rollover energy at $200$ keV.

\begin{figure*}[t]
\centerline{\psfig{file=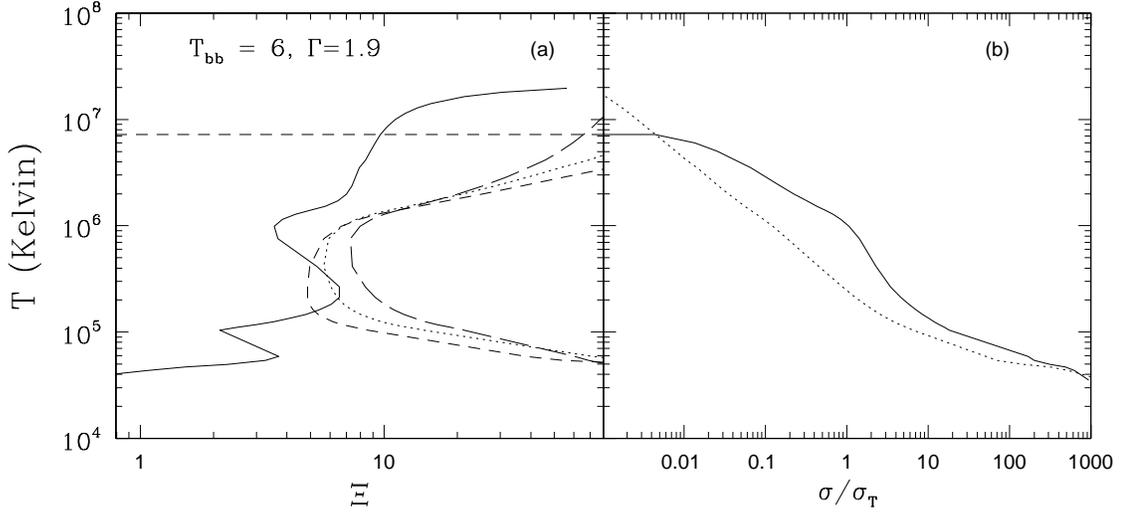,width=.99\textwidth,angle=-90}}
\caption{Same as Figure (\ref{fig:gbhc1}), but for the
AGN transition layer. (a): Solid curve shows the ionization
equilibrium curve. Using that equilibrium, the pressure equilibrium
arguments are used to estimate the corresponding $\Xi$. For the three
pressure equilibrium curves the parameters are: $\tau_x = 0.5$, 1, 2
and $P_0c/\fx = 10^{-3}$, $10^{-3}$ and $10^{-2}$ for the dashed,
dotted and long-dashed curves, correspondingly. A stable configuration
of the transition layer is achieved at location where the ionization
and pressure equilibrium curves cross. (b): the gas absorption
opacities. The solid curve is the X-ray opacity $\mean{\sigmax}$,
whereas the dotted curve depicts the UV-opacity $\mean{\sigmauv}$ }
\label{fig:agn1}
\end{figure*}

We show results of these two simulations in Figures (\ref{fig:agn1})
and (\ref{fig:agn3}). Notice that the ``cold'' equilibrium branch,
i.e., the region with $T\sim 10^5$ K is more stable than the island
state. None of the pressure equilibrium curves intercepts the island
state. The two curves with $\tau_x = 0.5$ and $1$ do intercept the
cold equilibrium state, but the more optically thick case with $\tau_x
= 2$ does not in Figure (\ref{fig:agn1}), whereas all the three curves
intercepts the cold state in Figure (\ref{fig:agn3}). On both Figures,
the horizontal lines are caused by the UV opacity exceeding the X-ray
opacity for low and high temperature in one of the simulations, thus
leading to the negative estimates of $\Xi$.  Physically, it means that
the UV pressure on the gas exceeds that of the incident X-rays, so
that the net radiation force points upward. A wind may be induced in
these temperature ranges (i.e., below $10^5$ and above $\sim 10^7$ K).
Clearly, more detailed future work is needed to investigate the
parameter space where the cold state is stable. However, its existence
is required by observations of X-ray reflection and fluorescent iron
lines in Seyferts, as we will enunciate in Chapter 7.

\begin{figure*}[t]
\centerline{\psfig{file=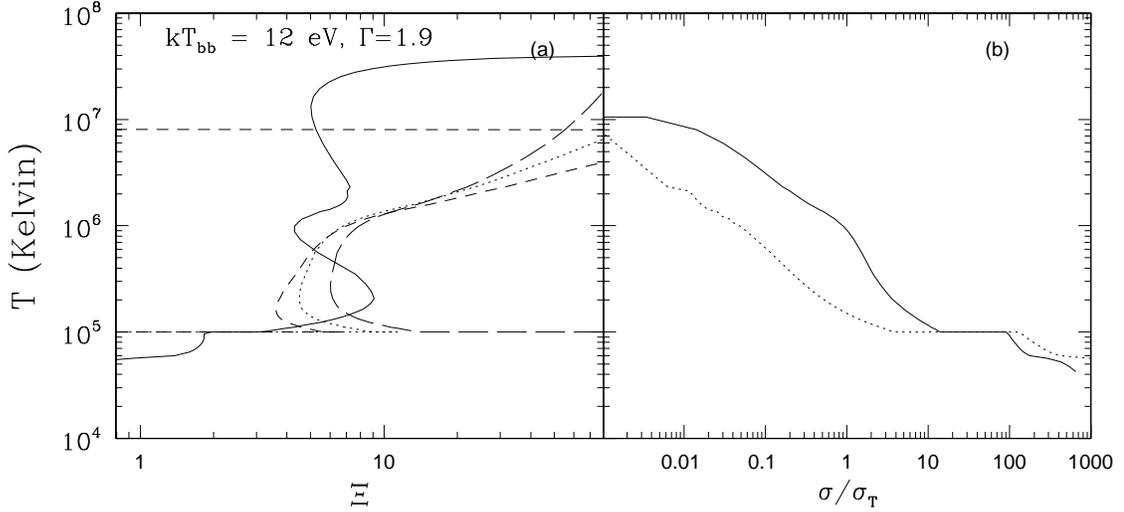,width=.99\textwidth,angle=-90}}
\caption{Same as Figure (\ref{fig:agn1}), but for the effective
temperature $kT_{\rm bb} = 12$ eV.}
\label{fig:agn3}
\end{figure*}

\section{The Origin of the Big Blue Bump}
\label{sect:obbb}

As was elaborated in \S \ref{sect:bbbs}, there is no consistent
explanation for the origin of the BBB, one of the most prominent
features in the AGN spectra. We believe that our theory of the
ionization pressure instability may offer a plausible explanation for
the BBB emission. As the ionization equilibria calculations show,
there is no stable solution for the transition region in the
temperature range $2-3\times 10^5 \simlt T \simlt 10^7$ Kelvin (see
Figure \ref{fig:agn1} and \ref{fig:agn3}). Furthermore, temperatures
below the effective temperature of the X-ray radiation are also
forbidden. The only low temperature solution permitted by the
stability analysis is the one with a temperature around $2\times
10^5$ Kelvin.

Notice that the Rosseland mean optical depth to the UV emission is of
order 1 to a few for this temperature range, as seen from simulations
(see panels (b) in figures \ref{fig:agn1} and \ref{fig:agn3}). The
radiation spectrum produced by the transition layer will be either a
blackbody spectrum, or a modified blackbody (with recombination lines
as well, of course).  This spectrum may explain the observed roll-over
energies in the BBB emission from Seyfert Galaxies (see references in
\S \ref{sect:bbbs}) rather naturally. Since the moderately optically
thick emission spectrum will saturate at photon energy of $\sim 2- 4
\times T$, $T\sim 2 \times 10^5$ provides an excellent match to the
observed roll-over energies of $\sim 40-80$ eV.

As one can check using Field's (1965) stability criterion, the
transition layer radiating via blackbody emission is thermally
stable. Furthermore, even the modified blackbody emission mechanism
stabilizes the instability. This consideration shows that the
transition layer will be even more stable in the cold state in the
realistic optically thick calculation, which would take into account
spectral reprocessing of the incident spectrum. Further, notice that
effective temperature $T_{\rm eff}$ cannot be seriously lower than
$T_{\rm eff}
\simeq 1 \times 10^5$ Kelvin. The point here is that the compactness
parameter in the two-phase patchy corona model cannot be much lower
than unity (see \S
\ref{sect:tpmodel} and equations \ref{tminsg} \& \ref{tminsr}).

Thus, the temperature of the BBB is fixed by the atomic physics, in
particular by the fact that many atomic species have ionization
potential close to 1 Rydberg, which corresponds to a temperature of $T
\simeq 1.5 \times 10^5$ K.  This temperature turns out to be
independent of the number of magnetic flares, and so it is independent
of the X-ray luminosity of the source, as found by Walter \& Fink
(1993) and Walter et al.  (1994).

\section{The X-ray Baldwin effect}\label{sect:xbaldwin}

Here we will present a simple argument to demonstrate how our theory
of the ionized transition layer may account for the recent observation
of the X-ray Baldwin effect. Namely, it has been found (Nandra et
al. 1997) that all AGNs that are very luminous in the X-ray band
(i.e., $\lx > 10^{44}-10^{45}$ erg/sec in the 2-10 keV window) show
little or no Iron line emission, in sharp contrast to lower luminosity
AGNs. This is not easily understood without invoking an instability,
as explained in the following citation of A. C. Fabian (1998), which
we received as a private communication:

''Nandra et al (1995, 1997) find no evidence for the iron line or any
reflection features in most quasars. In the second paper, it is shown
that the equivalent width of the iron line diminishes with source
luminosity above about $10^{44}$ erg/sec and it is suggested that the
disk is increasingly ionized, perhaps because the objects are closer
to the Eddington limit. This is puzzling because the disk must jump
from being `cold' to completely ionized, otherwise there would be
intermediate objects with even larger equivalent widths when the
surface iron in the disk is H or He-like (Matt, Ross \& Fabian
1994). There should at the same time be a deep broad iron edge which
is not seen. (There is strong iron absorption in the reflection
continuum in both a cold and an ionized disk, but it only shows up in
the latter case because the lack of oxygen and iron-L absorption below
the edge make the reflection continuum strong there.)

Possibly there is a jump between states caused by the way in which the
corona is energized. When the Eddington ratio is low the magnetic field
amplified by differential motions in the disk extracts most of the energy
released to well above the disk. There is then little thermal energy
release from within the disk and it is essentially cold and has a sharp
surface for reflection purposes. When however the ratio is increased there
may be a flip at some level to a somewhat thicker, radiation-pressure
supported (inner)  disk which has a fuzzy, highly-ionized surface (i.e.
the density drops gradually with height into the corona). Studies of the
behavior of disks in the Galactic Black Hole Candidates will be
instructive here.''
%

We think that the pressure ionization instability described in this
Chapter and Chapter 4 may be the process that explains the
disappearance of the line. First, note that the energy equilibrium in
most luminous AGN may push the transition layer over ``the edge'' of
the stable cold solution (i.e., $\sim 3\times 10^5$ K) to the unstable
region.  The disk effective temperature is
\begin{equation}
T_{\rm eff}\simeq 1 \times 10^5 M_8^{-1/4}\, \dm^{1/4}
\left[(1-f)\right]^{1/4} \ .
\label{tefd}
\end{equation}
The actual minimum temperature may even be higher than this estimate,
since the accretion disk itself may become effectively thin for
accretion rates $ \dm $ close to unity (see SZ94), and the emission
may become a modified blackbody emission. A particular accretion disk
model (with $\alpha$ and other parameters specified) and an exact
treatment of the transition layer ionization, pressure and energy
equilibria are needed to find the transition layer temperature, but it
is rather natural to expect that this temperature may be larger than
the disk effective temperature by a factor $\sim$ few. For example,
Ross, Fabian \& Minishige (1992) finds that the disk spectrum may be
for some conditions better approximated by a Wien spectrum with $T=
2.5 T_{\rm eff}$. Thus, for accretion rates close to the
Eddington-limited one, the transition layer temperature may be higher
than $3 \times 10^5$ K {\it irrespectively of the strength of the
X-ray coronal heating}, which would make it impossible for the layer
to be on the cold stable branch of the solution.

The next energetically stable solution exists above $T\sim 100$ eV
(see Figure \ref{fig:agn1}). It could be either the short ``island''
state, or could be the whole region of the curve upwards of $T\sim
100$ eV, since we found that for steeper X-ray illuminating spectra
the $\sim 200$ eV $- 1$ keV region may become stable. If the
transition layer indeed was at this temperature range, the
observations by Nandra et al. (1997) would be puzzling, since
fluorescent iron line emission for $\xi \sim 500-5000$ is even
stronger than it is for the cold stable state (Matt et al. 1993,
1996).

However, when the disk switches from a gas to a radiation-dominated
solution, and when the disk intrinsic flux $\fdisk$ becomes comparable
to or larger than the X-ray flux $\fx$ from the active region, the
pressure equilibrium for the transition layer may be of a different
nature than we found it to be in \S \ref{sect:morepeq}. Specifically,
in the latter case we neglected the contribution of the intrinsic disk
flux to the radiation pressure in the transition region, since
$\fdisk\ll \fx$ was shown to be the case. Here, however, the disk
intrinsic flux produces the main force on the gas in the transition
layer. In fact, one can neglect the ram pressure of the incident
X-radiation in this limit.

The unperturbed gas pressure in the transition region, $P_{\rm
gas,s}$, can be found by first finding the overall pressure required
to maintain the equilibrium (cf. equation \ref{p0}), and then
subtracting the radiation pressure due to the escaping disk flux
$\fdisk$:
\begin{equation}
P_{\rm gas,s} \simeq P_0 - \tau_d \fdisk/c = {\tau_x\over 2}
{\fdisk\over c}\, \left[\zeta - 2 {\tau_d\over\tau_x}\ {\rm ,}
\right]
\label{psg}
\end{equation}
where $\tau_d$ is the Rosseland mean total optical depth of the
transition layer to the escaping radiation, whereas $\tau_x$ is the
Thomson optical depth.  We have used standard accretion disk theory in
the formulation of SZ94 to arrive at this expression. The parameter
$\zeta$ is the parameter introduced by SZ94 to account for the
uncertainty in the vertical averaging of the radiation diffusion out
of the disk (see \S \ref{sect:adstr}). Different authors use $\zeta$
ranging from $2/3$ to $2$. In any event, this estimate shows that, in
the given case, the gas pressure may account for a tiny fraction of
the transition region pressure (negative $P_{\rm gas,s}$ in equation
\ref{psg} means that a wind will be induced, or the treatment needs to
be refined to take into account radiation anisotropy, etc). We then
have $P_{\rm gas,s} \ll
\fdisk$. Now, if $\fx$ is not too much smaller than $\fdisk$,
then it also implies $\Xi = (\fx/cP_{\rm gas,s}) \gg 1$. For high
values of $\Xi$, the only stable equilibrium is the Compton
equilibrium, and here the transition layer is completely ionized. The
latter fact can be the explanation for the absence of the iron lines
in very luminous AGN.

We should also note that this goes in line with the finding of no BBB
in high-luminosity AGNs of Zheng et al. (1996) and Laor et al. (1997),
since, just as we found in the case of GBHCs (see Chapter 4, figures
\ref{fig:sequence_of_spectra} and
\ref{fig:sequence-of-spectra-dove}), the completely ionized transition
layer with $\tau_x\simgt 1$ may reduce the reflected UV component.
Since we also think that X-rays do not contribute the majority of
power in luminous Radio Quiet AGN, the Big Blue Bump may be smeared and
reduced to invisibility on the background of the dominant thermal disk
emission.

Finally, it is also curious to note that since the disk effective
temperature scales as $M^{-1/4}$, AGNs with $M\sim 10^{10}\msun$ may
be cold enough that the transition layer (if the corona is full) once
again may exist on the cold stable equilibrium solution, and the iron
line may re-appear again, in line with the suggestions of
Prof. Fabian.

\section{Constraints on Magnetic Flares from X-ray reflection}
\label{sect:sconstraints}

Note that the spectral constraints limit the compactness parameter in
both AGNs and GBHCs. In particular, for AGNs, we know that $T_{\rm
eff}$ as given by equations (\ref{tminsg}) and (\ref{tminsr}) cannot
exceed $\sim 2\times 10^5$ K. If $T_{\rm eff}$ was larger, the stable
low temperature state would disappear, and the observations of a low
ionization degree reflector and a cold neutral iron line in AGNs would
be left unexplained by our theory. Thus, $l\sim 0.1 - $ few in AGNs.
Similarly, for GBHCs, the observed soft X-ray emission in the hard
state of GBHCs can be fitted by a blackbody with a temperature $\simeq
150$ eV (see equation \ref{tmincyg}). This low temperature is only
possible if $l$ is smaller than unity, and is as small as $0.1$. In
principle, a better spectrum calculation of the X-ray reflection and
the spectrum formation in the geometry of an active region is needed
in order to set the upper limit on $l$. In any event, it is notable
that the constraints on $l$ from AGNs and GBHCs observations are
rather similar. We will use this fact when modelling the global
spectral behavior of disks in Chapter 7, and will see that it leads to
observationally testable predictions.

	\chapter{Energy budget of the corona}
	\section{Energetics of the magnetically-fed Corona}\label{sect:chapter6}

We shall now turn to the question of the global energy transport by
magnetic flux tubes. This issue is very important, since observations
of Seyfert Galaxies show that probably as much $\sim 50$\% of the
accretion power must be channeled to the corona, and yet it has never
been shown that this can be accomplished by any particular energy
transport mechanism. Furthermore, as we saw in Chapter 4, observations
of GBHCs, if interpreted in the context of the two-phase model,
require even larger portion of the energy to be transported out of the
disk by magnetic flares, such that the global energy transport is
dominated by magnetic energy flux. In fact, the energy balance issue
is equivalent to the question of the normalization of the observed
X-ray spectra, and as such is as important a test of the two-phase
model as the shape of the spectrum itself. This test has never been
conducted before.

The time-averaged magnetic energy flux (from inside the disk to the
corona) $\fmag$ is
\begin{equation}
\fmag = v_b \langle {B^2\over 8\pi}\rangle {\rm ,}
\end{equation}
where $\langle B^2/ 8\pi\rangle$ is the volume average of the magnetic
field pressure in the disk, and $v_b$ is the buoyant rise velocity,
i.e., the average velocity with which a magnetic flux tube is rising
due to buoyancy. It is well known that a magnetic field inside the
accretion disk is an efficient mechanism for angular momentum
transport in the disk (e.g., Shakura \& Sunyaev 1973, Lightman \&
Eardley 1974, Hawley, Gammie \& Balbus 1995). Therefore, it is thought
that given a numerical value for the Shakura-Sunyaev
$\alpha$-parameter, one may constrain the {\it volume average}
magnetic field reasonably well.  This conclusion comes from the fact
that a magnetic field line resists in a known way the shearing
resulting from the differential rotation of the disk. The relevant
magnetic stress is $\mean{B_r B_{\phi}/ 8\pi}$, where $B_r$,
$B_{\phi}$ are the $r$ and $\phi$ components of the magnetic field,
respectively.  This volume average is expected to be of the order of
the average magnetic pressure in a turbulent medium. On the other
hand, the component of the stress tensor responsible for the momentum
transfer in the framework of the standard theory is $\alpha
\ptot$. Accordingly,
\begin{equation}
\langle{B^2\over 8\pi}\rangle\simlt \alpha \ptot\; .
\label{baverage}
\end{equation}
The $ < $ sign in this equation corresponds to a possible case when
the so-called turbulent viscosity is larger than the magnetic
viscosity, such that $\alpha$ is larger than that due to the magnetic
field alone.  Numerical simulations show that the turbulent and
magnetic viscosities are of the same order (e.g., Stone et al. 1996),
and we will assume that this is indeed true.

At the same time, with the standard prescription for viscosity, one
can show that the local total energy flux $\ftot$ (which is equal to
the sum of the magnetic and radiation energy fluxes) is equal to
$(9/8)
\alpha c_s \ptot$ (via equation 4.26 of Frank, King \& Raine 1992, for
example). The fraction of the power transported away from the disk by
magnetic fields is then
\begin{equation}
f\equiv {\fmag\over \ftot} \simlt {v_b\over c_s}\; .
\label{fratio}
\end{equation}
Thus, in order for the magnetic energy flux to amount to a significant
portion of the total energy flux, the buoyant rise velocity $v_b$
should be almost equal to the gas sound speed $c_s$. Vishniac
(1995a,b), however, found that $v_b$ cannot be as large as the sound
speed.  In fact, Vishniac (1995a,b) estimated that the magnetic energy
out-flux only accounts for a fraction $\sim \alpha$ of the radiative
energy transport. This conclusion may be expected to change in the
case of strong magnetic flux tubes, with pressure comparable to the
gas ambient pressure, because their Alfv\'en speed may be closer to
$c_s$ (Vishniac 1997, private communication). Still, the fluid
viscosity is not zero, which leads to a friction between rising flux
tubes and the fluid, and the problem may be further complicated by
interactions between the neighboring tubes as well as other factors
(e.g., Parker 1979, Vishniac 1995a,b). It does not appear reasonable
to us to suggest that $v_b$ can closely approach $c_s$, although it
cannot be completely ruled out at this time. The hard spectrum of Cyg
X-1, in any event, requires $f\simeq 0.8$ ans so can be accounted for
(by the two-phase model) only if $v_b\simeq c_s$ -- clearly an
unphysical situation.

A possible solution to this theoretical difficulty lies in the fact
that the above estimate of the volume average of the magnetic field
(equation
\ref{baverage}) is only correct for a diffuse magnetic field, and a
similar argument carefully applied to the case where most of the field
is localized to strong magnetic flux tubes allows the volume average
magnetic field to be much larger than $\alpha \ptot$, as we will now
show. The reason to suspect that there may be a difference in the
amount of the magnetic field resistance to the differential flow in
these two cases is the following fact. The diffuse and tangled
magnetic field will be strongly coupled to the fluid and thus will
definitely take part in the differential rotation of the fluid, so it
will be stretched and will contribute to the $\alpha$-parameter by
resisting this stretching. At the same time, a flux tube is an entity
of its own, which manifests itself in the fact that the tube can move
with respect to the fluid, e.g., be buoyant. Accordingly, the flux
tube may avoid the stretching by simply not following these motions of
the fluid that try to deform the tube. We need to turn to some
estimates to see if this is the case for accretion disks.

Consider a flux tube that is shaped as a torus with larger radius $a_0
\simlt H$, where $H$ is the disk vertical thickness, and the smaller
radius $a$ ($\pi a^2$ is then the cross sectional area of the
tube). Note that we are concerned here with a flux tube that is
immersed in the accretion disk gas rather than a tube that has already
risen to the top of the disk and is ready to produce a flare, because
it is the former flux tube that carry the energy to the corona. Let us
assume that without differential rotation (see below), the flux tube
would be at some equilibrium state.  In the accretion disk, the two
opposite sides of the torus may be at different radii: $R$ and $\simeq
R + a_0$. The gas at these radii moves with different Keplerian
velocities, and thus the flux tube will be experiencing a shear
force. The differential velocity $v_d$ between the two rings of matter
separated by distance $a_0$ in the disk can be written as $v_d\simeq
c_s (a_0/H)$. Now let us assume that the flux tube is not being
stretched by the differential motion, that is, it moves with some
average azimuthal velocity, as a {\it solid body}. There is a viscous
drag force $D$ on the flux tube in this case, caused by the friction
as the fluid flows by the tube. The magnitude of this force is
\begin{equation}
D \simeq C_d \rho v_d^2 a a_0/2 
\label{drforce}
\end{equation}
(e.g., Parker 1979, \S 8.7, and references there), where $C_d$ is the
drag coefficient.  For the flux tube not to be deformed by the drag
force, the tube magnetic tension should exceed this force. The flux
tube tension force $T$ is given by
\begin{equation}
T = {B^2\over 8\pi} \pi a^2 \; .
\label{tens}
\end{equation}
The ratio of these two forces is
\begin{equation}
{T\over D} \sim {2a\over a_0} C_d^{-1} {\pmg\over \ptot} \left (H\over
a_0\right )^2 \; ,
\label{tod}
\end{equation}
where $\ptot$ is the total disk pressure. For a relatively thick flux
tube, we have $4a/a_0\simlt 1$. The drag coefficient is uncertain in
this equation, since it depends on the level of the fluid viscosity
and many other model dependent factors. The value typically used for
this coefficient for conditions appropriate to accretion disks or
stars is $C_d\sim 1/4$ (following Vishniac 1995; Stella \& Rosner
1984; Sakimoto \& Coroniti 1989; Parker 1979). With this value of the
drag coefficient, equation(\ref{tod}) asserts that for flux tubes with
magnetic field pressure comparable to the equipartition value, and the
size $a_0$ smaller than the disk scale hight $H$, the magnetic field
tension is larger than the drag force applied to the flux tube by the
differential flow in the disk. The tube cannot be deformed by the flow
in this case, and instead is dragged around almost as a solid
body. The contribution of the flux tube to the momentum transfer is
reduced by the ratio of the drag force $D$ to the tension force $
T$. If all the magnetic field is in the form of strong flux tubes for
which the magnetic tension exceeds the drag force, then the limits on
the magnetic field volume average become
\begin{equation}
\langle {D\over T} \pmg\rangle \simlt \alpha \ptot \; ,
\label{fod}
\end{equation}
or, approximately, $\langle \pmg\rangle\simlt \alpha \ptot \langle
(T/D)\rangle$. Note that observations of the Sun suggest that as much
as $\sim 90$\% of the overall magnetic field is concentrated in strong
magnetic flux tubes, at least on the surface (Parker 1979, \S 10.1);
it therefore seems reasonable that most of the field in accretion
disks is contained in the flux tubes too. Depending on the exact value
of the typical flux tube size, $C_d$ and other uncertainties, the
volume average of the magnetic field can be considerably larger than
$\alpha \ptot$. We can estimate the ratio of the magnetic energy flux
$F_{\rm m}$ to the radiation energy flux as
\begin{equation}
{f\over 1-f}\, \simeq {v_b\over c_s} {1 + T\over D}\;{\rm ,}
\label{magflux}
\end{equation}
which is much easier to reconcile with the magnetic energy flux
required by observations, since now the buoyant rise velocity can be
comfortably below its absolute maximum value, i.e., the sound speed
$c_s$ and yet provide a magnetic energy flux exceeding the radiation
flux.

A simple physical analogy here is a sail on a ship. When the sail is
``on'', the force (due to wind) acting on the sail is many times
larger than it is in the case of the sail that is folded in. The
amount of this wind-sail interaction clearly depends not on the
overall mass of the sail, but on the state of the sail -- whether it
is open and positioned properly with respect to wind or whether it is
rolled in a tube. Similarly, with the same volume average magnetic
field one gets less or more interaction between differential flow and
the field depending on whether the field is uniform in space, or
contained within strong flux tubes, such that most of the flow simply
misses the tubes to interact with them.

\section{Radiation Pressure and Properties of a Single Flux Tube}
\label{sect:radp}

In our consideration of the magnetic fields in the previous section, we
did not explicitly separate the total pressure into the radiation
pressure $\pr$ and the gas pressure $\pg$. This approach is
practically always used in the literature (e.g., Galeev et al. 1979;
Vishniac et al. 1995a,b and further references cited therein) mainly
because of two reasons. The first one is that historically it is the
Solar magnetic field phenomena that stimulated much of the work on
magnetic fields in turbulent plasmas, and for the Solar interior
conditions the radiation pressure is everywhere much smaller than the
gas pressure. The second reason is a more pragmatic one: the problem
becomes virtually intractable if the radiation pressure dominates,
since one now has three interacting components instead of two. This
approach (of neglecting the radiation pressure dynamical effects) is
equivalent to the assumption that the radiation and particles move
together, as one fluid, even when radiation pressure dominates over
the gas pressure. It is clear that such an approach should indeed be
valid as long as the scales of interest in the disk are much larger
than the photon mean free path, since in this case the radiation is
strongly coupled to particles due to the large opacity. Below we will
attempt to quantify when such a ``one fluid'' approximation is valid
and when it is not, and what are the implications for the magnetic
flux tubes in the radiation dominated accretion disks.

To do so, we need to compare the time scale for the radiation
diffusion into the flux tube with a time scale important for
generation and maintenance of strong magnetic flux tubes. The
radiation diffusion time scale $t_{\rm d}$ can be estimated as $t_{\rm
d}\sim (a/c) n'_e \sigma_T a$, where $n'_e$ is the particle density
inside the flux tube, which we can assume to be of the order of the
disk particle density $n_e$, $\sigma_T$ is the Thomson cross section,
and $a$ is the flux tube cross sectional radius.

Turbulent motions of the fluid are believed to be the mechanism for
the magnetic field amplification (e.g., Vishniac 1995a,b).  Let
$u_{\rm t}$ be the typical turbulent velocity, and $\lambda_{\rm t}$
be the turbulent length scale (corresponding to the largest eddy
length scale).  The gas executes turbulent motions on the eddy turn
over time scale $t_{\rm t}\equiv \lambda_{\rm t}/u_{\rm t}$.  To see
if diffusion is faster than turbulent motions, we compare the time
scales $t_{\rm d}$ and $t_{\rm t}$:
\begin{equation}
{t_{\rm d}\over t_{\rm t}} \sim {a^2\over H \lambda_{\rm t}}\,
{u_{\rm t}\over c_s} {\tau_d c_s\over c}\; ,
\label{dift}
\end{equation}
where $c_s$ is the gas sound speed, $H$ is the vertical scale height
of the disk and $\tau_d$ is the disk Thomson optical depth. The scale
of the flux tube $a$ is likely to be of the order of the scale of
turbulent motions (Vishniac 1995a). Further, in the standard
Shakura-Sunyaev viscosity prescription, the turbulent velocity and
spatial scale are parameterized by $u_{\rm t} \lambda_{\rm t} = \alpha
c_s H$. Finally, in the radiation pressure dominated region of the
disk, the standard disk equations lead to $\tau_d c_s/c\simeq
\alpha^{-1}$, for arbitrary radii and accretion rates. Therefore, one
can see from Equation (\ref{dift}) that the ratio of the diffusion
time scale to the turbulent time scale is of the order unity. This
means that standard accretion disk equations permit radiation to
diffuse into the flux tubes in the radiation dominated region of the
disk. Moreover, we compared $t_{\rm d}$ with {\it one} eddy turn over
time scale, whereas generation of the field comparable with the
equipartition value is likely to take much longer, simply because one
turbulent eddy does not carry enough energy. Due to this it is almost
guaranteed that the diffusion of radiation into the flux tubes is much
faster than the field generation process in the standard accretion
disk theory.

What does this mean for a {\it single} magnetic flare? Since the
radiation easily diffuses inside the flux tube, the radiation pressure
inside the flux tube should be approximately equal to the ambient
radiation pressure. It is then only the ambient gas pressure that can
confine the flux tube side-ways, that is, $\pmg\leq \pg$.  In fact it is
even not clear if accretion disks will produce magnetic flares of the
same compactness parameter as it does in the gas-dominated case (see
\S \ref{sect:ints}), i.e., whether the spectrum from a single magnetic
flare will change or not.

Summarizing the ongoing discussion, it is clear that the magnetic
fields in the radiation-dominated disks are either mostly in a diffuse
form, or in the form of weak flux tubes, whose maximum pressure is
given by the gas pressure. Weakness of the flux tubes in the
radiation-dominated disk means that they will not behave as solid
bodies anymore (see equation \ref{tod}), and will be stretched by the
differential flow just as diffuse magnetic fields are. Thus, one
recovers the estimate $\mean{\pmg}\simlt \alpha\ptot$, and the amount
of energy deposited into the hard X-rays decreases as the accretion
disk becomes radiation-dominated.

\section{The Model Parameter Space}\label{sect:parsp}

\subsection{Dim State}\label{sect:dim}

We start by discussing very dim accreting disk systems, namely ones
that accrete at such a low accretion rate $\dm\simlt\dm_d$ that
magnetic flux tubes cannot provide enough energy for the emitting
regions to be compact. As was shown in \S \ref{sect:tpmodel},
bremsstrahlung, rather than inverse Comptonization becomes the
dominant emission mechanism when $l\simlt 0.01$. Setting $l$ to 0.01
in equation (\ref{comps}) gives us the estimate of the corresponding
accretion rate:
\begin{equation}
\dm_d\simlt 2\times 10^{-4}\, \alpha\, \zeta^{-1}\; .
\end{equation}
Below this accretion rate, the X-ray spectra should be different from
the standard Seyfert hard spectrum, since the two-phase model becomes
invalid. Note that the disk is gas-dominated for these low accretion
rates, and so the magnetic energy flux might be dominant over the
radiation transport, and it is possible that the X-ray component may
again be very prominent in the overall spectrum. However, studies of
magnetic flare emission mechanisms in the regime $l\ll 1$ need to be
done to test this situation further. We shall call this parameter
space `dim' accreting systems, and the most model independent
statement that we can make at this time is that their spectra should
be different than that of standard hard Seyfert spectra.  

\subsection{Hard State}\label{sect:hards}

Let us now move up in the accretion rate parameter space, such that
the compactness parameter of the magnetic flares is $l\simgt 10^{-2}$,
and the gas pressure dominates over that of the radiation in the disk,
i.e. $\dm_d\simlt \dm\simlt \dmr$, where $\dmr$ is
\begin{equation}
\dmr = \dm_0 \, (1-f)^{-9/8} = 2.2 \times 10^{-3} \left(
\alpha M_8\right)^{-1/8}\, (1-f)^{-9/8} {\rm ,}
\label{mr}
\end{equation}
for AGN, and 
\begin{equation}
\dmr =  1.6 \times 10^{-2} \left(
\alpha M_1\right)^{-1/8}\, (1-f)^{-9/8} {\rm ,}
\label{mrc}
\end{equation}
for GBHCs. Svensson \& Zdziarski [1994] showed that the transition
from gas to radiation dominated regime is affected by transferring a
fraction $f$ of the disk energy into the corona. Namely, they found
that this transition happens at $\dmr = \dm_0/(1-f)^{9/8}$, where
$\dm_0$ is the accretion rate when $\pr = \pg$ in the standard
theory. However, this approach assumes that the fraction $f$ is itself
a constant, which may not necessarily be the case. As we found earlier
in this Chapter, the fraction $f$ decreases when $\pr$ exceeds $\pg$.
If we are to use some guidance from observations of GBHC state
transitions, we would have concluded that $\dmr$ is not greatly
affected by the corona, since the state transitions seem to happen at
$\dmr\simeq 0.05$ (see \S \ref{sect:obgbhc}), and this (depending on
the exact value of $\alpha$, which may be quite small) is just a
factor $\sim 2-3$ higher than equation (\ref{mrc}) predicts for a a 10
$\msun$ blackhole. If we re-scale $\dmr$ from $0.05$ for GBHCs using
$\dmr\propto M^{-1/8}$, we obtain that $\dmr \simeq 0.007$ for Seyfert
Galaxies with $M\sim 10^8\msun$.

Our discussion in \S \ref{sect:chapter6} showed that it is possible
for magnetic buoyancy to expel out of the disk more energy in the form
of magnetic fields than the common radiation transport does in the
gas-dominated accretion disks. At the same time, the approximate
nature of the discussion may not yield an exact value for $f$. If we
now again try to use some guidance from GBHCs, whose broad band
spectra are better understood observationally than those of Seyferts,
we will see that $f$ must probably be as large as $0.7 - 0.9$ to
explain the hard spectra of some GBHCs (see Chapter 4).

To summarize, in the ``hard'' accretion rate parameter range, i.e.,
when $\dm_d\simlt \dm\simlt \dmr$, we expect that the spectrum is hard
for both Seyferts and GBHCs, with the latter spectrum being harder
than the former spectrum, due to the strongly ionized nature of the
X-ray reflection in GBHCs (see Chapter 4). Also, it is energetically
allowed to have the corona to dominate the overall luminosity of the
disk-corona system. Note that observationally, because of the almost
neutral transition layer in Seyferts, these sources should still have
$\luv/\lx\simgt 1$, the lowest value possible in the two-phase
corona-disk geometry (see e.g., Svensson 1996).

\subsection{Intermediate State}\label{sect:ints}

As the accretion rate increases above $\dmr$, the importance of X-ray
production by magnetic flares decreases, although we are unable to
describe this in a model independent fashion. Nevertheless, the trend
in the division of the emitted power between the disk and the corona
is clear: since the fraction of energy reprocessed in magnetic flares
is decreasing as $\dm$ increases, the importance of the intrinsic disk
emission in the overall spectrum increases, and it becomes the
dominant feature in the spectrum of an AGN or a GBHC.

Based on theoretical arguments alone, we cannot be certain about what
happens to the shape of the X-ray spectrum from corona/flares in
radiation-dominated disks. The uncertainty is present not only due to
our rather sketchy understanding of the physics of magnetic flares,
but also due to our ignorance of the numerical value of the viscosity
parameter $\alpha$ when $\pr\gg\pg$, which means we do not really know
the underlying accretion disk structure. The standard accretion disk
theory in the radiation-dominated case is unstable to viscous and
thermal perturbations (e.g., Frank et al. 1992), and so the form of
viscosity law in radiation dominated disks remains a highly
controversial issue.

One possibility, as suggested by, e.g., Lightman and Eardley (1974),
Stella \& Rosner (1984), Sakimoto \& Coroniti (1989), is that $\alpha$
scales as $\alpha= \alpha_g \pg/\ptot$, where $\alpha_g$ is a
constant. We consider it plausible that this happens in reality, since
observations of GBHCs accreting in the radiation-dominated regime show
that their accretion disk structure is stable, with the notable
exception of GRS~1915+105, which exhibits large amplitude oscillations
(e.g., Belloni et al. 1997a,b). This particular source is unusually
luminous, though, and may be accreting at $\dm \sim 1$, i.e., close to
the Eddington limit.

If $\alpha$ does decreases with $\dm$ as $\alpha= \alpha_g \pg/\ptot$
or a similar dependence, the decrease in $\alpha$ may actually
compensate for the decrease in the magnetic flux tube pressure
(bounded by $\pg$), so that the compactness parameter (equation
\ref{compm}) will not become smaller when radiation pressure exceeds
the gas pressure. Then, as long as the assumptions of the two-phase
patchy corona model (equations \ref{cond1},\ref{cond2}) are satisfied,
the magnetic flares will generate a hard X-ray spectrum as in the hard
state, even though the corona does not possess most of the power in
this case. It is thus sensible to refer to this parameter range as
''intermediate'' state. In simplest terms, the difference between the
hard and intermediate states is that the latter has fewer active
magnetic flares than the former at any time, whereas magnetic flares
themselves do not change substantially in their properties going from
one state to the other.

\subsection{Soft State}\label{sect:softs}

Let us now compare the X-ray flux from a magnetic flare $\fx$ and the
accretion disk intrinsic flux $\fdisk$ in the {\it radiation-dominated
case}. For $\fx$, this yields
\begin{equation}
\fx\simeq 6.9 \times 10^{14} l M_8^{-1} \dm^{-1} 
\left[1-f\right]^{-1}\; ,
\label{fxrad}
\end{equation}
whereas $\fdisk$ has the same form in both gas- and
radiation-dominated cases (equation \ref{fdisk}). The two-phase model
is valid as long as $\fx\gg\fdisk$, which leads to
\begin{equation}
\dm\,\left(1-f\right) \ll 0.19 \, l^{1/2}\equiv \dms\; .
\label{dms}
\end{equation}
In this equation, $f$ depends on the accretion rate itself. For
$\dm\geq \dmr$, $f$ should be less than a half, on both theoretical
(\S \ref{sect:radp}) and observational grounds (radiation-dominated,
luminous accretion disk systems always seem to emit most of their
energy at the disk temperature rather than in a hard power-law tail,
see \S \ref{sect:obgbhc}).

The significance of the equation (\ref{dms}) is that X-ray spectra
should steepen as the accretion rate approaches $\dms$. This is a
testable prediction: if magnetic flare physics, for some reason,
dictates a certain value for the compactness parameter, then equation
(\ref{dms}) suggests that the steepening of the X-ray spectrum should
happen at the same dimensionless accretion rate {\it independently} of
the blackhole mass $M$. Our treatment of the X-ray reflection in the
transition layer, and attempts to reconcile theory with observed
temperatures for the BBB and the disk thermal emission in Cyg~X-1,
point to a rather small compactness parameter $l\simeq 0.1$ (\S
\ref{sect:sconstraints}) in both AGN and GBHCs. Above this value, 
the reflected spectrum becomes too ``hot'' to explain observations,
while $l$ below $0.1$ seems to be ruled out by the emission mechanism
constraints (\S \ref{sect:tpmodel}) and the fact that in Cyg~X-1 the
compactness parameter $l\sim 0.1$. Thus, if we scale $l$ to 0.1, we
get the following estimate for the ``soft'' accretion rate $\dms$:
\begin{equation}
\dms = 0.06 \left({l\over 0.1}\right)^{1/2} \,\left[1-f_s\right]^{-1}
\; ,
\label{dmsa}
\end{equation}
where $f_s < 1/2$ is the $f$-fraction of power transferred to the
corona from the disk below it at $\dm = \dms$. Above $\dms$, not only
does most of the power come out as the disk's internal emission, but
the X-ray spectrum should steepen from its ``canonical'' hard values
for Seyfert 1 Galaxies and GBHCs. In equation (\ref{dmsa}), the
fraction $f_s$ may be eliminated via observations, since $f_s$
depends on the spectral shape, which then should make the estimate of
$\dms$ to be less model dependent.

\subsection{Very High State}\label{sect:vhs}

At even higher luminosities, i.e., above $\sim 0.2 \ledd$, the
``cold'' accretion disk structure may depart from the standard
accretion disk model considerably due to advection of energy into the
black hole (Abramowicz et al. 1988), and a breakdown of the thin disk
approximation should take place as well (see equation
\ref{hoverrr}). Note that we do not mean Advection-Dominated Accretion
Flows here (see references on Narayan \& Yi in Chapter 1), since the
existence of those depend on the key assumption that the ions are much
hotter than the electrons. This could be the case if the electrons and
ions interact only through Coulomb collisions, which seems to be a
rather unsafe and arbitrary assumption to us, especially when the
magnetic field pressure inside the disk, as is often invoked by ADAF
workers is close to the equipartition value (see also Bisnovatiy-Kogan
\& Lovelace 1997, Begelman \& Chiueh 1988).

However, {\it cold} advection-dominated flows (the ``slim'' disks of
Abramowicz et al. 1988) are almost certain to exist for $\dm \sim 1$,
since the importance of the advective flux compared to the radiation
flux in the vertical direction is approximately given by the ratio
$(H/R)^2$. This ratio approaches unity when $\dm \sim 1$, {\it
independently of the viscosity law}, at least in the framework of the
standard accretion disk theory.

Accordingly, we do not attempt to apply our model to systems with a
very high $\dm$, and leave this to future work. We refer to this
parameter space as the ``very high'' state, and expect it to exist for
$\dm \simgt 0.2$.  There are still some issues that can be considered
even if the basic disk structure is unknown in the very high
state. For example, as we argued in Chapter 4, the structure of the
X-ray skin is determined by the local conditions, i.e., the local
gravity, intrinsic disk and illuminating X-ray fluxes, and thus can be
solved for (in some approximation) even if we cannot describe the
physical conditions in the disk interior.

	\chapter{Classification of Accretion Disk States and Comparison
	to Observations}

\section{Theoretical and Observational Motivation}

	In this section we will attempt to merge the main results of
our work into a consistent picture that would describe the
observational appearance of accretion disks in AGNs and GBHCs. We
believe it can and should be done, because, most of the accretion
power is derived in the inner disk region, very far from the outer
boundary of the disk, so that the nature of the accretion flow further
away from the black hole than $\sim 10^3$ Gravitational radii is of
limited importance for the observed spectra, except maybe for setting
the overall accretion rate. The main physical processes determining
the observed spectra from accreting black holes are scale-free. For
example, the importance of Comptonization as the main emission
mechanism is determined by the compactness parameter $l$ (\S
\ref{sect:tpmodel}), which scales as $l\propto \fx R$, where $\fx$ is
the X-ray flux, and $R$ is the geometrical size of the emitting
region.  Since $R\propto M$, where $M$ is the mass of the black hole,
and $\fx\sim L/R^2 \propto M^{-1}$ (if $\lrel$ is more or less the
same for GBHCs and AGNs), compactness parameter $l$ does not depend on
the mass of the black hole (see also \S
\ref{sect:compactness}). Further, many physical quantities of interest in
the disk itself are either scale free, or weakly depend on $M$ (see \S
\ref{sect:adstr}). Therefore, any successful theory of accretion disks
should not only explain a class of accreting black holes, or a
particular state of those objects, but also be either applicable to
the rest of accretion disk systems, or provide a natural physical
reason as to why the theory may not be applied to those systems.

Second, observations show that the X-ray spectra of GBHCs and Seyfert
1 Galaxies are rather similar (e.g., Zdziarski et al. 1996). Further,
there are numerous suggestions in the observing literature that
so-called steep spectrum Seyfert Galaxies, and other objects with a
steep X-ray slope are similar to the soft (sometimes also called
``high'') state of GBHCs (e.g., Laor et al. 1997). Moreover, the trend
``softer in X-rays when brighter'' is often seen in both AGN and
GBHCs.  Lastly, we believe that, since multi-wavelength observations
of accreting black holes produced so much {\it new} information in
recent years, combining observational constraints from such different
objects as AGN and GBHCs will allow us to constrain or reject any
accretion disk model.

\section{Accretion Disk States}\label{sect:adstates}

\begin{figure*}
\centerline{\psfig{file=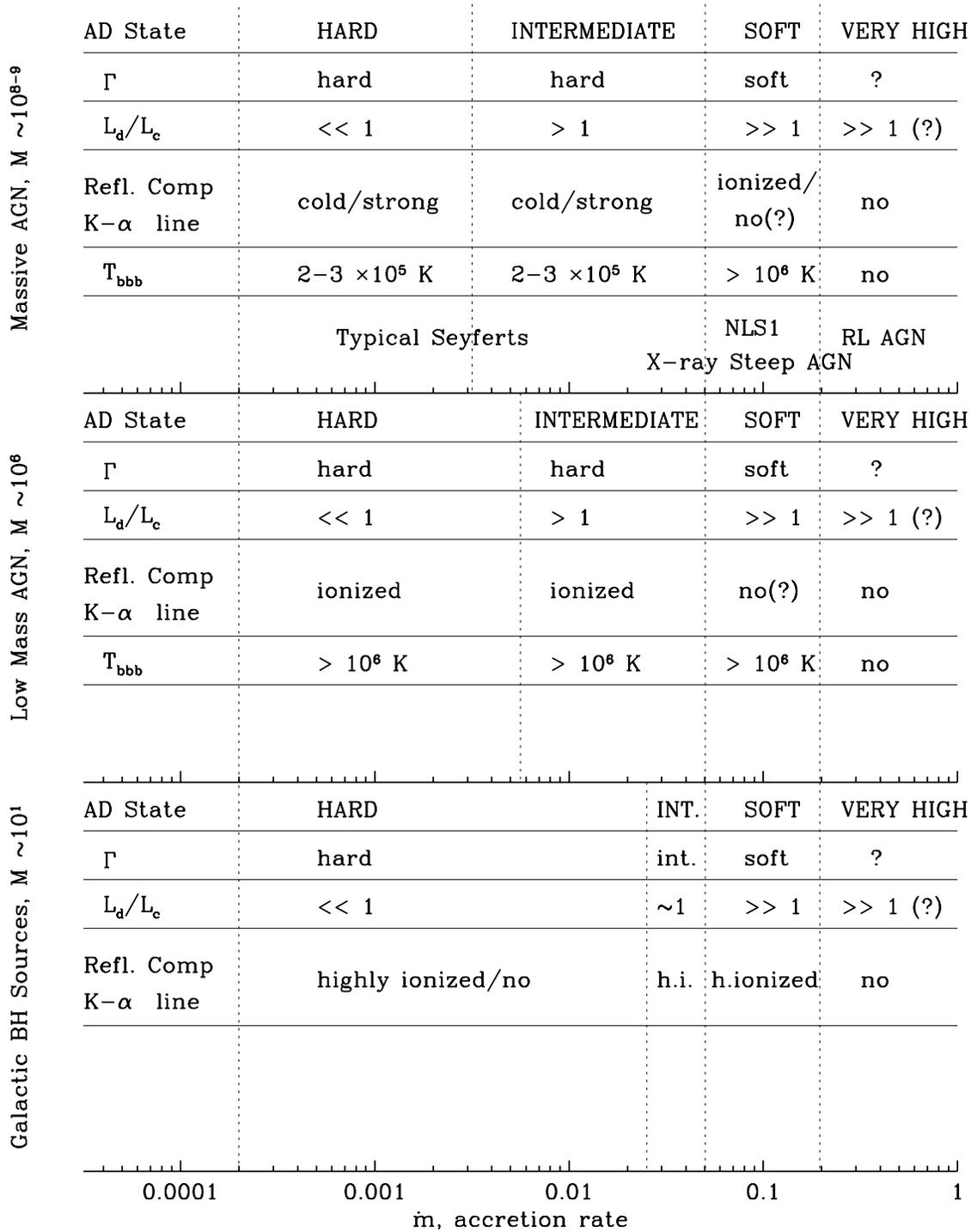,width=1.0
\textwidth,angle=0}}
\caption{Classification of Accreting BH States}
\label{fig:summary}
\end{figure*}

We have devised a diagram to summarize our results on the
classification of accretion disk states, which we show in Figure
\ref{fig:summary}. The $x$-axis of the diagram shows the dimensionless
accretion rate $\dm$. The ``$y$-axis'' of the diagram shows the black
hole mass $M$ in Solar mass units, although we had to introduce some
quantization in $M$ on the diagram in order to represent our results
in the most compact and observationally meaningful way. A natural
division in black hole mass parameter space would be between AGN that
are believed to have masses $M\sim 10^6 - 10^9\msun$ and GBHCs with
$M\sim 10\msun$. However, the pressure ionization instability in the
X-ray ionization calculations forced us to divide the AGN class
further, such that most massive and least massive objects are
considered separately. Each of these three classes of accreting BH is
shown in one of the panels in Fig. (\ref{fig:summary}).

Each panel consists of a table that lists the observationally
interesting and thus testable model predictions. The rows of the table
are divided into the four accretion disk states as described in \S
\ref{sect:parsp}, although we do not include here the dim accretion 
disk state, since more work is needed to understand this state (see
\S \ref{sect:dim}). Out of the four states presented in the table,
the least understood is the very high state (\S \ref{sect:vhs}).  Some
predictions still can be made concerning the ionization state of the
X-ray skin in this state, and thus we provided those. By a single
question mark we designated those issues that have not been considered
so far in the context of magnetic flares in accretion disks. Further,
we marked with a question mark in parenthesis issues for which we have
only very preliminary results/ideas.

The columns of the table have the following meanings. $\Gamma$ is the
X-ray $2-10$ keV photon spectral index. ``Comp. Refl. and
K$\alpha$-line'' represent the strength of the reflection component
and the iron line, respectively. These two were blended together since
they are created by the same physical process, e.g., X-ray
illumination of the transition layer, and thus their appearance is
controlled by the state of this layer. $T_{\rm bbb}$ is the
temperature of the ``Big Blue Bump'' (BBB). Our definition of the BBB
is motivated by theoretical rather than observational considerations:
we call BBB any UV or soft X-ray emission arising from the transition
layer due to its illumination by the incident X-rays from magnetic
flares. For massive AGN, our definition exactly coincides with the one
used in the observational literature, since the transition layer
emission appears in the right wavelength band. For low mass AGN and
GBHCs, the BBB defined here is most properly referred to as soft X-ray
excess (e.g.,
\S 5.2 of Zdziarski et al. 1998). Lastly, ``variability'' is the X-ray
variability properties of the system, for which we could only indicate
some rough trends.

\section{Observations of AGN and the Theory of Magnetic Flares}
\label{sect:obsagn}

In this section we will try to establish whether observations of
Seyfert 1 Galaxies fit within the framework of the model of accretion
disks with magnetic flares. We assume that most Seyfert 1 Galaxies
have relatively large masses, i.e., $\sim 10^8-10^9\msun$ and thus
belong to the uppermost panel of Figure (\ref{fig:summary}). We do not
consider Radio-Loud objects, since their X-ray emission may be
dominated by a jet. The dimensionless accretion rate is found via $\dm
= L/\ledd$, where $\ledd = 1.3\times 10^{46} M_8$ erg/sec is the
Eddington luminosity.

\subsection{X-ray Index and Relative Luminosity}\label{sect:xindex}

\subsubsection{Hard and Intermediate Seyferts}\label{sect:his}

We now discuss the first column (i.e., $\Gamma$) in Figure
(\ref{fig:summary}). We made no attempt to provide an exact value for
$\Gamma$, because of theoretical uncertainties. Namely, the theory is
too sketchy at this time to predict the exact geometry of the active
regions and to pinpoint other important parameters, most notably the
compactness parameter, and for this reason one actually uses
observations to deduce the correct values for these parameters.  So,
it would not be fair to say that the model firmly predicts $\Gamma$ to
be $\simeq 1.9$ for Seyfert 1 Galaxies, as is observed.  However,
the calculations of Stern et al. (1995), Poutanen \& Svensson (1996) show
that the X-ray index produced by the two-phase patchy corona-disk
model with perfectly reasonable (in the framework of magnetic flares)
geometries, such as a hemisphere or a sphere above the disk, is indeed
close to the observed distribution of these indexes in Seyferts. Note
that a large compactness parameter $l\sim 10 - 100$, that these
authors found to best match the data, may not be needed if the Thomson
optical depth of the AR is given by electrons rather than pairs, since
the large compactness was mainly required to provide a large enough
optical depth in pairs. Therefore, we can assume that the geometry is
such as to give the ``correct'' $\Gamma \simeq 1.9$ for typical
Seyferts, and then use the same geometry and compactness parameter to
study higher accretion rates {\it and} GBHCs.  Based on our
calculations (not completely self-consistent, yet; see Chapters 4 \&
5), we believe that the most likely geometry for an AR is a hemisphere
with the compactness parameter $l\sim 0.1$ (see \S \ref{sect:influx}),
sitting atop of a cold accretion disk,

As discussed in \S \ref{sect:parsp}, the magnetized accretion disk
model predicts that Seyferts with the typical hard X-ray spectra
should accrete at accretion rates below $\dms$. In order to find
$\dms$ from equation (\ref{dmsa}), we need to know the magnetic flare
compactness. According to \S \ref{sect:sconstraints}, $l\sim 0.1$, so
that we estimate $\dms\sim 0.06$ for Seyfert Galaxies as well as
GBHCs. This number was used to separate the intermediate and the soft
state in the diagram, but it is understood that it is only a
preliminary estimate.

One serious obstacle in carrying out a comparison of the theory and
observations is that the masses of AGNs are very hard to deduce, and
they are uncertain to a high degree. There exists no analog to the
mass function that has proven to be so useful in setting limits on the
mass of the Galactic blackhole candidates (e.g., Frank et al. 1992).
Nevertheless, variability studies may provide some help. For example,
the global compactness parameter has been estimated for a sample of
Seyfert Galaxies by Done \& Fabian (1989). In their estimate, they
assumed that the typical size of the emitting region is given by the
distance traveled by light during the shortest doubling time scale
observed for a given source. For an accretion disk this typical size
should be of order $\sim 10 R_g$, because most of the emitted
radiation is produced in the region of roughly this size (see, e.g.,
Frank et al. 1992). This makes it possible to relate the global
compactness parameter to $\dm$. Fabian (1994) notes that values
obtained by Done \& Fabian (1989) should be roughly halved, since they
assumed the gamma-ray continuum to persist up to $\sim 2$ MeV. With
all this in mind, we get a conversion factor from the global
compactness $l_g$ to the dimensionless accretion rate for a given
source:
\begin{equation}
\dm \simeq {10 m_e\over 2\pi m_p} \,{l_g\over 2}\simeq l_g/2000\; .
\label{fbd}
\end{equation}
In Table 1 of Done \& Fabian (1989), the maximum compactness is about
200, thus the maximum $\dm\sim 0.1$. Moreover, $80$\% of the sample
have $\dm < 0.02$, with the smallest values of order of $10^{-4}$.
These estimates do not include the contribution from the Big Blue
Bump, which may be a significant component in the bolometric
luminosity of Seyfert Galaxies. However, for the sources with the
highest estimates for the compactness, we found $L_{\rm uv} \sim
L_{\rm x}$, so that the inclusion of the emission at lower wavelengths
did not affect our conclusions significantly. To estimate $L_{\rm uv}$
we used $1375$ Angstrom fluxes reported by Walter \& Fink (1993),
whereas $L_{\rm x}$ was taken from Done \& Fabian (1989).

Sun \& Malkan (1989) fitted the multi-wavelength continua of quasars
and AGNs with improved versions of standard accretion disk
models. They found that low-redshift Seyfert Galaxies radiate at only
a few percent of their Eddington luminosities. Rush et al. (1996)
studied the soft X-ray (0.1-2.4 keV) properties of Seyfert
Galaxies. Their results indicate that $\sim 90$\% of sources in their
sample have a soft X-ray luminosity below $10^{44}$ erg/s (with the
mean value of order $\sim 10^{43}$ erg/s). If we assume the typical
Seyfert 1 spectrum above 2.4 keV, i.e. a power-law with intrinsic
photon index $\simeq 2$ and the cutoff at several hundred keV, then
total X-ray/gamma-ray luminosity of these objects can be a factor of
2-3 higher than the soft X-ray luminosity.  Nevertheless, if a typical
Seyfert Galaxy has a blackhole mass of $\sim 10^8$, then the average
bolometric luminosity of the Rush et al. (1996) sample is at or below
$\sim 1$\% of the Eddington luminosity. In addition, we found
estimates for the relative bolometric luminosity of several of the
most luminous Seyfert Galaxies in the literature. In particular, for
NGC5548, using the results of Kuraszkiewicz, Loska \& Czerny (1997),
we obtained $\dm\sim (4-16)
\times 10^{-3}$.

Summarizing, there is a large body of evidence that X-ray hard Seyfert
1 Galaxies accrete at a relatively low accretion rate, i.e., from
probably just below $\dm = 0.1$ to the very low values of $\sim
10^{-4}$.

\subsubsection{Steep Spectrum Seyferts (NLS1)}\label{sect:sssg}

Relatively recently, it was found that a subset of Seyfert Galaxies
have unusually steep soft X-ray spectra (for a review, see Pounds
\& Brandt 1996, PB96 hereafter, and Brandt \& Boller 1998). 
Common properties of the group include steep spectra, rapid
variability, strong Fe II emission and an identification with
narrow-line Seyfert 1 galaxies (NLS1: Seyfert 1 Galaxies that have
uncommonly narrow H$\beta$ line, e.g., FWHM $\simlt 10^3$
km/sec). PB96 speculated that the most likely explanation for the
steep X-ray spectrum is unusually high accretion rate. PB96 also
showed that soft X-ray (i.e., 0.1-2 keV) spectral index is strongly
correlated with the width of the H$\beta$ line for a sample of Seyfert
Galaxies.  Wandel
\& Boller (1998) suggested an explanation of the correlation based on
the simple idea that steeper X-ray spectrum implies a larger ionizing
UV luminosity, which translates into a larger broad line region size,
and thus a smaller velocity dispersion (since Keplerian rotation
velocity is $\propto 1/R^{1/2}$). They found that the masses of the
narrow-line Seyfert Galaxies tend to be lower that those of typical
broad H$\beta$ line Seyferts, and thus to have larger $\dm
\equiv L/\ledd$. In principle, larger values of $\Gamma$ in the soft X-ray
band may be partially caused by the soft X-ray excess rising steeply
towards lower photon energy, and therefore this correlation does not
have to hold for harder X-ray energies. However, Brandt, Mathur \&
Elvis (1997) found that the higher energy {\it ASCA} slopes (2-10 keV)
correlate with the H$\beta$ line as well. Thus, the narrow line
Seyfert Galaxies often have an intrinsic X-ray slope that is steeper
than that of normal Seyferts.

Laor et al. (1997) found the same correlation for a sample of quasars,
and suggested that NLS1 galaxies accrete at a higher fraction of the
Eddington accretion rate than normal Seyferts do. They used a simple
argument that the bulk motion of the broad line region is virialized,
and that the scaling of the BLR with luminosity is that found from a
reverberation line mapping of AGN (e.g., Peterson 1993).  In this case
larger luminosities $L$ correspond to larger BLR size, and thus
smaller H$\beta$ FWHM. Now, if $\Gamma$ is larger for higher $\lrel$,
then the observed relation (smaller H$\beta$ FWHM -- larger $\Gamma$)
ensues.  However, no reason for $\Gamma$ to become larger with
increasing $\lrel$ was given, except for a suggestion of Pounds et
al. (1995) that if the power released in the corona remained constant,
then the X-ray index would become steeper with increasing bolometric
luminosity. The latter suggestion is equivalent to assuming $f L =$
const, and was not supported by any theoretical considerations in
Pounds et al. (1995).  Our theory of accretion disk states may provide
a natural explanation for why the X-ray index becomes softer when
$\lrel$ increases. To repeat, this should occur for accretion rates
close to $\dms$, where the disk intrinsic flux becomes comparable with
the flux from magnetic flares, causing increased cooling of the active
regions and thus leading to steeper X-ray spectra.

There are also interesting observations of individual members of the
subclass. For example, Piro et al. (1997) report on the Seyfert 1
Galaxy E 1615+061, which is a candidate for the strongest variability
in X-rays. Its spectrum changed from a very steep high state
($\Gamma\sim 4$) in 1977 to a {\it two orders} of magnitude dimmer
state with a flatter spectrum with $\Gamma\sim 2$ (i.e., the typical
Seyfert photon index) as observed in 1985. The authors point out a
similarity with Galactic black hole transients and persistent sources,
where the spectrum becomes harder as the source luminosity
decreases. We can understand the behavior of this source if its high
luminosity state corresponds to the soft state, and the low luminosity
one to either intermediate or hard states.

Very recently, Becker (1997) has shown that steep X-ray slope AGNs are
not limited to relatively nearby, low luminosity NLS1 Galaxies, but
have a continuous redshift distribution out to a redshift of $z=2.5$.
He shows that $63\%$ of his sample of super-soft AGNs (mean photon
index is $\Gamma \simeq 3$) should be classified as quasars due to
their large optical luminosities. The soft X-ray luminosity in the
energy range $0.1-2$ keV can vary from $\sim 10^{43}$ to almost
$10^{46}$ erg/sec.  We believe these sources are accreting in the soft
regime, that is $\dm\geq \dms \sim 0.06$, and the substantial
difference in the luminosity is produced by a large difference in
blackhole masses in these sources. If the soft X-ray flux represents
some $\sim 10$ \% of a typical source bolometric luminosity (in the
soft state), then the dimmest sources may have $M\sim 10^7\msun$,
whereas the most luminous sources could be explained with a blackhole
mass of $M\sim 10^9\msun$. We thus believe that observations support
our theory of the X-ray spectrum formation in AGNs, if not
quantitatively, then qualitatively at least.

\subsection{Division of Power Between The UV and X-ray Components}
\label{sect:divpow}

Let us now consider the second column in Figure (\ref{fig:summary})
for the high mass AGNs. Here we show the theoretical predictions for
the division of power between the combined X-ray luminosity from
magnetic flares and the accretion disk bolometric luminosity. The
ratio of the two luminosities is by definition the fraction $f$ of
power supplied to the corona. According to the discussion in \S
\ref{sect:parsp}, $f$ is close to unity for the hard state, so that
most of power is reprocessed through the magnetic flares, and then $f$
decreases in the intermediate parameter space and the soft state,
although we were not able to find the exact function $f(\dm)$ at this
point due to computational uncertainties.

The observational situation on the division of power between the UV
and X-ray band in AGNs is not as clear as one would have hoped. The
largest difficulty here is the strength of the BBB, which lies in the
almost unobservable wavelength interval (because of the Galactic
absorption). Bolometric corrections for the UV flux may be quite large
(see discussion of this in Sincell \& Krolik 1997). The total power in
the optical-UV region $L_{\rm UV}$ cannot be determined accurately
under these circumstances. Nevertheless, the fact that there are
thousands of Seyferts and quasars makes it possible to study these
objects by statistical means, and get a rough observational picture
that way.

\begin{figure*}
\centerline{\psfig{file=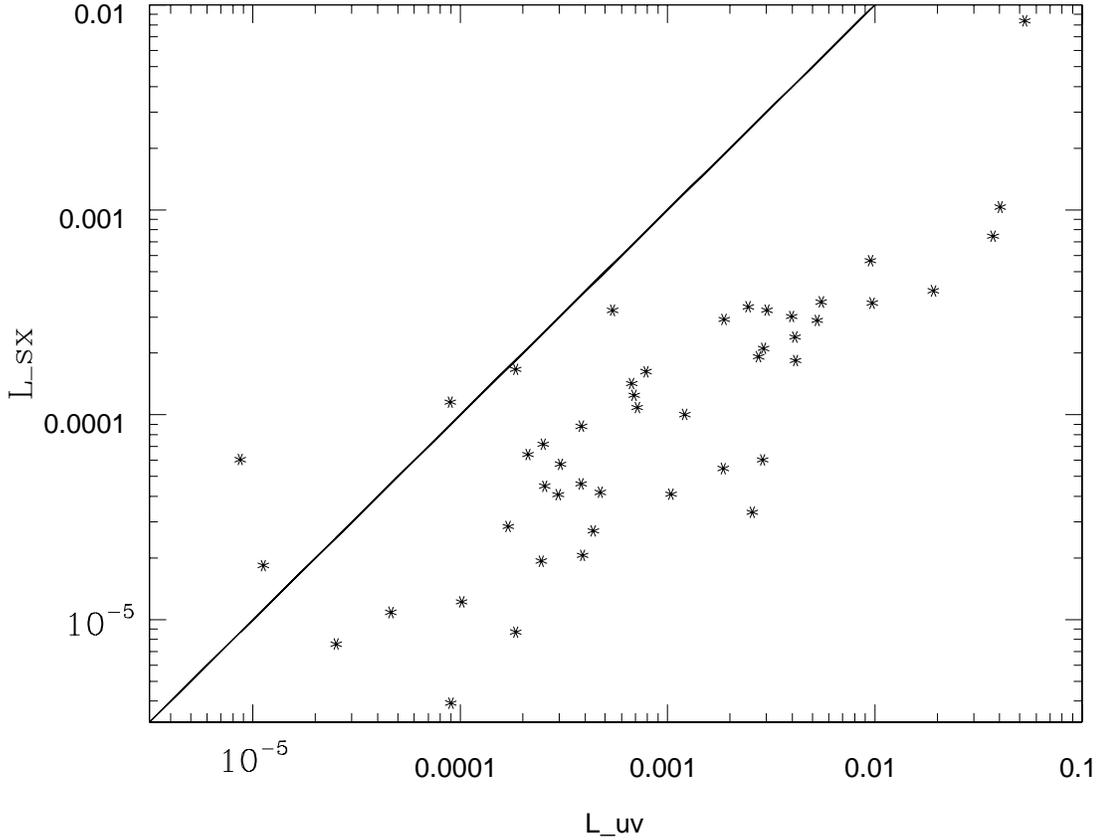,width=.97
\textwidth,angle=0}}
\caption{The soft X-ray luminosity $L_{\rm sx}$ versus UV-luminosity
$\luv$ for the radio-quiet objects from Walter \& Fink (1993) sample.
Note that for higher luminosity sources, the ratio of $\luv/L_{\rm
sx}$ is correspondingly higher, meaning that less power is reprocessed
through the corona. The solid line gives $L_{\rm sx} = \luv$. See text
for details.}
\label{fig:wf}
\end{figure*}

Walter \& Fink (1993) studied the soft X-ray bump (i.e., BBB in our
terminology here) of Seyfert Galaxies. One of their findings was that
the ratio of the UV to hard X-ray fluxes can vary by factors of a
hundred, and yet the X-ray index does not show any clear correlation
with the X-ray luminosity at 2 keV (see Figure 10 of Walter \&
Fink). This is unlike GBHCs, where the hard state has a hard X-ray
spectrum and a large ratio of hard to soft luminosity, whereas the
soft state has considerably lower ratio of the hard to soft
luminosities and, simultaneously, a much softer X-ray spectral
index. We think that the difference here is caused by the fact that
the hard and the soft states in Seyferts are separated by the
relatively large intermediate state, -- large in the sense that for
massive AGNs $\dmr\ll\dms$. Since we should expect that the normal
hard X-ray spectra Seyfert 1 Galaxies will come with a range of $\dm$,
they should have different $f$ as well, and thus
$\luv/\lx$. Physically, each individual flare in the intermediate
state may still produce the hard X-ray spectrum typical of Seyferts,
but the total number of flares at a given moment of time decreases
since $f$ decreases (as compared to the hard state). In GBHCs case,
however, the hard state almost borders the soft state, and thus the
spectrum is either hard in X-rays and hard in the sense of corona/disk
division of power, or it is soft in both these respects (see \S
\ref{sect:obgbhc}).

We have used the data listed in Tables 1 \& 2 of Walter \& Fink (1993)
to plot a phase portrait of Seyfert 1 Galaxies in Figure \ref{fig:wf}.
As an indication of the hard X-ray luminosity $\lx$ we have taken the
luminosity at 2 keV, found from 2 kev fluxes ($\nu F_{\nu}^{pl+tb}$ in
Walter \& Fink 1993). Similarly, the UV luminosity $\luv$ was
estimated using $\nu F_{1375}$ data. We excluded radio-loud sources
(i.e., sources \# 5, 23, 28, 45, 49, 50, 52 \& 55).  Notice that
higher luminosity sources appear to have larger $\luv/\lx$, which
qualitatively agrees with our theory. It is hard to see whether the
theory and the data agree well quantitatively, since the data have
large uncertainties (up to factor of $\sim $ few, not included in the
figure, since we do not know the bolometric corrections in any event),
and also we really need to plot the luminosities in terms of the
Eddington luminosity, which we have no way of knowing for a given AGN.

Another well-known observational fact for quasars is the correlation
between the optical to X-ray spectral slope $\alpha_{ox}$ and the
optical luminosity (Green et al. 1995), which should track the
bolometric luminosity well in the two-phase model where half of X-rays
are reprocessed in the UV range, and the disk thermal emission comes
out in that range as well. The optical to X-ray index $\alpha_{ox}$ is
not a real spectral index in this energy range, but is defined as the
index of an imaginary power-law connecting the observed optical and
X-ray emission.  Wilkes et al. (1994), and Green et al. (1995) show
that more luminous sources have larger $\alpha_{ox}$, i.e., more
luminous objects have comparatively less X-ray emission. This is again
in a qualitative agreement with our theory.

\subsection{The Reflection Component and The Iron Line}
\label{sect:rc_line}

The reflection component and the fluorescent iron line are always
present in the spectra of radio-quiet Seyfert Galaxies (Gondek et
al. 1996; Zdziarski et al. 1996; George \& Fabian 1991, and additional
references cited in Chapter 1). We will refer to the reflection
component as being ``cold'' if heavy metal ions are not strongly
ionized.  According to Zycki et al. (1994), the shape of the reflected
spectrum in indistinguishable from that produced by reflection off a
neutral cold medium, when the density ionization parameter $\xi\simlt
30$.  At around $\xi \simgt 100$, the X-ray albedo starts to increase.
Zycki et al. (1994) analyzed the {\it Ginga} data and fitted them with
their photoionized reflection model, which indicated that the quality
of the data is such that models with $0\leq \xi\simlt 200$ will
produce acceptable fits to the data. In other words, the data may not
accurately determine the ionization parameter in this range. Above
$\xi\sim 300$, however, the quality of the data is good enough to
distinguish between models with different values of the density
ionization parameter. Therefore, we will call the reflection off
ionized X-ray skin ``cold'' when $\xi\simlt 200$.

Matt, Fabian \& Ross (1993, 1996) investigated fluorescent iron line
emission for various ionization parameters, and came to the following
conclusions. For small ionization parameters ($\xi \simlt 100$), the
standard cold fluorescent line at 6.40 keV is produced. For $\xi
\simlt 100-500$, Auger destruction reduces the equivalent width of the
line to very low values. For $\xi \sim 500 - 5000$, Auger destruction
does not operate, so that ionized lines result (at 6.67 and 6.97 keV)
and are strong. For stronger ionization parameters, $\xi \simgt 5000$,
the iron is completely stripped and no fluorescent line is produced.
This situation is qualitatively similar to the appearance of the
reflection component with changing $\xi$. We will thus refer to the
line as being cold when $\xi\simlt 100$, ``ionized'' when $\xi \simgt
500$, and assume no line for ionization parameters larger than $\sim
5000$.

\begin{figure*}[t]
\centerline{\psfig{file=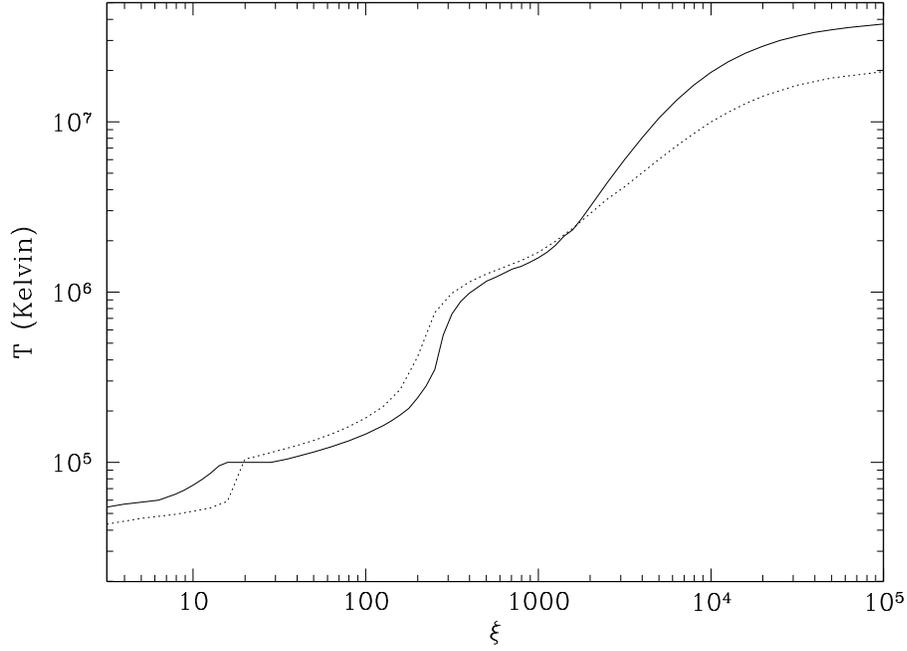,width=.8\textwidth,angle=-90}}
\caption{The transition layer temperature as a function of the
ionization parameter $\xi$ for the simulations shown in Figures
\ref{fig:agn3} (solid curve) and \ref{fig:agn1} (dotted).}
\label{fig:xi_agn}
\end{figure*}

In Figure (\ref{fig:xi_agn}) we plot the relation between the density
ionization parameter $\xi$ and the gas temperature for the simulations
shown in Figures (\ref{fig:agn3}) and (\ref{fig:agn1}). Several points
are to be noted. As discussed in Chapter 5, transition region in
typical Seyfert Galaxies resides on the lower equilibrium stable
branch of the solution, with $T\simeq 1-3\times 10^5$ K. Figure
(\ref{fig:xi_agn}) then shows that the fluorescent iron lines
corresponding to this temperature will be standard cold iron lines,
since $\xi$ ranges from few tens to $\sim 200$ for this state.  The
reflection component will also be cold, i.e., as if it resulted from a
neutral medium. This finding agrees well with the observations of
Seyferts (e.g., Zdziarski et al. 1996).

In the soft and very high states, the pressure ionization parameter
$\Xi$ may be large, because the gas pressure becomes a small fraction
of the intrinsic radiation pressure in the transition region (see \S
\ref{sect:xbaldwin}). In the soft parameter range, it may mean that
the transition layer will have a temperature $T$ above a 100 eV and
below the Compton temperature. Some preliminary tests with XSTAR
indicate that this region becomes {\it energetically} stable for a
steeper than standard $\Gamma \simeq 1.9$ hard Seyfert spectrum. The
pressure equilibrium for the transition layer has to be solved exactly
to determine the temperature of the transition layer in the soft
regime. Depending on the outcome of such a calculation, the iron line
and reflection component may be either ionized ($\xi \sim 500 - 5000$)
or be completely absent ($\xi
\simgt 5000$). The observational situation here is not clear, since the iron
line fitting for NLS1 galaxies is difficult at present due to the poor
statistics (see \S 5 of Brandt \& Boller 1998), but it seems that some
NLS1 do show evidence for ionized iron lines and reflection. For
example, Fiore et al. (1998) found a presence of a spectral feature
around $\sim 1$ keV in the spectrum of the steep X-ray spectrum quasar
PG1244+026. This feature seems to be common in such quasars (see
references in Fiore et al 1998). One of the possible explanations for
this feature (not present in typical Seyfert 1s), is that it is due to
reflection from a highly ionized accretion disk.

In the very high state, where the disk becomes even less dense, these
effects should be even more pronounced, and so we expect no iron line
and no reflection component. The latter is in agreement with Zdziarski
et al. (1995), who find no evidence for a reflection component in
Radio Loud Seyferts, that are commonly believed to accrete at higher
accretion rate than RQ Seyferts do (see also \S \ref{sect:indob}). The
absence of the fluorescent iron line is confirmed by observations of
the X-ray Baldwin effect (\S \ref{sect:xbaldwin}). Also, it is
interesting to note that the second X-ray steep quasar studied by
Fiore et al. (1998), NAB0205+024, did not show the spectral feature
around 1 keV. This second source showed about the same variability
time scale as PG1244+0264, but was a factor $\sim 10$ brighter in
X-rays. It is then possible that both sources have about the same
black hole mass (which likely is the factor setting variability time
scales, see \S \ref{sect:his}), but NAB0205+024's accretion rate $\dm$
is higher. If the latter source accretes in the very high state, then
the absence of the spectral feature around $1$ keV could be natural in
our theory, since we expect the reflector to be completely ionized in
this case.

\subsection{Is There A Big Blue Bump or not?}
\label{sect:obsbbb}

As enunciated in Chapter 5, the pressure ionization instability makes
the reflecting medium -- the transition layer -- in AGNs occupy either
the cold stable state, e.g., a narrow region in temperature $T\sim
1-3\times 10^5$ Kelvin, or a hotter completely ionized state.
Observations show that the transition layer picks the lower
temperature stable equilibrium, which is possible as long as the
ionizing X-ray flux exceeds the intrinsic disk flux. Thus, we expect
that hard and intermediate Seyferts will re-emit the X-rays within
this temperature range, with a complex spectrum somewhere between a
modified blackbody and blackbody emission, with a number of
recombination lines. We believe that this emission may explain the
observed Big Blue Bump of Seyfert Galaxies (see \S \ref{sect:obbb}).

However, recent work of Zheng et al. (1996) and Laor et al. (1997)
showed that quasars in their (different) samples do not show the steep
soft component below 2 keV. In other words, these authors have shown
that the BBB is not present in their samples, which represents a sharp
contrast with the findings of Walter \& Fink (1993), Walter et
al. (1994), Zhou et al. (1997) and Wang, Lu, \& Zhou (1998).  Here we
show how our theory may reconcile these observational findings with
one another.

As we detailed in Chapters 6 \& 7, typical Seyfert Galaxies should
accrete at relatively small rates, i.e., $\dm\simlt \dms$, which
correspond to a luminosity $L \simeq 5 \times 10^{44} M_8$ ergs/sec. In
the sample of Walter \& Fink (1993), very few objects have a UV
luminosity above a few $\times 10^{44}$ ergs/sec, whereas Zheng et
'sal. (1996) fit to the mean spectrum in their sample gives $L\simeq 8.5
\times 10^{45}$ ergs/sec. Similarly, almost all AGNs in the Laor et
al. (1997) sample have $L_{3000}> 10^{45}$ ergs/sec. Therefore, the
Walter
\& Fink (1993) sample contains Seyferts that are dimmer than sources
in the other two samples by factors of 10 to 100. Accordingly, sources
in the Walter \& Fink (1993) sample may accrete at a hard or
intermediate state $\dm \simlt \dms$, whereas AGNs of the two other
samples could accrete at the soft and very high state state.

As we already described in \S \ref{sect:rc_line}, for higher accretion
rates, i.e., for $\dm\simgt \dms$, the ionization instability will
drive the gas to higher temperatures, at least $\sim 10^6$ K, where
the transition layer energy equilibrium can become stable again.  The
emission spectrum from such a transition layer will probably peak at
least at a few $\times 10^6$ K, so that no emission will be observed
in the regular BBB energy window, i.e., around $50$ eV. Furthermore,
if the transition layer becomes completely ionized, {\it no}
observable emission due to spectral reprocessing will appear. Indeed,
if the reflecting medium is completely ionized, the reflection process
is given by Compton reflection only, which produces just a power-law
with an index basically equal to the incident X-ray index (e.g.,
\S 3.1 \& Fig. 1 in Zycki et al. 1994). This spectrum will be impossible
to disentangle from the coronal X-ray power-law up to $\sim 30$ keV,
where the Compton reflection component rolls over.

In addition, since in our theory the BBB is just the reflection of the
X-rays produced in magnetic flares off the surface of the accretion
disk, the hard Seyferts should have a prominent BBB, because a large
fraction of their bolometric luminosity comes out in X-rays. Seyferts
accreting in the intermediate range should have a BBB similar in shape
to hard Seyferts, but, since the fraction of energy produced by
magnetic flares (i.e., $f$) can be much smaller than unity, the
relative strengh of the bump may be much smaller. This could account
for the surprisingly large variation in the BBB normalization with no
obvious change in its shape from source to source in the Walter \&
Fink (1993) sample.

At the same time, the more distant and more luminous sources of Zheng
et al. (1996) and Laor et al. (1997) belong to the soft state, which
is characterized by a softer X-ray spectrum as well as a smaller
overall contribution of X-rays to the bolometric luminosity. Under
these conditions, not only is the shape of the BBB ``wrong'' to be
observed at $\sim 50 eV$, but the BBB normalisation is small as well,
so that the whole feature may be lost in the dominant accretion disk
thermal emission.  Note that this picture is also consistent with our
theory on the X-ray Baldwin effect.

\subsection{Observations of Individual Objects}\label{sect:indob}

Nandra et al. (1995) were able to study the X-ray emission of two
quasars at $z > 1$ in some detail (it is usually very difficult since
quasars are normally dim in X-rays compared to their UV emission).
They found that the X-ray spectrum of these two quasars was
substantially different from that in typical Seyfert 1 Galaxies. In
particular, the 13.6 eV to 300 keV integrated luminosities of these
two sources were only $\sim 0.15$ and $0.02$ of their respective
optical-UV luminosities, which were exceptionally high -- $2.4\times
10^{47}$ and $1.4\times 10^{48}$ erg/sec, whereas at least some
Seyfert Galaxies have broad-band X-ray luminosities comparable to
their optical-UV luminosity. These observations can be considered as a
confirmation of our theory which predicts comparatively less power in
X-rays for sources accreting at a very high $\dm$ (i.e., $f\ll 1$ for
such sources). Further, the optical-UV component in the two quasars
peaks and quickly rolls over at $E\sim 5 eV$, substantially below the
value than $\sim 50 eV$ that is appropriate for the BBB emission in
Seyferts. It is most likely that the optical-UV emission of these two
quasars is created by the accretion disk intrinsic emission. Notice
that in our theory, a similar component in typical Seyferts can be
smaller as well as larger than the BBB (reprocessed X-rays) component,
depending on the accretion rate. Further, Nandra et al. (1995) found
no reflection features or the fluorescent iron line emission in these
two quasars, which is consistent with our theory (\S\S
\ref{sect:xbaldwin} \&
\ref{sect:rc_line}).

\section{Iron Line Strength and the Temperature of the BBB}
\label{sect:ilbbb}

Poutanen, Svensson \& Stern (1997, PSS97) applied the iterative
scattering code of Poutanen \& Svensson (1996) to the two-phase corona
model for AGN. They tested different geometries for the active
regions, and concluded that slab coronae have difficulties reproducing
the observed distribution of X-ray spectral indexes and iron line
equivalent width (EW), and that localized active regions are therefore
favored by the data. Further, they showed that anisotropic scattering
effects in the corona are very important. In particular, they found
that the emission corresponding to the first scattering order is
highly anisotropic and directed mostly to the disk (see also Poutanen
\& Svensson 1996). This affects the EW of the iron K$\alpha$ line, if
the energy of the first scattering order is above the iron line
centroid energy, e.g., $6.4$ keV. PSS97 considered two values of the
accretion disk intrinsic emission temperature, $T_{\rm bb} = 5$ and
$50$ eV. The higher temperature case turned out to always produce
higher iron line EWs when compared to the lower $T_{\rm bb}$ case,
which was interpreted as being due to the fact that in the latter case
the first order scattering photons are not energetic enough to produce
Iron line fluorescence.

Observations of Seyferts show that the temperature in the active
regions should be $\simgt 150$ keV, since the spectral rollover is
needed above $\sim 200$ keV (Zdziarski et al. 1996). Figure 4 of PSS97
demonstrates that for the case with $T_{\rm bb} = 50 eV$, the AR
temperature needs to be $\sim 200-300$ keV in order to explain the
observed unexpectedly high EWs of $\sim 300$ or higher for some
Seyferts (e.g., Iwasawa et al. 1996). At the same time, with $T_{\rm
bb} = 5$ eV, this requisite temperature becomes larger than $500$ keV,
which is ruled out by the observed $\gamma$-ray spectra of Seyferts 1.

In the context of our work here, these results of PSS97 hint at an
interesting link between the iron line emission and the temperature of
the BBB, which can also be used as a check-point. Our modelling of the
X-ray reflection in AGN showed that the temperature of the reflecting
layer is likely to be $2-3\times 10^5$ Kelvin, which is not too much
lower than $50$ eV used by PSS97. In addition, note that the
reprocessed radiation from the transition layer is not exponentially
cutoff as a blackbody spectrum is, and includes recombination emission
at energies higher than $3 kT$. It is thus possible that those high
energy photons will be upscattered in the first scattering to high
enough energy to explain the high EWs of the Iron lines observed for
some Seyfert Galaxies. We believe that our results (in particular,
Chapter 5) combined with those of PSS97 represent a rather convincing
argument that the Big Blue Bump emission is intimately connected with
the X-ray reflection process near active magnetic flares, and not with
the disk intrinsic emission, since its effective temperature is too
low ($\sim 5$ eV).

\section{Low Mass AGNs}

We now wish to discuss the second entry in blackhole mass space, i.e.,
low mass AGNs, $M\sim 10^6 \msun$. The pressure ionization
instability, considered in Chapters 4 \& 5, makes those AGNs
special. Namely, the minimum temperature of the transition layer for
these sources is
\begin{equation}
T_{\rm min}\simeq 4.2\times 10^5\, l^{1/4} \,\alpha^{1/40} \,M_6^{-9/40}
\left[{\dm\over 0.005}\right]^{-1/20} \,\left(1-f\right)^{-1/40} \,
\left({q\over 10}\right)^{-1/4}\,
{\rm ,}
\label{tminlagn}
\end{equation}
(cf. equation \ref{tminsg}). Thus, the transition layer cannot reside
on the low temperature equilibrium state with $T\sim 2\times 10^5$ K,
and should climb to at least $T\simgt 10^6$ K. Without a better
treatment of the transition layer (i.e., with self-consistent
optically thick radiation transfer and the pressure balance), we
cannot predict whether the transition layer will saturate at the
island state with $T\sim 1-2\times 10^6$ K, or will go to the
completely ionized Compton equilibrium. In any event, we are confident
that the X-ray reflection and iron line formation are very different
in the low mass AGN as compared to higher mass Seyferts and Quasars,
which is an observationally testable prediction.

Of course, the observational difficulty here is the impossibility to
directly measure the AGN blackhole mass. However, if we consider nearby,
low luminosity Seyfert Galaxies, we might expect that their mass is
lower than that of more distant and luminous Seyferts.  There are
several potentially interesting observations one can mention here. 

NGC4151 is one of these low luminosity sources. Its X-ray $2-10$ keV
luminosity is variable within the range $(2-20) \times 10^{42}$
ergs/sec (Warwick et al. 1996) and no or little evidence for a
reflection component (Yaqoob et al. 1993). The iron fluorescent line
emission also seems to come from a ``slightly warm'' reflector, which
could be explained if the transition layer indeed occupied the island
state. NGC7172 (Ryde, Poutanen et al. 1997), seems to be similar to
NGC4151, with $L_{2-10}\,{\rm keV} \sim 10^{43}$ erg/sec and no
reflection component. Note that its X-ray spectral index is variable
and can at times be steeper than the typical Seyfert spectrum, which
could be explained if the transition layer is (as in the case of
GBHCs, see Chapter 4) highly ionized and thus the reprocessed spectrum
is hotter, which leads to a smaller cooling rate of the active
regions.

NGC4051 is another AGN deserving a mentioning here. It is a low
luminosity Seyfert Galaxy ($\nu F_{\rm 2 keV}^{pl+tb}$ given in Walter
\& Fink 1993 gives $L_x \sim $ few $\times 10^{42}$ erg/sec), and is the
only source in the Walter et al. (1994) sample that showed soft X-ray
variability on a time scale of one orbit. This source is also known to
exhibit quasi-periodic oscillations with a rather short period of
$\sim 1$ hour for an AGN (Papadakis \& Lawrence 1993, see also Iwasawa
et al. 1998). This faster than usual variability suggests that the
source is an atypically low mass AGN, which then explains why it is
rather dim as well.

NGC4051 was a distinctive source in the Walter et al. (1994) sample,
requiring special considerations when fitting its spectrum.  It is
remarkable that the lowest $\chi^2$ fit gave the cutoff energy of
$\simeq 150 eV$ (see Table 5 in Walter et al.), which is a factor of
$\sim 3$ higher than that in other Seyferts. Even the two other less
acceptable fits to this source clearly show an additional feature
around $\sim 0.5$ keV (see Fig. 2a and 2b in Walter et al. 1994), not
observed in other members of the sample. We believe that these
observations can be explained by our theory of the pressure ionization
instability, if we suggest that the transition layer (a whole or a
part of it) in this low mass AGN saturates at the island stable state
with $kT \sim 100-200 eV$.

\section{Comparison With Observations of GBHCs}\label{sect:obgbhc}

We now briefly compare our theory with observations of GBHCs.  A more
detailed comparison will be done in the future, when the transition
region structure is solved for using a better approximation. Further,
the observational situation with the reflection and iron line in GBHCs
is not as clear cut as for AGNs, so that observations may also need to
be improved to provide better constraints for the theory. As detailed
in Chapter 4, the broad-band spectra of GBHCs can plausibly be
explained by magnetic flares of the same compactness as in
Seyferts. Our preliminary conclusions are that the compactness
parameter is $l\simeq 0.1$, and the transition layer Thomson optical
depth is $2-3$. The iron line and edge are absent due to a complete
ionization of the X-ray skin, and the reflection component appears to
be weaker because it is more diffuse than the standard cold reflection
component typical for Seyferts.

One aspect of observations where Galactic sources provide very much
more valuable information than AGNs is the spectral states and
transitions, since some GBHCs have measured or well constrained BH
masses, which then tell us exactly where in $\dm$ space the
transitions happen. Grove, Kroeger \& Strickman (1997) and Grove
et. al. (1998) showed that GBHCs occupy at least four spectral states
in order of decreasing X-ray luminosity. In particular, the four
states can be classified as
\newline
(1) the {\it ultra-soft state}. When the X-ray luminosity (above $1$
keV) is at about the Eddington limit, the spectrum is dominated by the
so-called ultra-soft blackbody component with $kT\sim 1$ keV; a weak
hard tail is seen above $\sim 10$ keV, and rapid intensity variations
are present.
\newline 
(2) The {\it soft} state. At lower luminosities (typically $\sim 0.1
L_{\rm Edd}$), the spectrum again shows an ultra-soft component (with
$T$ somewhat lower than 1 keV) and a weak hard tail, but rapid
intensity variations are weak or absent.
\newline (3)
The {\it hard} state exhibits a single power-law spectrum with a
photon number index $\Gamma\sim 1.5-2$, and corresponds to a
luminosity in the range $10^{36-37.5}$ erg s$^{-1}$. 
\newline (4)
Finally, for the low luminosity, {\it quiescent} state, $L < 10^{34}$
erg s$^{-1}$, but its spectral shape is not very well known.

Note that the luminosity above 1 keV should be close to the disk's
bolometric luminosity since the temperature in the ultra-soft and soft
states is $1$ and $> 0.34$ keV, respectively (see Zheng et al. 1996),
and thus most of the blackbody power should lie above $1$ keV, whereas
in the hard state the intrinsic disk temperature can be much smaller,
i.e., $\simgt 0.1$ keV, but its contribution to the overall spectrum
is relatively small compared to that of the hard power-law (see, e.g.,
Gierlinski et al. 1997 and \S 4). Finally, several GBHCs have shown an
{\it intermediate state}, which has the luminosity and X-ray spectral
index between the values of these quantities for the hard and soft
state (references in Esin et al. 1997, astro-ph/9711167).

\begin{figure*}
\centerline{\psfig{file=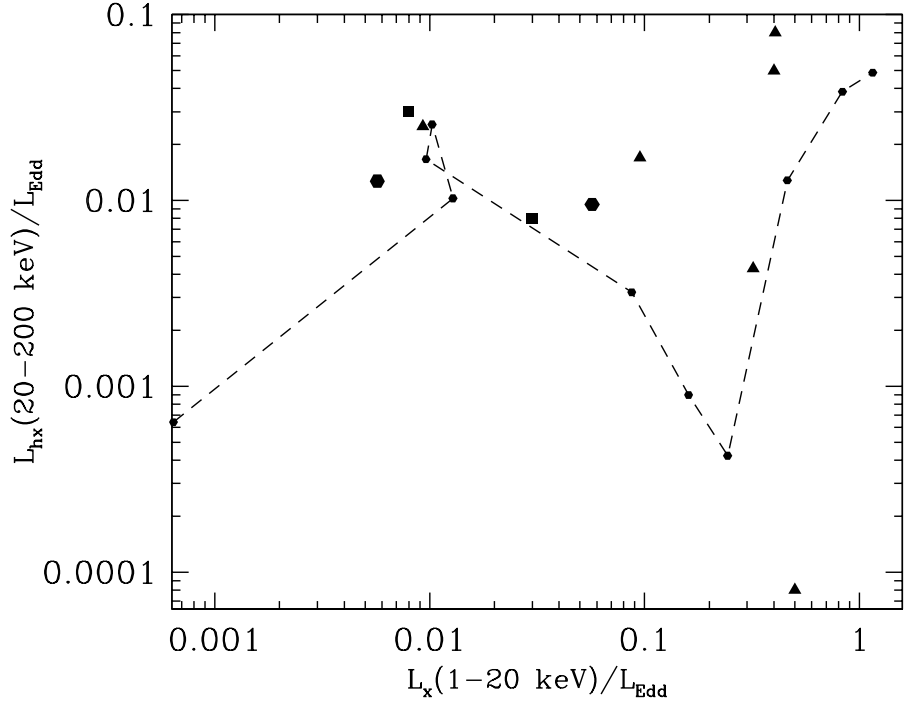,width=.8\textwidth,angle=0}}
\caption{The division of power between the hard X-ray luminosity 
$L_{\rm hx}$ (20 -- 200 keV) and the soft X-ray luminosity $L_{\rm x}$
(1 -- 20 keV) for GBHCs. Most of the data are from Barret et
al. (1996).}
\label{fig:gbhc_states}
\end{figure*}

Barret, McClintock \& Grindlay (1996) assembled a sample of GBHCs and
several other transient sources (neutron stars) in a $L_{\rm
hx}-L_{\rm x}$ phase space, where $L_{\rm hx}$ is the hard X-ray
luminosity in the range $20-200$ keV, and $L_{\rm x}$ is the X-ray
luminosity in the range $1-20$ keV. One of the striking results of
this exercise is that GBHCs always have relatively soft spectra when
they radiate at a high fraction of their Eddington luminosity.  Barret
et al. (1996) also plotted the evolutionary track of the GBHC
transient source GRS 1124-68 on the same $L_{\rm hx}-L_{\rm x}$ phase
diagram. During its evolution, this source occupied both a hard and a
soft state, and it spanned a broad range in luminosity. As found in
previous chapters, properties of accretion disks change with mass $M$
rather slowly, i.e., accretion disks with the same dimensionless
accretion rate $\dm$ but with different masses (by a factor of $\sim$
few, for example) should be very similar. Therefore, to test this
statement, we reproduce the data of Barret et al. in terms of $L_{\rm
x}/\ledd$ and $L_{\rm hx}/\ledd$ in Figure (\ref{fig:gbhc_states}),
adding some data for Cyg~X-1.  To find $\ledd$ for a given source, we
used estimates of the blackhole mass for that source given in Table 1
of Barret et al.

Figure (\ref{fig:gbhc_states}) shows that the spectra of GBHCs have
most of their power in the hard component up to $\dm \sim 0.04$, and
then there is a rather strong spectral transition. This is consistent
with our theory of magnetic energy transport, since the spectral
transitions take place exactly where the transition from the gas- to
the radiation-dominated disks should occur (within the uncertainties
in the exact value of $\dmr$, see equation \ref{rmc}). In other words,
the hard state of GBHCs corresponds to the most basic two-phase
corona-disk model. Further, notice that the hard state of GBHCs can
extend down to $\dm_d \sim 10^{-4}$ or lower, as long as the
assumptions of the two-phase model are satisfied. This seems to be
consistent with observations, since the hard state of GBHCs exists in
the luminosity range $10^{36-37.5}$ (see also the lowest luminosity
portion in the evolution of GRS 1124-68 in Figure
\ref{fig:gbhc_states}).

Above the gas- to radiation transition, our theory predicts the
existence of the intermediate state ($\dmr\leq \dm\leq\dms$). Note
that the exact value of $\dms$ -- the accretion rate where the X-ray
spectral index becomes steeper -- depends on the compactness parameter
$l$. For $l\simgt 20$, for example, the soft state would have never
been reached unless $\dm\simgt 1$. However, we should recall that the
compactness parameter is constrained from the X-ray reflection in
GBHCs (chapters 4 \& 5) to be of order $\sim 0.1$, or else the soft
X-ray part of the spectrum will not match observations. Now, if we use
that value in equation (\ref{dmsa}), we obtain that $\dms\simeq 0.06$,
e.g., for GBHCs the intermediate state is squeezed in the narrow
interval between the hard and the soft states. Practically, as soon as
the accretion rate becomes larger than the gas-to-radiation transition
value, the spectrum should be dominated by the disk emission {\it and}
the X-ray spectral index should increase, in line with observations of
GBHCs.

We will not try to discuss the quiescent and the ultra-soft states at
this time, due to uncertainties in the theory for these two regions.
In the case of the quiescent state, the compactness parameter may
become too low for the flares to be in the two-phase model set of
assumptions, and thus we would first need to develop a theory of
spectra formation for such magnetic flares. For the ultra-soft state,
the uncertainty in viscosity law and the accretion disk getting rather
thick (i.e., $H/R$ may approach 1) make the standard model unreliable,
and thus the theory developed here as well.

\subsection{Variability}

The issue of variability is a difficult and extensive one, since one
can consider variability in different energy ranges, correlation/lags
of one energy range with another, power density spectra, etc. We have
not yet considered these questions in sufficient detail, and thus will
only present a very sketchy discussion here.

Consider first the hard X-rays (i.e., produced intrinsically by the
magnetic flares). Due to spectral constrains (\S
\ref{sect:sconstraints}), the compactness parameter of magnetic flares
seems to be rather low, $l\sim 0.1-1$. This then implies that these
flares are relatively dim, so that to explain the observed X-ray
fluxes one needs as many as $\sim 10^3$ flares in the hard parameter
range (\S \ref{sect:number}), which means that the observed variations
of factor of $\sim 2$ in X-ray flux for many Seyfert Galaxies may not
be explained by statistical fluctuations in $N$. This implies that the
flares must behave in somewhat connected, global way.  This could
happen if a region of the accretion disk suddenly became very
efficient in producing flares. Alternatively, magnetic flux tubes that
break into the corona may have a quasi-stable configuration, and be
quiet until a little `push' makes them unstable. The push may be a
change in global magnetic field in the corona or a shock wave
there. In any event, there is a range of theoretical possibilities
here, and these need to be investigated in the future.

Another issue deserving considerable attention is the observed delays
of hard X-rays with respect to softer ones in GBHCs (e.g., Miyamoto,
et al. 1991, Kazanas \& Hua 1997). Whereas the observations of these
delays can be rather naturally explained by delays due to
Comptonization in an {\it extended} corona around the black hole (Hua
et al. 1997), we believe that the present theory may contain an
explanation as well. Our point here is that the observed (long) delays
between the hard and soft X-rays or other variability time scales do
not have to be directly related to the light crossing time of the
system.  This statement is true for magnetic flares because they are
controlled by the accretion disk, which is known to have several
variability time scales. For example, the disk thermal time scale is
considerably longer than one Keplerian rotation time scale (i.e., by
the factor of $\alpha^{-1}$; see, e.g., Frank et al. 1992). As we have
seen in Chapters 4$-$6, the accretion disk state influences the number
and properties of magnetic flares that are generated within the
disk. It is then likely that when a part of the disk produces a
``shot'' (consisting of many flares) in time history of the source,
the disk there is cooled rapidly by the loss of thermal disk energy to
the flares, so that the disk becomes cooler and thus more
gas-dominated, which should produce harder flares at a later time. In
such a scenario, the observed delays would be explained by the
spectral change during each shot.  We plan to investigate this
question in the future.

\section{Concluding Remarks}\label{sect:conclusions}

We have attempted to build a model of accretion disks with magnetic
flares acting as the main energy release mechanism. The accretion disk
structure was assumed to be close to the standard accretion disk
structure corrected for the additional energy out-flux (note that our
theory is weakly dependent on $\alpha$ through the estimate of the
compactness parameter of magnetic flares only). Magnetic flares are
schematically represented by active regions of the two-phase model,
i.e., these are regions of hot plasma with a Thomson optical depth of
order unity, that are heated by magnetic reconnection. Once the
Thomson optical depth is set (possibly by the mechanism described in
Chapter 3), the temperature of the active region is determined by the
balance between heating and cooling due to reprocessed soft radiation
from the disk. The compactness parameter of the active regions cannot
be determined from theoretical considerations at this time, and so is
chosen to be consistent with observations and some physical
constraints, i.e., $l\sim 0.1$.

We then considered the X-ray reflection in the accretion disk
atmosphere in the regions (which we called transition regions) close
to active magnetic flares. We found that there exists an ionization
instability, such that there are two stable solutions. For AGNs, both
solutions are possible. We further speculated, based on observations,
that in reality it is the low temperature solution that is applicable
to AGN. In this case, the temperature of the transition region is
within the range $1-3\times 10^5$ Kelvin, with an ionization parameter
$\xi$ of few tens to a hundred, consistent with the observations of
Seyfert 1 Galaxies.  The relatively narrow temperature range of the
cold stable solution appears to be in a good agreement with the
observed rollover energy in the BBB of Seyfert 1 Galaxies, and thus we
believe that the reprocessing of X-radiation from magnetic flares may
be the origin of the BBB. We also qualitatively showed that in AGNs
working at a high fraction of the Eddington accretion rate, the
transition layer may be forced to go into the hot completely ionized
stable state, which then accounts for the recently discovered
disappearance of the iron line in AGNs with an X-ray luminosity above
$\sim 10^{45}$ erg/sec. The same ionization instability may be
responsible for the absence of the BBB in the high luminosity AGNs.

We also found that for GBHCs, the ionization instability can saturate
only at the hot stable branch, which is characterized by a complete
ionization of heavy metals in the transition layer, so that it
achieves Compton equilibrium with the local radiation field. We
attempted to model the influence of this effect on the X-ray spectra
by introducing a completely ionized layer of material situated between
the hot corona (active region) and the cold disk below it, in contrast
with the usual {\it ad hoc} assumption that the disk surface is
cold. It was found that an increase in the Thomson depth of the
transition layer leads to an increase in the disk X-ray albedo, and
correspondingly harder X-ray spectra. While a future detailed
numerical modeling of the transition layer and the active region
spectra is needed to investigate parameter space of the problem
carefully, it is possible that {\it the same geometry and compactness}
of magnetic flares may explain the X-ray spectra of Seyfert 1 galaxies
and GBHCs.

The global energetics of the corona was discussed in Chapter 6, where
we tried to model the energy flux into the corona due to magnetic flux
tubes that rise out of the disk. We found that if magnetic fields in
the accretion disk are mostly diffuse, then the magnetic energy flux
can never be dominant over the radiation flux (and more realistically
is negligibly small, unless $\alpha$ is close unity). At the same
time, the spectra of GBHCs in the hard state, and at least some
Seyfert Galaxies require most of the radiation to be produced by
magnetic flares. We found that this observational fact can be
explained if most of the magnetic field in the disk is confined to
magnetic flux tubes, similar to the fields at the Solar surface. If
the magnetic field pressure (in the tubes) reaches a noticeable
fraction of the ambient pressure, then the tubes are in the ``solid
body limit'', so that they can avoid stretching by the differential
flow of the gas in the disk. This then reduces the disk viscosity
compared with the case of the same volume averaged diffuse field,
resulting in a larger disk optical depth and a smaller radiation flux.

In radiation-dominated disks, the radiation easily diffuses into the
flux tubes, and thus magnetic pressure may be only as large as the
{\it gas} pressure. As the accretion rate increases, the tubes then
lose their ability to resist stretching due to Keplerian differential
flow, and start to behave as diffuse fields. This means that the
magnetic field energy transport into the corona weakens, and the
overall accretion disk spectrum becomes dominated by the disk thermal
emission rather than by flares. We found a good agreement of this
picture with observations of Seyfert Galaxies and the state
transitions in GBHCs, which allowed us to classify the observed
spectral states of both types of objects in terms of the dimensionless
accretion rate.

We believe that our preliminary results are very encouraging, and that
they warrant further theoretical and observational studies of the
model and its predictions. We find a particular satisfaction in the
thought that the way in which accretion disks around massive and very
massive compact objects choose to produce X-rays may well be similar
to what happens on the Sun and other stars.

\section{Summary of Additional Graduate Work not Included in This Thesis}

Here we mention other work completed while at the graduate school at
the University of Arizona, which could have been included in this
dissertation for purposes of completeness, but was omitted due to
irrelevance to the project on accretion disks with magnetic flares.

During the summers of 1994 and 1995, I worked with Dr. Edward
E. Fenimore of the Los Alamos National Laboratory on constraints from
time histories of observed GRB spectra on the expanding shells in the
cosmological models of GRBs. We found that the expanding shells need
to be patchy, i.e., most of the shell area has to be non-emitting, or
else the time history will be rather smooth and will not reproduce the
observed chaotic variations (Fenimore, Madras \& Nayakshin 1996). This
presents an important constraint, since it increases the requisite
amount of energy from the GRB central source. We also found the
minimum Lorentz factor of the bulk motion that may be consistent with
{\it both} chaotic time histories and observations of photons with
energy up to several GeV. We concluded that the minimum
$\gamma$-factor is relatively low, i.e. $\sim 50-100$, even for some
extreme bursts. This result is to be reported in a future paper.

Until several years ago, it was believed that the X-ray/gamma-ray
spectra of all types of AGNs are best explained by non-thermal pair
cascades (e.g., Svensson 1994 and references there). However, the pair
distribution in such models was often treated approximately, as a sum
of a Maxwellian distribution and a power-law. To overcome this
deficiency, we worked out a Fokker-Plank approach to find the electron
distribution exactly (Nayakshin \& Melia 1998). The approach works for
time-dependent and static situations as well, and is highly efficient,
i.e., it allows one to solve for photon spectra and exact electron
distributions in several minutes on a modern workstation. We have
investigated the parameter space, i.e., what happens to the electron
distribution and spectra under different conditions.  We believe this
work has a considerable long-term value, since the analytical and
numerical methods developed are not tied to any particular situation,
and can be applied to many Astrophysical problems where the exact
particle distribution is crucial.

	\sloppy
        \bibliography{references.tex}
	\end{document}